\documentclass[paper,11pt]{JHEP}
\usepackage[centertags]{amsmath}
\usepackage{amsfonts} 
\usepackage{amssymb} 
\usepackage{amsthm}
\usepackage{graphicx}

           \let\p=\pi      
  \let\t=\tau      

     \let\L=\Lambda

\newcommand{\Z}{{\mathbb Z}}
\newcommand{\R}{{\mathbb R}}

\newcommand{\be}{\begin{equation}}
\newcommand{\eeq}{\end{equation}}
\newcommand{\bea}{\begin{eqnarray}}
\newcommand{\eea}{\end{eqnarray}}
\newcommand{\ba}{\begin{array}}
\newcommand{\ea}{\end{array}}

\newcommand{\ft}[2]{{\textstyle\frac{#1}{#2}}}
\newcommand{\eq}[1]{Eq.~(\ref{#1})}

\newcommand{\ii}{\mathrm{i}}
\newcommand{\ee}{\mathrm{e}}

\newcommand{\Tr}{\mathrm{Tr}\,}
\newcommand{\tr}{\mathrm{tr}\,}
\newcommand{\diag}{\mathrm{diag}\,}

\newcommand{\cK}{{\mathcal{K}}}
\newcommand{\cF}{{\mathcal{F}}}
\newcommand{\cG}{{\mathcal{G}}}
\newcommand{\cM}{{\mathcal{M}}}
\newcommand{\cN}{{\mathcal{N}}}

\newcommand{\cT}{{\mathcal{T}}}
\newcommand{\cV}{{\mathcal{V}}}
\newcommand{\cW}{{\mathcal{W}}}

\newcommand{\re}{\mbox{Re}\,}

\newcommand{\one}{{\rm 1\kern -.9mm l}}
\newcommand{\bone}{\mathbf{1}}
\newdimen\tableauside\tableauside=1.0ex
\newdimen\tableaurule\tableaurule=0.4pt
\newdimen\tableaustep
\def\sp#1#2{\big[^{#1}_{#2}\big]}
\def\phantomhrule#1{\hbox{\vbox to0pt{\hrule height\tableaurule
width#1\vss}}}
\def\phantomvrule#1{\vbox{\hbox to0pt{\vrule width\tableaurule
height#1\hss}}}
\def\sqr{\vbox{%
 \phantomhrule\tableaustep
\hbox{\phantomvrule\tableaustep\kern\tableaustep\phantomvrule\tableaustep}%
 \hbox{\vbox{\phantomhrule\tableauside}\kern-\tableaurule}}}
\def\squares#1{\hbox{\count0=#1\noindent\loop\sqr
 \advance\count0 by-1 \ifnum\count0>0\repeat}}
\def\tableau#1{\vcenter{\offinterlineskip
 \tableaustep=\tableauside\advance\tableaustep by-\tableaurule
 \kern\normallineskip\hbox
   {\kern\normallineskip\vbox
     {\gettableau#1 0 }%
    \kern\normallineskip\kern\tableaurule}%
 \kern\normallineskip\kern\tableaurule}}
\def\gettableau#1 {\ifnum#1=0\let\next=\null\else
 \squares{#1}\let\next=\gettableau\fi\next}
\newcommand{\Yfund}{\tableau{1}}
\newcommand{\Ysymm}{\tableau{2}}
\newcommand{\Yasymm}{\tableau{1 1}}
\newcommand{\reps}[4]{(\mathbf{#1},\mathbf{#2},\mathbf{#3},\mathbf{#4})}
\newcommand{\repst}[3]{(\mathbf{#1},\mathbf{#2},\mathbf{#3})}
\newcommand{\Smod}{S_{\rm mod}}

\title{Stringy instanton corrections to $\mathcal N=2$ gauge couplings}
\author{\parbox{13.5cm}{Marco Bill\`o$^1$, Marialuisa Frau$^1$,
Francesco Fucito$^2$, Alberto Lerda$^3$, Jose F. Morales$^2$
and Rubik Poghossian$^{2,4}$}
\\
~\\
~\\
$^1$Dipartimento di Fisica Teorica, Universit\`a di Torino\\
and I.N.F.N. - sezione di Torino \\
Via P. Giuria 1, I-10125 Torino, Italy\\
\vspace{0.3cm}
$^2$I.N.F.N. - sezione di Roma Tor Vergata
\\
Via della Ricerca Scientifica, I-00133 Roma, Italy\\
\vspace{0.3cm}
$^3$Dipartimento di Scienze e Tecnologie Avanzate, Universit\`a del Piemonte
Orientale\\
and I.N.F.N. - Gruppo Collegato di Alessandria - sezione di Torino\\
Viale T. Michel  11, I-15121 Alessandria, Italy\\
\vspace{0.3cm}
$^4$Yerevan Physics Institute, Alikhanian Br. 2,
0036 Yerevan, Armenia\\
\vspace{0.5cm}
\email{billo,frau,lerda@to.infn.it;
fucito,morales,poghosyan@roma2.infn.it}
}

\abstract{We discuss a string model where a conformal four-dimensional
$\mathcal{N}=2$ gauge theory receives corrections to its gauge kinetic functions
from ``stringy'' instantons. These
contributions are explicitly evaluated by exploiting the localization
properties of the integral over the stringy instanton moduli space. The model we
consider corresponds to a setup with D7/D3-branes in type I$^\prime$ theory
compactified on $\cT_4/\mathbb{Z}_2\times \cT_2$, and possesses a perturbatively computable 
heterotic dual. In the heteoric side the corrections to the quadratic gauge 
couplings are provided by a 1-loop threshold computation and, under the duality map, 
match precisely the first few stringy instanton effects in the type I$^\prime$ setup. 
This agreement represents a very non-trivial test of our approach to
the exotic instanton calculus.}
\preprint{DFTT/03/2010\\ROM2F/2010/02}

\begin{document}

\section{Introduction and motivations}
\label{sec:intro}

It has been recently found \cite{Blumenhagen:2006xt}-\nocite{Ibanez:2006da}\cite{Florea:2006si}
that certain classes of D-brane instantons 
arising in intersecting brane models can generate effective interactions 
at energies that are not linked to the gauge theory scale, and 
for this reason they are usually called ``stringy'' or ``exotic'' instantons.
This feature is very welcome in
the search of semi-realistic string scenarios for the TeV physics, where a
hierarchy between various Majorana masses and Yukawa couplings is expected. 
It is therefore of the greatest importance to devise efficient and reliable techniques 
to determine quantitatively such exotic non-perturbative corrections through
their explicit realization at the string level. This consideration is one of the main
motivations behind the present work. 

In Refs.~\cite{Argurio:2007vqa,Bianchi:2007wy} explicit models with
stringy instantons were constructed; since then much work has
been done extending and exploiting these results \cite{Ibanez:2007rs}-\nocite{Blumenhagen:2007zk,Aharony:2007pr,Blumenhagen:2007bn,Camara:2007dy,Ibanez:2007tu,Petersson:2007sc,Blumenhagen:2007sm,Cvetic:2008ws,Billo':2008sp,Billo':2008pg,Ibanez:2008my,
Argurio:2008jm,Angelantonj:2009yj,Bianchi:2009bg,Buican:2008qe,Billo':2009gc,Billo:2009di,Fucito:2009rs}\cite{Petersson:2010qu} (for a recent 
exhaustive review on the subject see Ref.~\cite{Blumenhagen:2009qh}).
Even if the effects of exotic and gauge instantons are quite different from each other,
in both cases they can be obtained from 
Euclidean branes entirely wrapping some cycle of the internal space. Depending on 
whether this cycle coincides or not with the one wrapped by the space-filling
D-branes on which the gauge theory is defined, such Euclidean branes correspond 
to gauge or exotic instantons, respectively.

In the simplest cases, four-dimensional gauge instantons can be realized with
bound states of space-filling D3-branes and point-like D(--1)-branes (or
D-instantons) 
\cite{Witten:1995gx,Douglas:1995bn}. Indeed, in these systems
there are four directions in which the string coordinates may have mixed
Neumann-Dirichlet (ND) boundary conditions, and
the massless sector of open strings
having at least one endpoint on the D(--1)'s is in one-to-one
correspondence with the moduli (positions, sizes and gauge orientations) of the
four dimensional gauge instanton solution.
Actually, also the effective action on the moduli
space, the rules of the instanton calculus and the profile of the classical solution
can be explicitly obtained using D(--1)/D3-brane systems 
\cite{Green:1997tv}-\nocite{Green:1998yf,Green:2000ke}\cite{Billo:2002hm}.

In the exotic cases, the gauge and instantonic branes intersect non-trivially in the internal
space or carry different magnetic fluxes, and the open strings stretching between
them have extra ``twisted'' directions besides the four ND ones along the
space-time. This twist lifts some of their massless excitations,
and some instanton moduli (specifically those related to sizes and gauge orientations)
disappear from the spectrum. Their supersymmetric fermionic partners remain
massless though, and when integrated out they can, under certain conditions,
lead to the effective interactions we alluded to above.

A very simple example of this phenomenon occurs in the D(--1)/D7 brane system,
which exhibits the world-sheet features of exotic instantons since
mixed open strings have eight ND directions.
By adding O7-planes, this system can be embedded in type I$^\prime$ string theory
compactified on a 2-torus $\cT_2$, 
a setup which possesses a computable perturbative heterotic
dual
\cite{Berkooz:1996iz}-\nocite{Kiritsis:1997hf,Aldazabal:1997wi,Lerche:1998nx,
Gutperle:1999xu,Gava:1999ky}\cite{Kiritsis:2000zi}. 
If the D7-branes are distributed democratically over the four orientifold fixed points
on $\cT_2$, they support a maximally supersymmetric gauge
theory in eight dimensions with gauge group SO(8). In this gauge theory
a D(--1)-brane represents a non-perturbative point-like configuration that
has been recently identified \cite{Billo':2009gc} with
the zero-size limit of the eight-dimensional octonionic instanton 
solution found long ago in Refs.~\cite{Grossman:1984pi,Grossman:1989bb}.

The non-perturbative contributions of D-instantons to the effective action on
the D7-branes can be explicitly computed as integrals over the moduli space via
localization techniques, in analogy with what is done for usual gauge instantons
\cite{Nekrasov:2002qd}, though with an exotic moduli spectrum. All D-instanton
numbers correct the quartic gauge couplings of the eight-dimensional gauge
theory \cite{Billo:2009di}, and this whole series of terms can be compared to
those obtained in the dual heterotic string theory, where they correspond to
world-sheet instantons describing the wrapping of the heterotic string on
$\cT_2$ \cite{Bachas:1997mc,Bachas:1997xn}. 
The success of this comparison provides a very non-trivial check of both the
type I$^\prime$/heterotic duality and the correctness of this approach to the
exotic instanton calculus \cite{Billo:2009di}. Similar techniques can be used
also in non-conformal settings and for exotic instantons with fewer number of
super-symmetries \cite{Fucito:2009rs}, although the heterotic counterpart of the
induced interactions in these cases is far from clear (see also
Ref.~\cite{Petersson:2010qu} for related recent work).

An interesting feature of these eight-dimensional gauge theories is their
similarity with the four-dimensional ${\mathcal N}=2$ super Yang-Mills theories:
indeed, the eight-dimensional prepotentials and the correlators of the chiral
ring satisfy Matone-type relations for arbitrary $\mathrm{SO}(N)$ gauge groups
\cite{Fucito:2009rs}; this observation points to the existence of some direct
relation between the eight-dimensional effective action and some underlying
Seiberg-Witten curve, connected presumably to an F-theory description (see, for
example, Refs.~\cite{Sen:1996vd,Banks:1996nj,
Lerche:1998nx} for earlier results
in this direction).

In this paper we investigate the exotic calculus in a four-dimensional setup. We
consider a perturbatively conformal $\cN=2$ gauge theory that, on the
one hand, admits a brane realization where exotic instantons generate a
whole series of corrections to the quadratic gauge couplings, while on the
other hand it possesses a calculable heterotic dual against
which these corrections can be checked (see \cite{Bianchi:2007rb} for a recent
test of  
four fermionic couplings in the six-dimensional version of this type I/heterotic
dual pair). 
This allows to provide a test of the exotic instanton calculus as reliable as
the eight-dimensional one described above, but in a four-dimensional context.

The gauge theory we consider is realized on the world-volume of D7-branes at an
O7 fixed-point within a D7/D3-brane system of type I$^\prime$ compactified on
$\cT_4/\Z_2\times \cT_2$. This is a T-dual variant of the first example of a
consistent ${\mathcal N}=2$ open string compactification in which all tadpoles
cancel \cite{Bianchi:1990tb,Gimon:1996rq}. In Section~\ref{sec:Imodel} we
describe in detail the four-dimensional model, which actually admits different
realizations corresponding to different consistent distributions of branes, and
show how the conformal $\cN=2$ theory we are interested in arises. Then, we
determine the holomorphic quadratic gauge couplings of the low-energy effective
theory \cite{Dixon:1990pc}-\nocite{Kaplunovsky:1995jw}\cite{deWit:1995zg}
starting, in Section \ref{sec:prepI}, with the perturbative terms (limited to
1-loop by supersymmetry). 
The theory, however, admits also non-perturbative corrections produced by brane
instantons. These can be Euclidean 3-branes wrapped on $\cT_4/\Z_2$, namely the
same cycle wrapped by the D7-branes supporting the gauge theory, or
D-instantons. In the first case, they correspond to ordinary gauge instantons
and might yield corrections weighted by powers of $\exp(-{8\pi^2}/{g^2})$, where
$g$ is the Yang-Mills coupling. The D-instanton corrections, instead, are
weighted by powers of
$\exp(-{\pi}/{g_{\mathrm{s}}})=\exp(-{4\pi^2}/{g^2\,\cV_4})$, where
$g_{\mathrm{s}}$ is the string coupling and $\cV_4$ the volume of $\cT_4/\Z_2$;
they represent non-perturbative exotic contributions which are the subject of
the analysis in Sections \ref{sec:Dmod} and \ref{sec:loc}. In particular, in
Section \ref{sec:Dmod} we show that the spectrum of moduli supported by
D-instantons is such that they can affect the quadratic gauge couplings of the
D7-branes, and in Section \ref{sec:loc} we compute these corrections by carrying
out the integrations over the exotic moduli space by means of localization
techniques analogous to those used for ordinary instanton calculus; the formulas
are rather involved, but we have been able to get explicit results up to $k=3$
D-instantons. Section \ref{sec:prephet} introduces the heterotic dual model and
describes the computation of the 1-loop thresholds from which the holomorphic
quadratic gauge couplings can be deduced. Upon using the duality map, we show
that the type I$^\prime$ and the heterotic results perfectly agree. We take this
as a highly non-trivial test of the correctness of our D-instanton computation.
A summary of our main findings and some considerations regarding possible
developments can be found in the conclusive Section \ref{sec:concl}. Finally, in
the six appendices we have gathered many technical results needed to reproduce
the computations in the main part of the paper.

\section{A $\mathcal{N}=2$ conformal model from an orbifold of type I$^\prime$}
\label{sec:Imodel}
We consider type IIB string theory compactified on a 6-torus
$\cT_2^{(1)}\times \cT_2^{(2)} \times \cT_2^{(3)}$ 
and modded out by $\mathbb Z_2 \times \mathbb Z_2$ where the generators of the
two $\Z_2$ groups are
\begin{equation}
\Omega' = \Omega\,(-1)^{F_L}\,I^{(3)}~~~\mbox{and}~~~
\hat g = I^{(1)}I^{(2)}~,
\label{parities}
\end{equation}
with $\Omega$ the word-sheet parity, $F_L$ the space-time left-fermion number 
and $I^{(i)}$ the reflection along the coordinates of $\cT_2^{(i)}$.
This compactification preserves eight supercharges, {\it i.e.} $\mathcal N=2$ 
supersymmetry in four dimensions.

Type IIB string theory compactified on $\cT_2^{(3)}$ and modded out by $\Omega'$
is usually called type I$^\prime$ and is dual to a torus compactification of the
heterotic SO(32) string with Wilson lines breaking the gauge group to
$\mathrm{SO}(8)^4$. For this set-up, the D-instanton corrections to the
\emph{quartic} gauge prepotential on D7-branes were computed in
Ref.~\cite{Billo:2009di} and checked against the dual heterotic results
\cite{Lerche:1998nx,Gutperle:1999xu,Gava:1999ky,Kiritsis:2000zi}, finding
perfect agreement. In this paper we consider instead a K3 compactification of
the type~I$^\prime$ theory in the orbifold limit represented by
$(\cT_2^{(1)}\times \cT_2^{(2)})/\mathbb Z_2$, where $\mathbb Z_2$ is generated
by $\hat g$, and analyze the \emph{quadratic} gauge couplings on stacks of
D7-branes. The compactification of the unoriented string on a $\cT_4/\mathbb
Z_2$ orbifold was considered long ago in
Refs.~\cite{Bianchi:1990tb,Gimon:1996rq}, and the global constraints imposed by
the tadpole cancellation condition were solved in that case. Thus, upon
compactification on $\cT_2^{(3)}$, our present set-up can be seen as the T-dual
version of that model, for which the quadratic gauge couplings on D9-branes were
recently considered in Ref.~\cite{Camara:2008zk}.

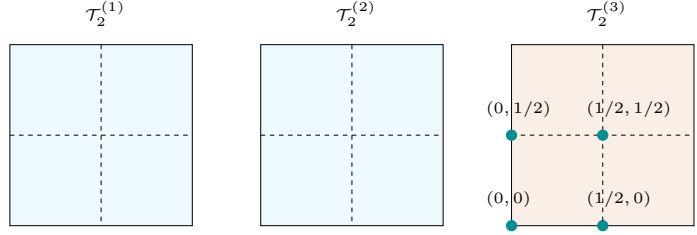
\begin{figure}[htb]
\begin {center}
 \begin{picture}(0,0)%
\includegraphics{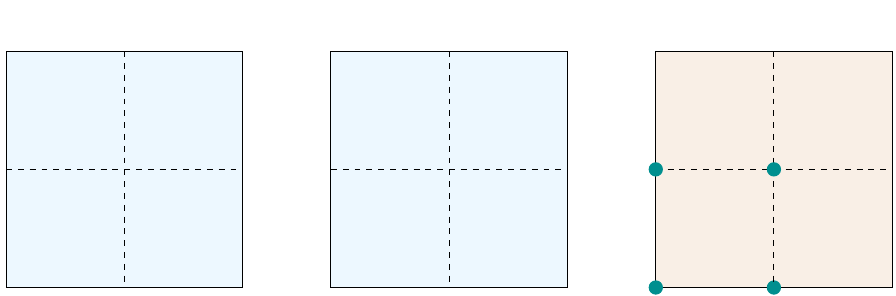}%
\end{picture}%
\setlength{\unitlength}{1243sp}%
\begingroup\makeatletter\ifx\SetFigFont\undefined%
\gdef\SetFigFont#1#2#3#4#5{%
 \reset@font\fontsize{#1}{#2pt}%
 \fontfamily{#3}\fontseries{#4}\fontshape{#5}%
 \selectfont}%
\fi\endgroup%
\begin{picture}(13613,4466)(585,-4002)
\put(2161,209){\makebox(0,0)[lb]{\smash{{\SetFigFont{6}{7.2}{\familydefault}{
\mddefault}{\updefault}$\cT_2^{(1)}$}}}}
\put(7111,209){\makebox(0,0)[lb]{\smash{{\SetFigFont{6}{7.2}{\familydefault}{
\mddefault}{\updefault}$\cT_2^{(2)}$}}}}
\put(12061,209){\makebox(0,0)[lb]{\smash{{\SetFigFont{6}{7.2}{\familydefault}{
\mddefault}{\updefault}$\cT_2^{(3)}$}}}}
\put(12061,-1636){\makebox(0,0)[lb]{\smash{{\SetFigFont{6}{7.2}{\familydefault}{
\mddefault}{\updefault}$(1/2,1/2)$}}}}
\put(10081,-1636){\makebox(0,0)[lb]{\smash{{\SetFigFont{6}{7.2}{\familydefault}{
\mddefault}{\updefault}$(0,1/2)$}}}}
\put(10081,-3436){\makebox(0,0)[lb]{\smash{{\SetFigFont{6}{7.2}{\familydefault}{
\mddefault}{\updefault}$(0,0)$}}}}
\put(12061,-3436){\makebox(0,0)[lb]{\smash{{\SetFigFont{6}{7.2}{\familydefault}{
\mddefault}{\updefault}$(1/2,0)$}}}}
\end{picture}%
\end{center}
\caption{The location of the 4 O7-planes in $\cT_2^{(3)}$ is
 identified by a 2-vector $\vec\alpha $ whose components can take the values
 $0$ and $1/2$, if the torus is parameterized with ``flat'' coordinates ranging
 from $0$ to $1$ (see Appendix \ref{app:notations} for our notations and
 conventions).}
\label{fig:O7}
\end{figure}

The action of $\Omega'$ selects 4 O7-planes, located at the invariant points of
the torus $\cT_2^{(3)}$ with respect to the $I^{(3)}$ reflection. These
points are labeled by a 2-vector $\vec\alpha$ as indicated in Fig.~\ref{fig:O7}.

\begin{figure}[htb]
\begin {center}
 \begin{picture}(0,0)%
\includegraphics{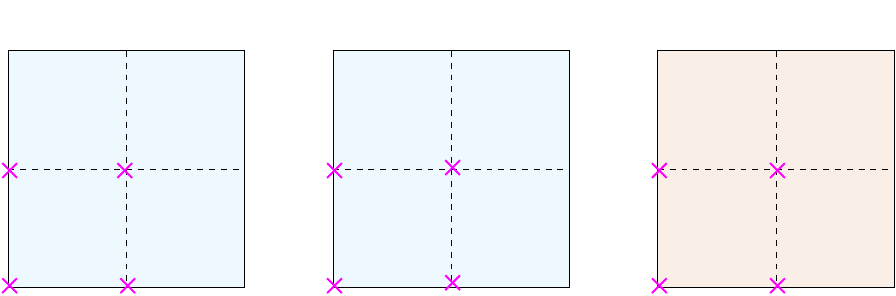}%
\end{picture}%
\setlength{\unitlength}{1243sp}%
\begingroup\makeatletter\ifx\SetFigFont\undefined%
\gdef\SetFigFont#1#2#3#4#5{%
 \reset@font\fontsize{#1}{#2pt}%
 \fontfamily{#3}\fontseries{#4}\fontshape{#5}%
 \selectfont}%
\fi\endgroup%
\begin{picture}(13645,4473)(553,-4009)
\put(2161,209){\makebox(0,0)[lb]{\smash{{\SetFigFont{6}{7.2}{\familydefault}{
\mddefault}{\updefault}$\cT_2^{(1)}$}}}}
\put(7111,209){\makebox(0,0)[lb]{\smash{{\SetFigFont{6}{7.2}{\familydefault}{
\mddefault}{\updefault}$\cT_2^{(2)}$}}}}
\put(12061,209){\makebox(0,0)[lb]{\smash{{\SetFigFont{6}{7.2}{\familydefault}{
\mddefault}{\updefault}$\cT_2^{(3)}$}}}}
\end{picture}%
\end{center}
\caption{The location of the 64 O3-planes in $\cT_2^{(1)}\times \cT_2^{(2)}
\times \cT_2^{(3)}$
is identified by a 6-vector $\vec\xi$ whose components again take values $0$ or
$1/2$.} 
\label{fig:O7O3}
\end{figure}
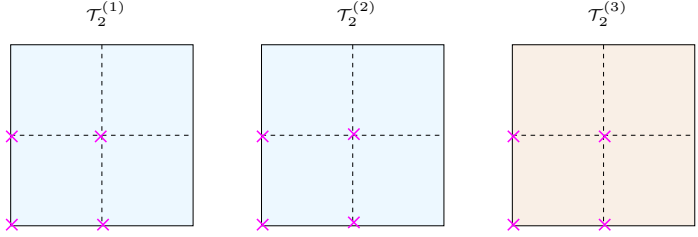

Similarly, $\Omega' \hat g$ preserves 64 O3-planes, located 
at the fixed points of the inversions in all three
tori $\cT_2^{(i)}$ which we will denote by a 6-vector $\vec\xi$ (see
Fig.~\ref{fig:O7O3}).

The (dimensionless) volume $\cV$ of the internal compactification manifold is
given by
\begin{equation}
\label{vol6}
\cV = T_2^{(1)} T_2^{(2)} T_2^{(3)}~,
\end{equation}
where $T_2^{(i)}$ is the K\"ahler modulus%
\footnote{As usual, the K\"ahler moduli $T_2^{(i)}$ 
are complexified into $T^{(i)}=T_1^{(i)}+\ii T_2^{(i)}$ by the
B-field along the $i$-th torus; see Appendix \ref{app:notations} for our
conventions.}
of the torus $\cT_2^{(i)}$, whose complex structure we denote by $U^{(i)}$. 
Since the third torus plays a distinguished r\^ole, in the following
we will write simply $T$ and $U$ in place of $T^{(3)}$ and $U^{(3)}$.
The low-energy effective super-gravity action for the above orientifold
compactification is best expressed in terms of $U^{(i)}$ and of the complex
fields 
$t^{(i)}$, whose imaginary parts $t_2^{(i)}$ are given by%
\footnote{The real parts $t_1^{(i)}$ are related, instead, to suitable RR
potentials.}  
\cite{Grimm:2004uq,Lust:2004fi}
\begin{equation}
\label{deft}
 t_2^{(1)}= \ee^{-\phi_{10}}\,T_2^{(2)} T_2~,~~~
 t_2^{(2)}= \ee^{-\phi_{10}}\,T_2^{(1)} T_2~,~~~
 t_2 \equiv  t_2^{(3)}= \ee^{-\phi_{10}}\,T_2^{(1)}T_2^{(2)}~,
\end{equation}
where $\phi_{10}$ is the ten-dimensional dilaton.

The four-dimensional Planck mass $M_{\mathrm{Pl}}$, which represents
the natural UV cut-off in the low-energy effective theory, is
\begin{equation}
M_{\mathrm{Pl}}^2=\frac{1}{\alpha'}\,\ee^{-2\phi_{10}}\,\mathcal{V}
=\frac{1}{\alpha'}\,t_2\,\lambda_2\,T_2~,
\label{mpl1}
\end{equation}
where
\begin{equation}
\label{axiodil}
\lambda = C_0 + \ii \ee^{-\phi_{10}} \equiv \lambda_1 + \ii \lambda_2
\end{equation}
is the usual axio-dilaton field. 
The tree-level bulk K\"ahler potential $K$ of our theory can be written as 
\begin{equation}
\label{ktotI}
K = -\log\big(\lambda_2\big) -\sum_{i=1}^3\log\big(t_2^{(i)}\,U_2^{(i)}\big)~.
\end{equation}
As we will briefly recall in the next subsection, the cancellation of the RR
tadpoles requires the
presence of D7-branes transverse to $\cT_2^{(3)}$ and of
D3-branes transverse to the internal 6-torus, with a specific action of
$\Omega'$, $\hat g$ and $\Omega' \hat g$ on their Chan-Paton (CP) factors.
In this framework the modulus $t_2$ 
defined in (\ref{deft}) basically corresponds to the tree-level coupling of the
gauge theory on the 
D7-branes, while $\lambda_2=\ee^{-\phi_{10}}$ describes the gauge coupling on
the D3-branes.
Notice that the orientifold projections (\ref{parities}) 
are compatible also with D-instantons and Euclidean E3-branes wrapped
on $\cT_2^{(1)}\times \cT_2^{(2)}$, which must therefore be added to our model
giving rise to non-perturbative corrections.

\subsection{Tadpole cancellation constraints}
\label{subsec:tad}
Let us denote the number of D7-branes in each fixed point $\vec\alpha$ by
$N_{\vec\alpha}$, and the number of D3-branes in each fixed
point $\vec\xi$ by $M_{\vec\xi}$.  Open string states connecting the various
branes will be described by CP matrices with index 
range $N_{\vec\alpha}$ or $M_{\vec\xi}$ depending on whether the string ends on
a D7- or on
a D3-brane respectively.
The $\mathbb Z_2\times \mathbb Z_2$ generators (\ref{parities}) act on these CP
indices by means of unitary matrices $\gamma$. More precisely, we denote the
$(N_{\vec\alpha}\times N_{\vec\alpha})$ matrix
representing the generator $\hat g$ on the D7-branes at the fixed
point $\vec\alpha$ by $\gamma_{\vec\alpha}(\hat g)$, and use the same
notation, \emph{mutatis mutandis}, for the other orbifold generators and for the
CP
indices of the D3-branes. All these unitary matrices square to the identity 
since they represent $\mathbb Z_2$ generators, and thus for all of them we have
\begin{equation}
\label{C}
\gamma^{-1} = \gamma~,~~~
\gamma^* = \gamma^T~.
\end{equation}
Moreover, in any representation (both on the D7's and on the D3's), the group
relations
require that
\begin{equation}
\label{gamgr}
\gamma(\Omega') \, \gamma(\hat g) = \gamma(\Omega'\hat g)~.
\end{equation}

The tadpole constraints arise from the analysis of the IR divergences
in the exchange channel of the Klein bottle amplitude and
of the annuli and M\"obius diagrams with boundaries on D7- and/or on D3-branes. 
In the RR sector such divergences signal the propagation of massless
RR forms, and hence the presence of unphysical tadpoles that should be canceled
globally for consistency. In our model (see App.~\ref{subapp:tadpoles} for
details) this
cancellation is achieved if, following Ref.~\cite{Gimon:1996rq}, we take the 
$\big(N_{\vec\alpha} \times N_{\vec\alpha}\big)$ matrices $\gamma_{\vec\alpha}$ 
to be of the form 
\begin{equation}
\gamma_{\vec\alpha}(\Omega') = \begin{pmatrix} ~\one~ & ~0~ \\ 
~0~& ~\one~\end{pmatrix}~,~~~
\gamma_{\vec\alpha}(\hat g) =\gamma_{\vec\alpha}(\Omega'\hat g)=
\begin{pmatrix} ~0~ & \ii\,\one \\ 
-\ii\,\one
& ~0~\end{pmatrix}~,
\label{gamma7}
\end{equation}
the $\big(M_{\vec\xi} \times M_{\vec\xi}\big)$
matrices $\gamma_{\vec\xi}$ to be of the form 
\begin{equation}
\gamma_{\vec\xi}(\Omega')
= \gamma_{\vec\xi}(\hat g) =\begin{pmatrix} ~0~ & \ii\,\one \\ 
-\ii\,\one
& ~0~\end{pmatrix}~,~~~
\gamma_{\vec\xi}(\Omega'\hat g) = \begin{pmatrix} ~\one~ & ~0~ \\ 
~0~& ~\one~\end{pmatrix}~,
\label{gamma3}
\end{equation}
and then if we require that
\begin{equation}
\label{global}
\sum_{\vec \alpha} N_{\vec\alpha} =32~~~~\mbox{and}~~~~
\sum_{\vec\xi  } M_{\vec\xi}=32~.
\end{equation}
When these conditions are satisfied, all RR tadpoles are canceled globally. 
However, it is possible also enforce a more stringent constraint and
\emph{locally} cancel 
the RR charge carried by each O7-plane if we require that
\begin{equation}
\label{local7}
N_{\vec\alpha} =8~,
\end{equation}
{\it i.e.} if we place exactly 4 dynamical%
\footnote{Here we follow the same terminology introduced in
Ref.~\cite{Berkooz:1996iz}. Therefore,
when the D7-brane CP indices take $N_{\vec\alpha}$ values, we say that there are
$N_{\vec\alpha}/2$ {dynamical} D7-branes since half of the CP indices can be
regarded as images of the others under the orientifold parity $\Omega'$.
Likewise, when the D3-brane CP indices
take $M_{\vec\xi}$ values, we say that there are $M_{\vec\xi}/2$ 
{dynamical} ``half'' D3-branes since a further half of the CP indices can be
regarded
as images under the orbifold parity $\hat g$.}
D7-branes on top of each O7-plane.
Since there are 64 O3-planes but only 16 dynamical ``half'' D3-branes 
as indicated by (\ref{global}), it is impossible to cancel the RR charge locally
at each O3 location; however, we can at least cancel the O3-charge in the last
torus by choosing 
\begin{equation}
\sum_{\vec\xi_4} M_{\vec\xi}=8
\end{equation}
with the sum running over all $\vec\xi=(\vec \xi_4,\vec \xi_2)$ for any fixed
$\vec\xi_2$, {\it i.e.}
over all O3-planes on top of the O7 specified by $\vec\xi_2$.
This is the choice we make from now on. Thus, on each O7-plane we put 4
dynamical D7-branes and 4 ``half'' D3-branes. The latter can then be distributed
over the 16 orbifold fixed points 
that are common to a given O7-plane, leading to different possibilities which
will
be briefly mentioned in the next subsection.

\subsection{A conformal set-up}
\label{subsubsec:confI}
Let us focus on one of the O7-planes, say for example on
the one at $\vec\alpha=(0,0)$, and on the 4 dynamical D7-branes located there.
The latter
support open string excitations
whose CP factors are $(8\times 8)$ matrices $\Lambda$ subject to the following
conditions
\begin{equation}
\gamma_{\vec\alpha}^{*}(\Omega')\,\Lambda^T\,
\gamma_{\vec\alpha}^T(\Omega') =  \varepsilon_{\Omega'}\,\Lambda~,\qquad
\gamma_{\vec\alpha}^{*}(\hat g)\,\Lambda\,
\gamma_{\vec\alpha}^T(\hat g) = \varepsilon_{\hat g}\, \Lambda~, 
\label{invCP7}
\end{equation}
where $\varepsilon_{\Omega'}$ and $\varepsilon_{\hat g}$ are the eigenvalues of
$\Omega'$ and
$\hat g$ on the oscillator part of the corresponding states, in such a way that
these
are invariant under the $\mathbb Z_2 \times \mathbb Z_2$ orientifold.
For instance, for the massless vector $V_\mu$ (represented
by the state $\psi_{-\frac{1}{2}}^\mu |0\rangle$ with $\mu=0,\ldots,3$)
and the massless complex scalar $\varphi$ (represented by the state
$\psi_{-\frac{1}{2}}^{(3)}|0\rangle$ along the torus $\cT_2^{(3)}$),
we have $\varepsilon_{\hat g}=-\varepsilon_{\Omega'}=1$. On the other hand, for
two massless complex scalars $h^{(1)}$ and $h^{(2)}$ along the directions of
$\cT_2^{(1)}\times \cT_2^{(2)}$ (represented by 
the states $\psi_{-\frac{1}{2}}^{(1)}|0\rangle$ and
$\psi_{-\frac{1}{2}}^{(2)}|0\rangle$) we have
$\varepsilon_{\hat g}=\varepsilon_{\Omega'}=-1$. Then, using (\ref{gamma7}) the
CP 
structure of the various massless fields selected by (\ref{invCP7}) turns out to
be
\begin{equation}
V_\mu = \begin{pmatrix} \,A\, & \,S\, \\ 
-S& \,A\,\end{pmatrix}~,~~~\varphi = \begin{pmatrix} \,A\, & \,S\, \\ 
-S& \,A\,\end{pmatrix}~,~~~ h^{(1)}= \begin{pmatrix} A_1\, & A_2 \\ 
A_2\,& -A_1\end{pmatrix}~,~~~h^{(2)} = \begin{pmatrix} A_1\, & A_2 \\ 
A_2\,& -A_1\end{pmatrix}
\label{massless7}
\end{equation}
where $A$, $A_1$ and $A_2$ are $(4\times 4)$ antisymmetric matrices, and $S$ is
a $(4\times 4)$ symmetric matrix. We therefore see that the vector $V_\mu$ and
the scalar $\varphi$ are in
the adjoint representation of U(4), embedded in SO(8), while the two scalars 
$h^{(1)}$ and $h^{(2)}$ are in the antisymmetric representation $\Yasymm$ of
U(4) plus its conjugate 
$\overline{\Yasymm}$, again embedded in SO(8).
Adding the corresponding fermions from the R sector, the massless
spectrum of the 7/7 strings consists of one $\mathcal N=2$ vector multiplet in
the adjoint representation of U(4) schematically given by 
\begin{equation}
\Phi(x,\theta) \sim \varphi(x) + \theta^2\,F(x)+~\mbox{fermions}
\label{vector}
\end{equation}
where $F$ is the gauge field-strength, one hyper-multiplet in the $\Yasymm$
representation and one hyper-multiplet in the conjugate $\overline{\Yasymm}$
representation.

Now let us consider the 7/3 open strings stretching between D7- and D3-branes.
In this case the massless excitations correspond to twisted states 
with mixed Neumann-Dirichlet boundary conditions
along the directions of $\cT_2^{(1)}$ and $\cT_2^{(2)}$, and organize in
4 hyper-multiplets (one for each ``half'' D3-brane) transforming in the
fundamental representation $\Yfund$
of U(4). To see this, let us consider $m$ ``half'' D3-branes located at a given
orbifold fixed point $\vec\xi$. In order to survive the orbifold%
\footnote{The orientifold projection $\Omega'$ does not impose any restriction
but only
identifies states of the 7/3 sector with states of opposite orientation
belonging to
the 3/7 sector.} projection, the CP factor $\Lambda$ 
of the massless states of the 7/3 sector must satisfy the following constraint
\begin{equation}
\gamma_{\vec\alpha}^{*}(\hat g)\,\Lambda\,\gamma_{\vec\xi}^T(\hat g)
=\varepsilon_{\hat g}\,\Lambda~~~\mbox{with}~~\varepsilon_{\hat g}=1~,
\label{lambda73}
\end{equation}
which, upon using (\ref{gamma7}) and (\ref{gamma3}), is solved by
\begin{equation}
\Lambda = \begin{pmatrix} X_1 & \,X_2 \\ 
-X_2& \,X_1\end{pmatrix}
\label{lambdafun}
\end{equation}
with $X_1$ and $X_2$ being generic $(4\times m)$ matrices. Thus, these mixed
states transform as $m$ hyper-multiplets in the fundamental representation of
U(4). In our model, of course, we have $m=0$ for 12 fixed points and $m=1$ for 4
fixed points contributing in total 4 hyper-multiplets.
Nothing changes in this respect, if the ``half'' D3-branes are distributed
differently among the various orbifold fixed points.

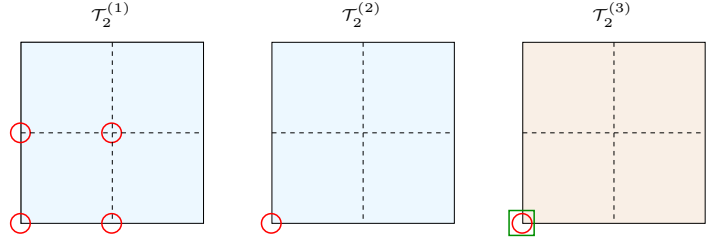
\begin{figure}[htb]
\begin {center}
 \begin{picture}(0,0)%
\includegraphics{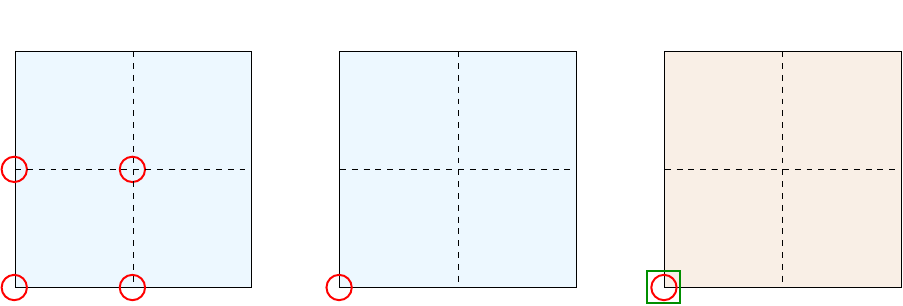}%
\end{picture}%
\setlength{\unitlength}{1243sp}%
\begingroup\makeatletter\ifx\SetFigFontNFSS\undefined%
\gdef\SetFigFontNFSS#1#2#3#4#5{%
 \reset@font\fontsize{#1}{#2pt}%
 \fontfamily{#3}\fontseries{#4}\fontshape{#5}%
 \selectfont}%
\fi\endgroup%
\begin{picture}(13643,4578)(555,-4102)
\put(2161,209){\makebox(0,0)[lb]{\smash{{\SetFigFontNFSS{6}{7.2}{\familydefault}
{\mddefault}{\updefault}$\cT_2^{(1)}$}}}}
\put(7111,209){\makebox(0,0)[lb]{\smash{{\SetFigFontNFSS{6}{7.2}{\familydefault}
{\mddefault}{\updefault}$\cT_2^{(2)}$}}}}
\put(12061,209){\makebox(0,0)[lb]{\smash{{\SetFigFontNFSS{6}{7.2}{\familydefault
}{\mddefault}{\updefault}$\cT_2^{(3)}$}}}}
\end{picture}%
\end{center}
\caption{Brane locations in our model. The square denotes the orientifold fixed
point 
$\vec\alpha=(0,0)$ where the 4 D7-branes are located, while the circles denote
the positions
of the 4 ``half'' D3-branes.}
\label{fig:D7D3}
\end{figure}

On the contrary, what changes according to the configuration of D3-branes is the
theory on the world-volume of the latter. If the 4 D3-branes are all located at
the same fixed point, we have a gauge theory with group U(4) and a matter
content similar to the one discussed above for the D7-branes. If, instead, 3
D3-branes are located at one fixed point and and the fourth D3 is at a different
one, we have a gauge theory with group $\mathrm{U}(3)\times\mathrm{U}(1)$, and
so and so forth. The case in which the 4 D3-branes are all in 4 different fixed
points, thus giving rise to a theory with a $\mathrm{U}(1)^4$ symmetry, is of
particular interest since it is this configuration which admits a simple
perturbative heterotic dual. Thus, from now on we will restrict our analysis to
this case only. The theory we consider is therefore the one living on the 4
D7-branes on top of one of the orientifold O7-planes, with the 4 D3-branes
placed at four different orbifold fixed points, as shown for example in Fig.
\ref{fig:D7D3}. 

The gauge group is $\mathrm{U}(4) \times \mathrm{U}(1)^4$, with the latter
factors representing flavor symmetries from the point of view of the theory on
the D7-branes. The massless content of this $\mathcal N=2$ model is summarized
in Tab. \ref{tablespec}, where in the last column we have indicated also the
$\mathrm{U}(1)_{D7}$-charge of the various multiplets. Notice that while the
adjoint fields are clearly neutral, the charge of 7/7 hyper-multiplets is, in
absolute value, twice the charge of the 7/3 hyper-multiplets. This fact can be
easily understood, since the 7/7 fields correspond to open strings with two
charged endpoints on the D7-branes, as opposed to the 7/3 fields which have only
one charged endpoint on the D7-branes.
\begin{table}[htb]
\begin{center}
\begin{tabular}{|c|c|c|c|c|c|c|c|}
\hline
$\phantom{\vdots}\mathcal N=2$ rep. 
&sector 
&$\varepsilon_{\Omega'}$
&~$\varepsilon_{\hat g}$~  
&CP factor 
&~$\#$~ 
&U(4)
&$q_{\mathrm{U}(1)}$\\
\hline\hline
& & & & & & &\\
vector & 7/7 &$-$ &$+$ & \begin{small}
$\begin{pmatrix} \,A\, & \,S\, \\ 
-S& \,A\,\end{pmatrix}$ \end{small}
&1 &adj.  &0\\
& & & & & & &\\
hyper & 7/7 &$-$ &$-$ & \begin{small}$\begin{pmatrix} A_1\, & A_2 \\ 
A_2\,& -A_1\end{pmatrix}$ \end{small}& 
1 & $\Yasymm$  & $-2$\\
& & & & & & &\\
hyper & 7/7 &$-$ &$-$ & \begin{small}$\begin{pmatrix} A_1\, & A_2 \\ 
A_2\,& -A_1\end{pmatrix}$ \end{small}& 
1 &$\overline{\Yasymm}$  & $+2$ \\
& & & & & & &\\
hyper & 7/3 & \begin{small} undef. \end{small}
& $+$ & \begin{small}$\begin{pmatrix} X_1 & \,X_2 \\ 
-X_2& \,X_1\end{pmatrix}$ \end{small}& 
4& $\Yfund$  & $-1$\\
& & & & & & &\\
\hline
\end{tabular}
\caption{Massless spectrum on the world-volume of the 4 D7-branes at one of the
orientifold
fixed points of our model.}
\label{tablespec}
\end{center}
\end{table}

It is not difficult to check that this model is conformal. Indeed, the 1-loop
$\beta$-function coefficient for a $\mathcal{N}=2$ theory with gauge group $G$ 
is given by
\begin{equation}
b= 2 \Big[T(G) - \sum_r n_r T(r)\Big]~,
\label{beta}
\end{equation}
where the index $T(r)$ of a representation $r$ of $G$ is defined by
\begin{equation}
T(r) \,\delta_{AB} = \tr(T_A(r)T_B(r))~, 
\label{indexT}
\end{equation}
$T(G)$ stands for $T(\mathrm{adj})$ 
and $n_r$ is the number of hyper-multiplets
transforming in the representation $r$. 
In our case (see Tab.~1) we have $G = \mathrm{U}(N)$ with $N=4$, and
\begin{equation}
\label{nmult}
n_{_{\Yasymm}} ~=~n_{_{\overline{\Yasymm}}} = 1~,~~~
n_{_{\Yfund}} = 4~.
\end{equation}
Since the relevant indices are%
\footnote{The following formulas actually refer to the $\mathrm{SU}(N)$ part of
the gauge group. Later
we will consider also the U(1) factor.}
\begin{equation}
\label{indices}
T(G)= N=4~,~~~
T(\Yasymm)=T(\overline{\Yasymm})=\frac{N-2}{2}=1~,~~~
T(\Yfund)=\frac{1}{2}~,
\end{equation}
the $\beta$-function coefficient vanishes:
\begin{equation}
\label{bais}
b = 4 -n_{_{\Yfund}}=0~.
\end{equation}

As we will see in the next section, the quadratic effective action for 
the D7 gauge fields contains also a double-trace structure $(\tr F)^2$, 
which clearly arises only in the U(1) sector and renormalizes 
separately from the Yang-Mills term $\tr F^2$. 
The 1-loop $\beta$-function coefficient for the double-trace coupling, 
which we denote as $b'$, can be deduced from the coefficient of the
$\beta$-function for the
U(1) factor of the gauge group, which in turn is computed from the charges of
the
various multiplets. Such abelian $\beta$-function coefficient is given by
\begin{equation}
\beta_{\mathrm{U(1)}} =- \sum_r n_r \,q_r^2\,d(r)
\label{betau1}
\end{equation}
where $q_r$ is the U(1) charge of the hyper-multiplet in the representation $r$
whose dimension is $d(r)$. Inserting in this expression the U(1) charges and
multiplicities given in Tab.~\ref{tablespec}, one finds
\begin{equation}
\beta_{\mathrm{U(1)}} = \big(4-n_{_{\Yfund}}\big)\,N - 4N^2=
\big(4-n_{_{\Yfund}}\big)\,\tr\one - 4\big(\tr\one)^2~.
\label{betau12}
\end{equation}
Our specific model is precisely of this type, 
with $N=4$ and $n_{_{\Yfund}}=4$. Thus, from (\ref{betau12}) we easily deduce 
that the U(1) contribution to the $\beta$-function of the single-trace term is
the
same as the non-abelian one (\ref{bais}) and vanishes in our model, 
while the contribution $b'$ to the double-trace term is
\begin{equation}
b'=-4~.
\label{baa}
\end{equation}

This concludes our analysis of the properties dictated by the (massless)
spectrum of string excitations on the D7 branes. In the next section we turn to
the structure of the 
interaction terms in the low-energy effective action, starting from the
perturbative contributions. 

\section{Type I$^\prime$ gauge effective action: perturbative part}
\label{sec:prepI}
The tree-level action for the $\mathcal N = 2$ Super Yang-Mills theory discussed
in the previous section can be obtained by computing disk scattering amplitudes
among the various massless excitations of the open strings with at least one
end-point on the D7-branes and then taking the field theory limit $\alpha'\to
0$. Alternatively, at least for the pure Yang-Mills part we can consider the
Dirac-Born-Infeld (DBI) action for a D7-brane, namely
\begin{equation}
S_{\mathrm{DBI}}= \frac{2\pi}{(2\pi\sqrt{\alpha'})^8}\,\int d^8x
~\ee^{-\phi_{10}}\,
\sqrt{\big(G_{MN}+2\pi\alpha'F_{MN}\big)}
\label{DBI}
\end{equation}
with $G_{MN}$ being the world-volume metric and $F_{MN}$ the gauge
field-strength
($M,N=0,\ldots,7$), and then compactify it to four dimensions on
$\big(\cT_2^{(1)}\times\cT_2^{(2)}\big)/\mathbb Z_2$.
In this way, after promoting the field-strength to be non-abelian and
rescaling the four-dimensional metric to the flat one, we obtain, at the
quadratic level,
\begin{equation}
S_{\mathrm{tree}}= \frac{1}{2g^2}\,\int d^4x ~\tr(F_{\mu\nu}^2)~,
\label{Stree}
\end{equation}
where $\mu,\nu=0,\ldots,3$, and the Yang-Mills coupling constant $g$ is given by
\begin{equation}
\frac{4\pi}{g^2}= t_2
\label{gt}
\end{equation}
with $t_2$ defined in \eq{deft}. Now we turn to the 1-loop terms.

\subsection{1-loop contributions}
\label{subsec:oneloopI}

To derive the 1-loop threshold corrections to the quadratic
gauge couplings on the D7-branes we use the background field method. 
The 1-loop amplitudes are extracted from the second derivatives 
of the weigthed partition function of open strings with at least one endpoint on
the D7-branes 
in presence of a constant magnetic field on the D7-brane world-volume (see 
Ref.~\cite{Bianchi:2000vb} for previous studies of $F^2$-amplitudes in
${\mathcal N}=2$ brane set-ups).

For concreteness we switch on a magnetic field $\mathcal H$ along, say,
the directions 2 and 3, {\it i.e.} 
\begin{equation}
F_{23}=-F_{32}=\mathcal H~,~~~F_{\mu\nu}=0 ~~\mbox{for}~~\mu,\nu\not=2,3~,
\label{backF}
\end{equation}
and taking value only in the Cartan directions of the gauge group.
Furthermore, we suppose again to have $N$ dynamical D7-branes and set $N=4$ in
the end. 
As discussed in Section~\ref{subsubsec:confI}, in the real
basis for the CP indices the adjoint of $\mathrm{U}(N)$ is embedded into
$\mathrm{SO}(2N)$,
so that Cartan subalgebra of $\mathrm{U}(N)$ is represented by skew-diagonal
matrices.
However, with a complex change of basis we diagonalize them and hence bring
our Cartan magnetization in the form
\begin{equation}
\label{fdiag}
\frac{\ii}{2\pi\alpha'}\,\mathrm{diag}\big(h_1,h_2, \ldots,
h_N,-h_1,-h_2,\ldots,-h_N\big)\equiv\frac{\ii}{2\pi\alpha'}\,\mathrm{diag}
\big(h_i\big)~.
\end{equation}
Here we have introduced the index $i=1,\ldots,2N$ running
over all D7-brane labels; the fundamental and antifundamental
indices of $\mathrm{U}(N)$, taking values $1,\ldots, N$,
will be denoted instead by
$I$ and $\bar I$ respectively. Thus, we have $i=I$ for $i=1,\ldots,N$ and 
$i= \bar I + N$ for $i=N+1,\ldots,2N$, so that $h_{\bar I} = -h_I$. The
$\mathrm{U}(N)$ field strength
$\mathcal H$ corresponds to the $(N\times N)$ upper block in (\ref{fdiag}),
namely
\begin{equation}
\label{Fdiag}
\mathcal H =  \frac{\ii}{2\pi\alpha'}\,\mathrm{diag}\big(h_1,h_2, \ldots,
h_N\big) \equiv \frac{\ii}{2\pi\alpha'}\, \mathrm{diag}(h_I)~.
\end{equation}
In presence of the magnetization (\ref{fdiag}), a 7/7 open string stretching
between the $i$-th and $j$-th D7-brane is twisted by an angle
\begin{equation}
\label{twistnu}
\nu_{ij} = -\frac{1}{\pi}\left(\arctan h_i -\arctan h_j\right)
\sim -\frac{h_i - h_j}{\pi} + O(h^3)~,
\end{equation}
and the spectrum of physical excitations changes correspondingly.

The 1-loop effective action of the D7-branes can be deduced from the 1-loop
vacuum energy in the background (\ref{fdiag}). 
For the 7/7 open strings this vacuum energy has the following schematic form
\begin{equation}
\int_0^\infty \frac{d\tau_2}{2\tau_2}\,\sum_{i,j}
\Tr_{(h_i,h_j)}\Bigg(\frac{1+\Omega'}{2}\,\frac{1+\hat g}{2}\,
\frac{1+(-1)^F}{2}\,
q^{L_0-\frac{c}{24}}\Bigg) = \mathcal A_{7/7}(h) +\mathcal M_{7/7}(h)
\label{D7loop}
\end{equation}
where $\frac{1+(-1)^F}{2}$ is the GSO projector, $q=\ee^{-\pi \tau_2}$, 
and the trace $\Tr_{(h_i,h_j)}$ is computed over the spectrum of 7/7
open strings with boundary conditions determined by the values $(h_i,h_j)$. 
In the right hand side of (\ref{D7loop}) we have distinguished, as usual, an
annulus
contribution $\mathcal A_{7/7}(h)$ and a M\"obius strip contribution $\mathcal
M_{7/7}(h)$,
which is non-vanishing only if $h_j = -h_i$ due to the presence of $\Omega'$
inside the trace \cite{Billo':2009gc}.

Our model contains also $m=4$ ``half'' D3-branes
at the same fixed point of $\cT_2^{(3)}$ of the D7-branes, so that also the
D7/D3 strings
can have massless modes contributing to the low-energy effective action. At
1-loop we should
therefore take into account also annuli with one boundary on a magnetized
D7-brane 
and the other on one of the D3-branes, corresponding to the amplitude
\begin{equation}
\label{mixedpg}
\begin{aligned}
\mathcal{A}_{7/3}(h) + \mathcal{A}_{3/7}(h)
& =\int_0^\infty \frac{d\tau_2}{2\tau_2}\,\sum_{i,a}
\Tr_{(h_i,a)}\Bigg(\frac{1+g}{2}\,
\frac{1+(-1)^F}{2}\, q^{L_0-\frac{c}{24}}\Bigg) 
\end{aligned}
\end{equation}
where $a$ labels the CP indices of the D3-branes, taking $2m$ values.

The amplitudes (\ref{D7loop}) and (\ref{mixedpg}) are computed in
App.~\ref{subapp:olm}
and the result is given in Eqs.~(\ref{aijres}), (\ref{mijres}) and
(\ref{73sineg}). 
All in all, the total 1-loop effective action turns out to be
\begin{equation}
\label{s1loop}
\begin{aligned}
S_{\mathrm{1-loop}} &= \phantom{\Bigg(}\!\!\mathcal A_{7/7}(h) +\mathcal
M_{7/7}(h) + 
\mathcal{A}_{7/3}(h)
+\mathcal{A}_{3/7}(h)\\
&= -\frac{V_4}{8\pi^2} \Big[(4 - m)\,\tr \mathcal H^2 -4 \,(\tr \mathcal H)^2
\Big]
\int_0^\infty  \frac{d\tau_2}{2\tau_2} \,W(\tau_2)+ O(h^3)~.
\end{aligned}
\end{equation}
with
\begin{equation}
W(\tau_2) = 
\sum_{\vec w\in\Z^2}\ee^{-\pi \tau_2\,\frac{|w^1+w^2 U|^2T_2}{U_2}}
\label{wt}
\end{equation}
representing the sum over winding states on $\cT_2^{(3)}$ (see \eq{defYtr}).
Notice that the expression in square brackets has the same structure appearing
in \eq{betau12}, 
and that the coefficient of the single-trace term is the correct
$\beta$-function coefficient 
for this model (see \eq{bais}), since the number $m$ of D3-branes equals 
the number $n_{_{\Yfund}}$ of fundamental hyper-multiplets.
In our conformal case, {\it i.e.} $m=n_{_{\Yfund}}=4$, there is no running for
the single-trace coupling, but there is a non-vanishing 1-loop contribution
proportional to $(\tr \mathcal H)^2$. Promoting $\mathcal H$ to a full dynamical
field $F_{\mu\nu}$,
this contribution in the end reads 
\begin{equation}
\label{s1loopdx}
S_{\mathrm{1-loop}} = \frac{1}{8\pi^2} \int d^4x\, (\tr F)^2 \int_0^\infty 
\frac{d\tau_2}{\tau_2}\,W(\tau_2)+ 
O(F^3)~.
\end{equation}
It is important to stress that this 1-loop action is entirely due to zero-mode
states wrapping around $\cT_2^{(3)}$ and giving rise to the winding sum
$W(\tau_2$. 
The contributions of the massive string states, instead, exactly cancel as a
consequence 
of the fact that in ${\mathcal N}=2$ theories the $F^2$-terms are ``BPS
saturated'' quantities 
(see {\it e.g.} Ref.~\cite{Bianchi:2000vb} for an 
extension of this result to more general brane setups). This property makes the 
quadratic gauge couplings reliable variables to follow under the
non-perturbative
type I$^\prime$/heterotic duality.

The integral over the modular parameter $\tau_2$ in (\ref{s1loopdx}) can be
evaluated following the
methods of Ref.~\cite{Dixon:1990pc}, as reviewed for example in the appendix of
Ref.~\cite{Billo:2007sw}, and, after regularizing the divergences,
the result (up to moduli independent constants) is
\begin{equation}
\begin{aligned}
\int_0^\infty  \frac{d\tau_2}{\tau_2}\,W(\tau_2) & = -\log(\alpha'\mu^2)
-\log\Big(\frac{U_2\,|\eta(U)|^4}{T_2}\Big)\\
& = -\log\Big(\frac{\mu^2
t_2}{M_{\mathrm{Pl}}^2}\Big)-\log\Big(\lambda_2\,U_2\,|\eta(U)|^4\Big)
\end{aligned}
\label{it}
\end{equation}
where in the second step we introduced  as UV cut-off the four-dimensional 
Planck mass $M_{\mathrm{Pl}}$ (\ref{mpl1}). Thus, the 1-loop action
(\ref{s1loopdx}) becomes
\begin{equation}
\label{s1loop1}
S_{\mathrm{1-loop}} = \frac{1}{32\pi^2} \int d^4x\, (\tr F)^2 \Bigg[
\!-4\log\Big(\frac{\mu^2 t_2}{M_{\mathrm{Pl}}^2}\Big)-4\log\Big(\lambda_2\,U_2\,
|\eta(U)|^4\Big)\Bigg] + O(F^3)~.
\end{equation}
{From} this explicit result, we can read the $\beta$-function coefficients and
the 1-loop
threshold corrections to the gauge couplings, as explained in App.~\ref{app:hol}
(see in particular
\eq{gstringd}). The absence of single-trace quadratic terms implies
that $b=0$ in agreement with \eq{bais}, and 
\begin{equation}
\Delta = 0~.
\label{da0}
\end{equation}
On the other hand, for the double-trace structure we see that $b'=-4$ in
agreement with \eq{baa}. The 1-loop threshold follows then from Eq.
(\ref{gstringd}) and reads
\begin{equation}
\Delta' = -4\log\Big(\lambda_2\,U_2\,|\eta(U)|^4\Big)~.
\label{dbaa}
\end{equation}
Notice that this threshold is invariant under the target-space modular
trasformations acting on
$U$, but it is not invariant under the $\mathrm{Sl}(2,\Z)$ transformations of
the axio-dilaton
$\lambda$. This lack of invariance signals the necessity of
non-perturbative
corrections which, as we will show in Sections~\ref{sec:Dmod} and \ref{sec:loc},
are induced
by D-instantons. 
\subsection{Holomorphic gauge couplings}
\label{DKLI}
The moduli dependence of string loop corrections to the gauge kinetic terms of a
supersymmetric effective quantum field theory like ours is best described in
terms of {\emph{holomorphic}} couplings, as explained in
Refs.~\cite{Dixon:1990pc,Kaplunovsky:1995jw,deWit:1995zg} and briefly reviewed
in App.~\ref{app:hol}. These Wilsonian functions, in general, have the following
structure
\begin{equation}
f = f_{(0)} + \frac{1}{4\pi}\,f_{(1)}+f_{\mathrm{n. p.}}
\label{f01}
\end{equation}
where the subscripts $_{(0)}$ and $_{(1)}$ refer, respectively, 
to the tree-level and 1-loop contributions, while the last term accounts for
possibile non 
perturbative corrections. 

In our specific theory, there two such functions: one for the usual single-trace
Yang-Mills
term $\tr F^2$, which we will denote by $f$, 
and one for the double-trace term $(\tr F)^2$, called $f'$ in the following. 
At tree level we have
\begin{equation}
f_{(0)}= -{\ii}\,t~,~~~f'_{(0)}=0~,
\label{f0}
\end{equation}
as one can see from \eq{gt} and the fact that no double-trace term is present in
the tree-level
action (\ref{Stree}). The 1-loop contributions, instead, can be obtained from
the formulas (see also
\eq{fa12})
\begin{equation}
\re f_{(1)} = \Delta +{\Delta}_{\mathrm{univ}} + b\,\widehat{K}~,~~~
\re f'_{(1)} = \Delta' +{\Delta}_{\mathrm{univ}} + b'\,\widehat{K}~,
\label{f1a}
\end{equation} where  
\begin{equation}
\widehat {K} =-\log\big(\lambda_2\,U_2\big)~.
\label{kalerI} 
\end{equation}
is related to the K\"ahler metric
of the adjoint scalar fields, while $\Delta_{\mathrm{univ}}$
is a universal effect due to the mixing of the dilaton with the compactification
moduli
\cite{Dixon:1990pc,Kaplunovsky:1995jw}. Since both $\Delta$, $\Delta'$,
$\Delta_{\mathrm{univ}}$ and
$\widehat{K}$ contain non-holomorphic terms, the relations (\ref{f1a}) imply
that
all such terms should compensate each other for consistency to yield
holomorphic 
expressions for $f_{(1)}$ and $f'_{(1)}$.
In type II or type I theories (as opposed to heterotic models) the universal
correction
${\Delta}_{\mathrm{univ}}$ is actually of $O(g_{\mathrm{s}})$, $g_{\mathrm{s}}$
being the
string coupling, and thus it does not contribute to the coupling functions at
1-loop. 
This can be seen for example from the explicit calculation of the corrections 
to the K\"ahler potential performed in Ref.~\cite{Berg:2005ja}. This same
observation
was used in Ref.~\cite{Billo:2007py} to obtain the K\"ahler metrics of twisted
matter fields from instantonic annulus diagrams in agreement with the explicit
perturbative string derivation presented in Ref.~\cite{Bertolini:2005qh}. Thus,
for our
purposes here, we can drop the ${\Delta}_{\mathrm{univ}}$ term from the various
formulas.
Recalling the expression (\ref{kalerI}) for $\widehat K$ and using the results
for $b$, $b'$,
$\Delta$ and $\Delta'$ obtained in the previous subsection, from \eq{f1a} we
finally get 
\begin{equation}
\re f_{(1)} = 0~,~~~
\re f'_{(1)}  = -4\log\big(|\eta(U)|^4\big)\phantom{\vdots},
\label{fIaa}
\end{equation}
in agreement with the holomorphy requirements. In the next sections we will
study the non-perturbative
corrections $f_{\mathrm{n. p.}}$ and $f'_{\mathrm{n. p.}}$ to the coupling
functions induced
by D-instantons.

\section{D-instantons and their moduli spectrum}
\label{sec:Dmod}
We now discuss the effects of instantonic branes on the system
described so far. There are two types of branes that are point-like with
respect to the four-dimensional uncompact space and can be put on the O7-planes
in a supersymmetric fashion, namely 
\begin{itemize}
\item extended Euclidean 3-branes (or E3-branes) wrapping
$\cT_2^{(1)} \times \cT_2^{(2)}$; 
\item point-like D(--1)-branes that are completely localized 
in all directions.
\end{itemize}
The E3-branes represent ordinary gauge instantons for the field
theory living on the D7-branes; indeed in the E3/D7 system there are precisely
four directions with mixed Dirichlet-Neumann boundary conditions and the
spectrum of the physical excitations of the open strings with at least one
endpoint on the Euclidean branes is in full agreement with that of the ADHM
construction for gauge instantons
\cite{Witten:1995gx,Douglas:1995bn,Billo:2002hm}. On the other hand the
D(--1)-branes describe truly stringy (or exotic) instanton configurations for
the D7-brane gauge theory \cite{Billo':2009gc}. In fact, in this case between
the instantonic and the space-filling branes there are eight directions with
mixed boundary conditions, and the corresponding spectrum of moduli is not the
conventional one. 

In this paper we only discuss the contributions produced by the D(--1)-branes,
leaving the study of the E3-branes to a future work. In particular we will show
that fractional D-instantons located at orbifold fixed points have the right
content of zero-modes to correct non-perturbatively the gauge kinetic function
of the $\mathcal N=2$ U(4) theory discussed in Sect.~\ref{sec:prepI}, and later
will check the result against the dual heterotic string calculation. Again we
focus on the four D7-branes located at one of the orientifold fixed points, and
place on them a number of fractional D-instantons. However, since there are also
four D3-branes distributed in four different orbifold fixed points, we have to
distinguish between two possibilities, depending on whether the D-instantons are
at the same position of one of the D3-branes or are at an empty fixed point. 

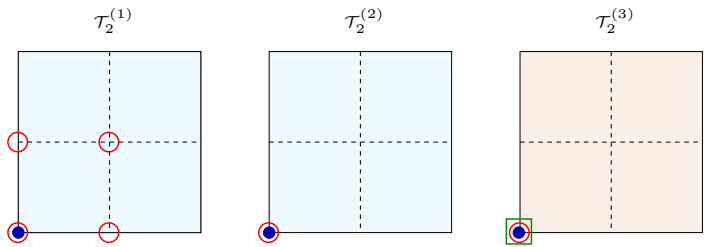
\begin{figure}[htb]
\begin {center}
 \begin{picture}(0,0)%
\includegraphics{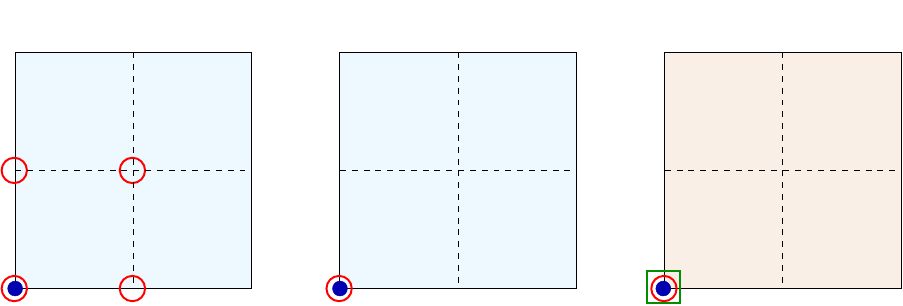}%
\end{picture}%
\setlength{\unitlength}{1243sp}%
\begingroup\makeatletter\ifx\SetFigFontNFSS\undefined%
\gdef\SetFigFontNFSS#1#2#3#4#5{%
 \reset@font\fontsize{#1}{#2pt}%
 \fontfamily{#3}\fontseries{#4}\fontshape{#5}%
 \selectfont}%
\fi\endgroup%
\begin{picture}(13751,4620)(447,-4144)
\put(2161,209){\makebox(0,0)[lb]{\smash{{\SetFigFontNFSS{6}{7.2}{\familydefault}
{\mddefault}{\updefault}$\cT_2^{(1)}$}}}}
\put(7111,209){\makebox(0,0)[lb]{\smash{{\SetFigFontNFSS{6}{7.2}{\familydefault}
{\mddefault}{\updefault}$\cT_2^{(2)}$}}}}
\put(12061,209){\makebox(0,0)[lb]{\smash{{\SetFigFontNFSS{6}{7.2}{\familydefault
}{\mddefault}{\updefault}$\cT_2^{(3)}$}}}}
\end{picture}%
\end{center}
\caption{A possible arrangement of the D(--1)/D3/D7 system in case \emph{a)}.
 The empty square denotes the fixed point $\vec\alpha$ of $\cT_2^{(3)}$
 occupied by the  D7-branes, the empty circles the four fixed points $\vec\xi$
 in $\cT_2^{(1)}\times\cT_2^{(2)}\times\cT_2^{(3)}$ occupied by the D3's and
 the filled circle the one among these latter where $k$ D(--1)'s are positioned.
 There are four inequivalent possibilities for the D(--1) location.} 
\label{fig:D7D3D-1_a}
\end{figure}

In the first case (case \emph{a)} in the following), schematically represented
in Fig.~\ref{fig:D7D3D-1_a}, there is one orbifold fixed point, say $\vec\xi$,
occupied both by the D(--1)'s and by one D3; therefore we can find massless
excitations not only in the spectrum of the $(-1)/(-1)$ and $(-1)/7$ open
strings, but also in that of the $(-1)/3$ strings stretching between the
D-instantons and the D3-brane located in that point. Since there are four
different D3-branes, this situation can be realized in four different but
completely equivalent ways. 

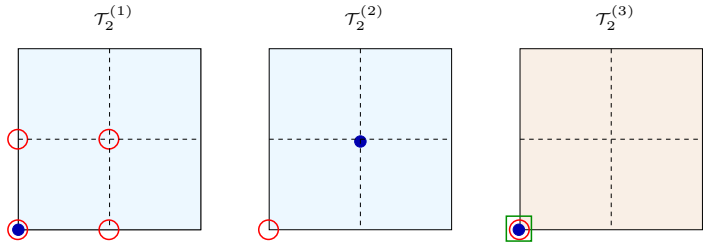
\begin{figure}[htb]
\begin {center}
 \begin{picture}(0,0)%
\includegraphics{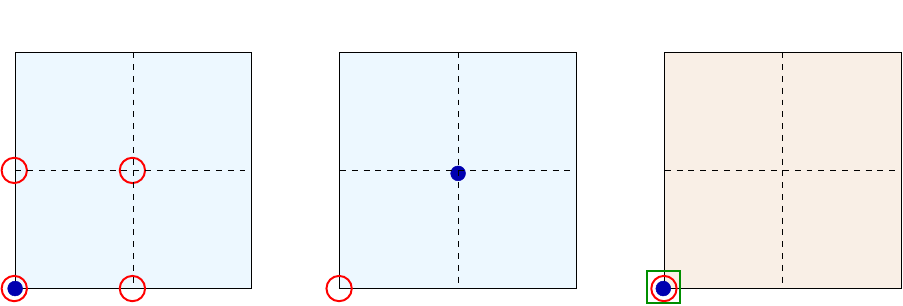}%
\end{picture}%
\setlength{\unitlength}{1243sp}%
\begingroup\makeatletter\ifx\SetFigFontNFSS\undefined%
\gdef\SetFigFontNFSS#1#2#3#4#5{%
 \reset@font\fontsize{#1}{#2pt}%
 \fontfamily{#3}\fontseries{#4}\fontshape{#5}%
 \selectfont}%
\fi\endgroup%
\begin{picture}(13751,4620)(447,-4144)
\put(2161,209){\makebox(0,0)[lb]{\smash{{\SetFigFontNFSS{6}{7.2}{\familydefault}
{\mddefault}{\updefault}$\cT_2^{(1)}$}}}}
\put(7111,209){\makebox(0,0)[lb]{\smash{{\SetFigFontNFSS{6}{7.2}{\familydefault}
{\mddefault}{\updefault}$\cT_2^{(2)}$}}}}
\put(12061,209){\makebox(0,0)[lb]{\smash{{\SetFigFontNFSS{6}{7.2}{\familydefault
}{\mddefault}{\updefault}$\cT_2^{(3)}$}}}}
\end{picture}%
\end{center}
\caption{A possible arrangement of the D(--1)/D3/D7 system, case \emph{b)}. 
This time the D(--1)'s occupy a fixed point where no D3's are sitting. There
are twelve inequivalent possibilities.} 
\label{fig:D7D3D-1_b}
\end{figure}

In the second case (case \emph{b)} in the following), represented in
Fig.~\ref{fig:D7D3D-1_b},
only the $(-1)/(-1)$ and $(-1)/7$ open strings can support massless moduli
because the $(-1)/3$ strings have always a non-vanishing stretching energy
due to the non-zero space separation between their endpoints. Since in our model
there are twelve orbifold fixed points that are not occupied by D3-branes, this
case can be realized in twelve different but completely equivalent ways.

In order to select which moduli survive the orientifold and orbifold
projections, it is necessary to specifiy how the discrete parities
$\Omega'$ and $\hat g$ act on the CP indices of the D-instantons. Extending the
consistency arguments of Ref.~\cite{Gimon:1996rq} to our case, we can show that
this action can be represented by matrices $\Gamma_{\vec\xi}$ having the same
form as the matrices $\gamma_{\vec\alpha }$ acting on the CP indices of the
D7-branes and introduced in \eq{gamma7}, namely
\begin{equation}
\Gamma_{\vec\xi}(\Omega') = \begin{pmatrix} ~\one~ & ~0~ \\ 
~0~& ~\one~\end{pmatrix}~,~~~
\Gamma_{\vec\xi}(\hat g) =\Gamma_{\vec\xi}(\Omega'\hat g)=
\begin{pmatrix} ~0~ & \ii\,\one \\ 
-\ii\,\one
& ~0~\end{pmatrix}~.
\label{gamma-1}
\end{equation}
Clearly this implies that the number of CP indices for the D(--1)'s must be an
even integer, say $2k$, so that the various blocks in (\ref{gamma-1}) are
$(k\times k)$ matrices. Adopting the same terminoloy used for the D7- and the
D3-branes, we say that this case corresponds to having $k$ ``half''
D-instantons. Let us
also note that the physical moduli organize in representations of the
Lorentz symmetry group, which in our local system is broken to
\begin{equation}
\label{lor}
 \mathrm{SO}(4)\times \widehat{\mathrm{SO}}(4) \times \mathrm{SO}(2) = 
 \mathrm{SU}(2)_+\times \mathrm{SU}(2)_- \times 
 \widehat{\mathrm{SU}}(2)_+\times \widehat{\mathrm{SU}}(2)_- \times
 \mathrm{SO}(2)
\end{equation}
by the presence of O7/O3-planes and D7/D3-branes. In the following, we will
refer collectively to the $\mathrm{SU}(2)$ factors above as
$\mathrm{SU}(2)^4$. The subscript $\pm$ refers to the fact that the irreducible
factors $\mathrm{SU}(2)_\pm$ inside $\mathrm{SO}(4)$ rotate (anti)-self-dual
tensors.

Let us now give some details. 

\paragraph{$(-1)/(-1)$ strings:} This is the neutral sector since it comprises
states that do not transform under the U(4) gauge group. A generic modulus in
this sector has a Chan-Paton matrix structure $\Lambda$ which must fulfill the
invariance conditions
\begin{equation}
\Gamma_{\vec\xi}^{*}(\Omega')\,\Lambda^T\, 
\Gamma_{\vec\xi}^T(\Omega') =  \varepsilon_{\Omega'}\, \Lambda~~~~\mbox{and}~~~~
\Gamma_{\vec\xi}^{*}(\hat g)\,\Lambda\,
\Gamma_{\vec\xi}^T(\hat g) = \varepsilon_{\hat g} \Lambda~, 
\label{invCP}
\end{equation}
where $\varepsilon_{\Omega'}$ and $\varepsilon_{\hat g}$ are the eigenvalues of
$\Omega'$ and
$\hat g$ on the oscillator part of the corresponding state. 

The physical zero-modes are easily obtained by dimensionally
reducing the $\mathcal N = 1$ supersymmetric gauge theory from ten to zero
dimensions. In the NS sector, the ten real bosonic moduli
split in different sets according to their transformation properties
under the Lorentz group (\ref{lor}). Adopting an ADHM
inspired notation, we label them as follows. There are two complex scalars
$B_{\ell}$ ($\ell=1,2$) associated to the four
real string states $\psi_{-1/2}^\mu\,|0\rangle$ along the space-time
directions; they transform as a vector of $\mathrm{SO}(4)$, i.e., in the
$\reps 2211$ of $\mathrm{SU}(2)^4$ and have $\varepsilon_{\Omega'} =
\varepsilon_g=1$.
Two complex scalars $B_{\dot \ell}$ ($\dot\ell=3,4$) are associated to 
$\psi_{-1/2}^{(1)}\,|0\rangle$ and $\psi_{-1/2}^{(2)}\,|0\rangle$ along
the directions of $\cT_2^{(1)}\times \cT_2^{(2)}$; they transform 
in the $\reps 1122$ and have $\varepsilon_{\Omega'} = 1$, $\varepsilon_{\hat g}
= -1$.
Finally, there is one complex scalar $\chi$ associated to the string excitation 
$\psi_{-1/2}^{(3)}\,|0\rangle$ along $\cT_2^{(3)}$, which is
a vector of $\mathrm{SO}(2)$ and a singlet of $\mathrm{SU}(2)^4$, and thus it
has  
$\varepsilon_{\Omega'} = -1$, $\varepsilon_{\hat g} = 1$ since $\Omega'$ 
contains the reflection in the $\cT_2^{(3)}$
directions.

All these bosonic fields are matrices, and must satisfy the
invariance constraint (\ref{invCP}) with $\Lambda$ replaced in turn by
$B_\ell$, $B_{\dot \ell}$ and $\chi$, with the values of $\varepsilon_{\Omega'}$
and $\varepsilon_{\hat g}$ specified above and listed in Tab.~2.
Using the explicit form (\ref{gamma-1}) for the matrices
$\Gamma_{\vec\xi}$, we see that this requires that
\begin{equation}
B_{\ell}= \begin{pmatrix} \,S\, & \,A\, \\ 
-A& \,S\,\end{pmatrix}~,~~~
B_{\dot\ell}= \begin{pmatrix} \,S_1\, & \,S_2\, \\ 
S_2& -S_1\end{pmatrix}~,~~~
\chi =\begin{pmatrix} 
\,A\, & \,S\, \\ 
-S& \,A\,\end{pmatrix}
~,
\label{modulins}
\end{equation}
where $S$, $S_1$ and $S_2$ are symmetric $(k\times k)$ matrices, and $A$ is an
anti-symmetric $(k\times k)$ matrix. Thus the scalars $B_\ell$ and
$\chi$ transform in the adjoint representation of $\mathrm{U}(k)$ (the first
embedded in the symmetric representation of $\mathrm{SO}(2k)$, the second in the
anti-symmetric one), while the scalars $B_{\dot\ell}$ transform in the symmetric
representation $\Ysymm$ of $\mathrm{U}(k)$ plus its conjugate
$\overline{\Ysymm}$.

A similar analysis can be performed in the R sector of the $(-1)/(-1)$ strings.
Here we have sixteen fermionic moduli which we can group into four sets $M_{\dot
\alpha a}$, $M_{\alpha  \dot a}$, $N_{ \alpha  a}$ and $N_{\dot \alpha  \dot
a}$ with $\alpha,\dot\alpha,a,\dot a$ labelling the spinor
representations of the four $\mathrm{SU}(2)$'s. We have denoted by $M$'s and
$N$'s the components with positive and negative $\mathrm{SO}(2)$ chiralities,
respectively, that correspond to eigenvalues plus and minus under $\Omega'$.
On the other hand, under $\hat g$ all modes carrying an index $\dot a$ pick up a
minus sign.
The resulting $\varepsilon_{\Omega'}$ and $\varepsilon_{\hat g}$ eigenvalues
have to be inserted into the constraint (\ref{invCP}) and determine the form of
the CP matrices.
We notice that $M_{\dot \alpha  a}$, $M_{\alpha  \dot a}$ have the same
eigenvalues as
$B_\ell$ and $B_{\dot \ell}$, and therefore they share the form of their 
CP factors. The other two sets of fermions satisfy again (\ref{invCP}) with
polarizations of the form
\begin{equation}
N_{\dot \alpha \dot a}= \begin{pmatrix} \,A_1\, & \,A_2\, \\ 
\,A_2\,& -A_1\end{pmatrix}~,~~~
N_{  \alpha a}=\begin{pmatrix} 
\,A\, & \,S\, \\ 
-S& \,A\,\end{pmatrix}
~,
\label{modulir}
\end{equation}
where $A$, $A_1$ and $A_2$ are anti-symmetric $(k\times k)$ matrices, and $S$ is
a symmetric $(k\times k)$ matrix. Thus, $N_{\dot \alpha \dot a}$ transform in
the
anti-symmetric representation $\Yasymm$ of $\mathrm{U}(k)$ plus its conjugate
$\overline{\Yasymm}$, while $N_{\alpha a}$ transform in the adjoint of
$\mathrm{U}(k)$.

For later convenience we have summarized the above properties of the neutral
moduli in Tab.~2, where in the last column we have indicated also their
scaling length dimensions. 
\begin{table}[htb]
\begin{center}
\begin{tabular}{|c|c|c|c|c|c|c|}
\hline
\begin{small} {\phantom{\vdots}}moduli\end{small}
&\begin{small} $\mathrm{SU}(2)^4$ \end{small}
&\begin{small} $\varepsilon_{\Omega'}$ \end{small}
&\begin{small} $\varepsilon_{\hat g}$ \end{small}
&\begin{small} CP factor \end{small} 
&\begin{small} $\mathrm{U}(k)$  \end{small}
&\begin{small} dimensions \end{small}
\\
\hline\hline
& & & & & & \\
{\phantom{\vdots}}
$\begin{array}{c} B_\ell \\  M_{\dot \alpha  a} \end{array}$
& $\begin{array}{c}\reps 2211 \\ \reps 1221\end{array}$
& $+$ & $+$
&\begin{small}
$\begin{pmatrix} \,S\, & \,A\, \\ 
-A& \,S\,\end{pmatrix}$ \end{small}
& adjoint  
& $\begin{array}{c}L^1\\L^{1/2}\end{array}$\\
& & & & & & \\
{\phantom{\vdots}}
$\begin{array}{c} B_{\dot\ell} \\ M_{\alpha  \dot a} \end{array}$
& $\begin{array}{c}\reps 1122 \\ \reps 2112\end{array}$
& $+$ & $-$
&\begin{small}
$\begin{pmatrix} \,S_1\, & \,S_2\, \\ 
S_2& -S_1\end{pmatrix}$ \end{small}
& $\Ysymm+ \overline{\Ysymm}$ 
& $\begin{array}{c}L^1\\L^{1/2}\end{array}$\\
& & & & & & \\
{\phantom{\vdots}}
$\begin{array}{c} N_{\dot \alpha \dot a}
\end{array}$
& $\begin{array}{c}\reps 1212 
\end{array}$
& $-$ & $-$
&\begin{small}
$\begin{pmatrix} \,A_1\, & \,A_2\, \\ 
A_2& -A_1\end{pmatrix}$ \end{small}
& $\Yasymm+ \overline{\Yasymm}$ 
& $\begin{array}{c}L^{-3/2}
\end{array}$\\
& & & & & &\\
{\phantom{\vdots}}
$\begin{array}{c}   N_{\alpha a} \\
\bar \chi \end{array}$
& $\begin{array}{c}\reps 2121 
\\ \reps 1111\end{array}$
& $-$ & $+$
&\begin{small}
$\begin{pmatrix} \,A\, & \,S\, \\ 
-S& \,A\,\end{pmatrix}$\end{small}
& adjoint 
& $\begin{array}{c}L^{-3/2}\\
L^{-1}\end{array}$ \\
& & & & & &\\
\hline
\end{tabular}
\label{tab:2}
\caption{The spectrum of neutral moduli aring from $(-1)/(-1)$ open strings
stretching between two D-instantons.}
\end{center}
\end{table}

It is useful to remark that the components of the moduli 
$B_\ell$ and $M_{\dot \alpha a} $ along the identity
play a distinguished 
r\^ole in the computation of D-instanton induced interactions;
in fact they do not interact with other moduli and correspond to
the 4-dimensional super-coordinates $x$ and $\theta$
\cite{Green:2000ke,Billo:2002hm}.
Due to their Chan-Paton structure, the moduli $B_{\dot\ell}$ and
$M_{\dot \alpha a} $, instead, have no trace part; their components $\hat x$
and $\hat\theta$ along $\sigma_3$ (in block-diagonal terms) do appear in the
moduli action, and thus cannot generically play the r\^ole of
supercoordinates in the internal directions along the orbifold. We will however
see later that effectively they behave as such when we compute certain types of
instanton-induced interactions.

\paragraph{$(-1)/7$ strings:} This is the charged sector that accounts for open
strings stretching between the $k$ D-instantons and the four D7-branes, whose CP
factors are $(2k\times 8)$ matrices. Since there are eight directions with mixed
Dirichlet-Neumann boundary conditions, in the NS sector it is not possible to
construct vertex operators of conformal weight one, and thus
there are no physical bosonic moduli in the spectrum. On the other hand, in the
R sector we do find physical moduli. These are fermionic scalars $\mu'$ that, in
order to survive the orbifold projection, must satisfy the following relation
\begin{equation}
\Gamma_{\vec\xi}^{*}(\hat g)\,\mu'\,
\gamma_{\vec\alpha }^T(\hat g) = \varepsilon_{\hat
g}\mu'~~~\mbox{with}~~\varepsilon_{\hat g}=+1~.
\label{gmu'}
\end{equation}
We do not get any further condition by applying the orientifold parity
$\Omega'$, since it exchanges the string orientation and thus leads to suitable
identifications between states of the $(-1)/7$ sector with those of the $7/(-1)$
one. Recalling the explicit form (\ref{gamma-1}) and (\ref{gamma7}) of the
matrices $\Gamma_{\vec\xi}(\hat g)$ and $\gamma_{\vec\alpha }(\hat g)$, we can
easily see that the above constraint implies that
\begin{equation}
\mu'= \begin{pmatrix} X_1 & \,X_2 \\ 
-X_2& \,X_1\end{pmatrix}
\label{mu'}
\end{equation}
where $X_1$ and $X_2$ are generic $(k\times 4)$ matrices. Thus, these fermionic
moduli organize into a complex scalar transforming in the
fundamental/anti-fundamental representation $(\Yfund,\overline{\Yfund})$ of the
symmetry group $\mathrm{U}(k)\times \mathrm{U}(4)$. 
Their properties are summarized in Tab.~3.
\begin{table}[htb]
\begin{center}
\begin{tabular}{|c|c|c|c|c|c|}
\hline
\begin{small} {\phantom{\vdots}}moduli\end{small}
&\begin{small} $\mathrm{SU}(2)^4$ \end{small}
&\begin{small} $\varepsilon_{\hat g}$ \end{small}
&\begin{small} CP factor \end{small} 
&\begin{small} $\mathrm{U}(k)\times \mathrm{U}(4)$ \end{small}
&\begin{small} dimensions \end{small} \\
\hline\hline
& & & & & \\
{\phantom{\vdots}}$\begin{array}{c}\mu'
\end{array}$  
& $\reps 1111$ & +
&\begin{small}
$\begin{pmatrix} X_1 & \,X_2 \\ 
-X_2& \,X_1\end{pmatrix}$ \end{small}
& $(\Yfund,\overline{\Yfund})$ 
& $\begin{array}{c}L^{1/2}
\end{array}$\\
& & & & & \\
\hline
\end{tabular}
\label{tab:3}
\caption{The spectrum of charged moduli arising from $(-1)/7$ open strings
stretching between $k$ D-instantons and four D7-branes.}
\end{center}
\end{table}

\paragraph{$(-1)/3$ strings:} Let us finally consider the flavored sector of the
instanton moduli space which arises from the open strings connecting the
D-instantons with the half D3-branes. As we have explained at the beginning
of this section, this sector exists only when the $\mathrm{D}(-1)$'s and the
D3's occupy the same fixed point. In our model, this happens in case
\emph{a)} considered in Fig.~\ref{fig:D7D3D-1_a}, with just one half
D3-brane at the fixed point of the D(--1)'s. The CP factors of the $(-1)/3$
moduli are
then $(2k\times 2)$ matrices transforming in some representation of
$\mathrm{U}(k)\times \mathrm{U}(1)$.
It will be useful in the following to consider the generalized case with $m$
half D3-branes suppporting a U$(m)$ symmetry, with $2k\times 2m$ CP factors;
the configuration \emph{a)} corresponds to $m=1$. In the case \emph{b)}
represented
in Fig.~\ref{fig:D7D3D-1_b}, the D3/D(--1) moduli are absent and we may say that
this case corresponds to $m=0$.
As usual in D(--1)/D3 systems, in the NS sector one finds two complex variables
$w_{\alpha}$ which transform as a chiral spinor with
respect to the $\mathrm{SO}(4)$ acting on the ND directions, namely belong
to the $\reps 2111$ of $\mathrm{SU}(2)^4$ in our language.
In order to survive the orbifold projection, they must satisfy the following
constraint:
\begin{equation}
\Gamma_{\vec\xi}^{*}(\hat g)\,w_{\alpha}\,
\gamma_{\vec\xi}^T(\hat g) = \varepsilon_{\hat
g}\,w_{\alpha}~~~\mbox{with}~~\varepsilon_{\hat g}=+1~.
\label{gw}
\end{equation}
Recalling Eqs. (\ref{gamma3}) and (\ref{gamma-1}), one can easily conclude that
\begin{equation}
w_{\alpha}= \begin{pmatrix} Y_1 & \,Y_2 \\ 
-Y_2& \,Y_1\end{pmatrix}~.
\label{w}
\end{equation}
{From} this we deduce that the moduli $w_{\alpha}$ transform in the fundamental
representation under $\mathrm{U}(k)$ and the anti-fundamental
under $\mathrm{U}(m)$.

In the R sector we find eight fermionic moduli which can be organized in 
two spinors, $\mu_a$ and $\mu_{\dot a}$, of of opposite chiralities with respect
to the internal $\widehat{\mathrm{SO}}(4)$. 
In particular, $\mu_a$ transforms in the $\reps 1121$ of $\mathrm{SU}(2)^4$ 
and is invariant under $\hat g$ like $w_\alpha$, with which it shares the same
CP structure (\ref{w}).
Instead, $\mu_{\dot a}$ belongs to $\reps 1112$ and is odd under $\hat g$,
leading to
\begin{equation}
\Gamma_{\vec\xi}^{*}(\hat g)\,\mu_{\dot a}\,
\gamma_{\vec\xi}^T(\hat g) = \varepsilon_{\hat g}\,\mu_{\dot
a}~~~\mbox{with}~~\varepsilon_{\hat g}=-1~.
\label{gmu}
\end{equation}
This is solved by taking
\begin{equation}
\mu_{\dot a}= \begin{pmatrix} Y_1 & \,Y_2 \\ 
\,Y_2& -Y_1\end{pmatrix}~,
\label{mu}
\end{equation}
implying that the $\mu_{\dot a}$'s transform in the fundamental representation
both of $\mathrm{U}(k)$
and of $\mathrm{U}(m)$. All this is
summarized in Tab.~4, where again the last column contains
the length dimensions of the various moduli.
\begin{table}[htb]
\begin{center}
\begin{tabular}{|c|c|c|c|c|c|}
\hline
\begin{small} \!{\phantom{\vdots}}moduli\end{small}
&\begin{small} $\mathrm{SU}(2)^4$ \end{small}
&\begin{small} $\varepsilon_{\hat g}$ \end{small}
&\begin{small} CP factor \end{small} 
&\begin{small} $\mathrm{U}(k)\times \mathrm{U}(m)$ \end{small}
&\begin{small} dimensions \end{small}
\\
\hline\hline
& & & & &\\
$\begin{array}{c} w_\alpha \\ \mu_a \end{array}$
& $\begin{array}{c} \reps 2111 \\ \reps 1121 \end{array}$
& $+$ 
&\begin{small}
$\begin{pmatrix} Y_1 & \,Y_2 \\ 
-Y_2& \,Y_1\end{pmatrix}$ \end{small}
& ($\Yfund,\overline{\Yfund})$ 
&$\begin{array}{c}L^{1}\\L^{1/2}\end{array}$ \\
& & & & & \\
$\begin{array}{c} \mu_{\dot a} 
\end{array}$
& $\reps 1112 $
& $-$
&\begin{small}
$\begin{pmatrix} Y_1 & \,Y_2 \\ 
\,Y_2& -Y_1\end{pmatrix}$ \end{small}
& $(\Yfund,\Yfund)$ 
& $\begin{array}{c}L^{1/2}
\end{array}$ \\
& & & & &\\
\hline
\end{tabular}
\label{tab:4}
\caption{The spectrum of flavored moduli aring from $(-1)/3$ open strings
stretching between $k$ D-instantons and one ``half'' D3-brane at the same fixed
point.} 
\end{center}
\end{table}

\section{D-instanton corrections from localization formul\ae}
\label{sec:loc}
The effective action on the D7-branes gets corrected by D-instantons. 
In our $\cN = 2$ setup these corrections are encoded in a prepotential
function of the chiral superfield $\Phi(x,\theta)$
containing the adjoint scalar $\varphi$ and the gauge field strength $F$ of the
U(4)
gauge group (see (\ref{vector})), and can be expressed as follows
\begin{equation}
\label{StoF}
 S_{\mathrm{n.p.}} \sim \int d^4x\,d^4\theta ~\cF_{\mathrm{n.p.}}(\Phi)~~+~
\mathrm{c.c.}
\end{equation}
The prepotential $\cF_{\mathrm{n.p.}}(\Phi)$ receives contributions from
D-instanton
configurations of type \emph{a)} and \emph{b)}, described in
Figs.~\ref{fig:D7D3D-1_a}
and \ref{fig:D7D3D-1_b}, corresponding , respectively, to 
instantons sitting on a fixed point occupied by a D3-brane or empty.
Taking into account the multiplicity of these configurations, we can write
\begin{equation}
\label{prepsum}
 \cF_{\mathrm{n.p.}}(\Phi) = 12\, \cF^{(m=0)}(\Phi) 
 + 4\, \cF^{(m=1)}(\Phi)~.
\end{equation}

The prepotentials $\cF^{(m)}$ can be expressed as an integral over
the ``centered'' moduli space $\widehat\cM_{k,m}$ of
the instantonic branes as follows
\begin{equation}
\label{prepphidef}
\cF^{(m)}(\Phi) =  \sum_k  q^k   \int d\widehat\cM_{k,m} ~
\ee^{-\Smod(\widehat\cM_{k,m},\Phi)}~.
\end{equation}
with $q=\ee^{\pi\ii \lambda}$. Here $(-\pi\ii k\lambda)$ is the classical action
of $k$ fractional 
half D-instantons, while $\Smod(\widehat\cM_{k,m},\Phi)$ is the action
describing the interactions of
the centered moduli ({\it i.e.} all moduli except $x$ and $\theta$)
among themselves and with the superfield $\Phi$; all these
interactions occur via disk diagrams%
\footnote{In principle one should include also annuli and M\"obius diagrams with
a boundary on the D-instantons \cite{Blumenhagen:2006xt}. In the present
model, such diagrams do not contribute: the D(-1)/D7 and D(-1)/O7 amplitudes are
related to the running of the quartic coupling on the D7's, which vanish in our
model \cite{Billo':2009gc}; the D(-1)/D3 and D(-1)/O3 amplitudes also
vanish.}
with at least part of their boundary
attached to D(--1) branes.

The moduli $x$ and $\theta$ play the r\^ole of the
$\cN = 2$ super-coordinates and appear in $\Smod(\widehat\cM_{k,m},\Phi)$ only
through
the superfield $\Phi(x,\theta)$.  This implies that during the calculation 
we can take $\Phi$ to be constant, and promote it to a full fledged dynamical
field only in the end. However, even with this
position, the integrals in \eq{prepphidef} remain rather cumbersome, and can be
explicitly performed only for very low instanton numbers, typically $k=1$.
Substantial progress can be achieved following
the seminal observation \cite{Nekrasov:2002qd} that, after suitable deformations
of the moduli action, 
the integrals localize around isolated points in the instanton moduli space,
and that an explicit result for the prepotential can be obtained 
after turning off in a controlled way the deformations.
This idea has already been made systematic and applied with success in 
several interesting contexts
\cite{Bruzzo:2002xf,Nekrasov:2003rj,Marino:2004cn,Billo:2009di,Fucito:2009rs}. 
Here we put it at work for a system
of $k$ D(--1)-instantons, $m$ D3-branes and $N$ D7-branes in presence
of O7- and O3-planes, and present explicit computations for the relevant cases
with $N=4$ and $m=0,1$, up to $k=3$.

We first take $\Phi=\diag(a_1,\ldots, a_N,-a_1,\ldots, -a_N)$ where $a_u$ are
constant
expectation values along the Cartan directions of $\mathrm{U}(N)$, and then
consider the $\epsilon$-deformed instanton partition function
\begin{equation}
\label{ipf0}
Z^{(m)}(a,\epsilon) =\sum_k q^k \,Z^{(m)}_k(a,\epsilon) 
= \sum_k q^k \int \!d\cM_{k,m} ~\ee^{-\Smod^{\epsilon}(\cM_{k,m},a)}~.
\end{equation}
Here we have conventionally set $Z^{(m)}_0(a,\epsilon)=1$, and introduced
$\Smod^{\epsilon}$ which
is obtained by deforming $\Smod$ with Lorentz breaking terms parameterized by
four parameters $\epsilon_I$ describing rotations along the four Cartan
directions of 
$\mathrm{SO}(4) \times \widehat{\mathrm{SO}}(4)$.
{From} the string perspective, these deformations can
be obtained by switching on suitable RR background fluxes on the D7-branes, as
shown in
Refs.~\cite{Billo:2006jm,Billo:2009di}. Notice that integrals in \eq{ipf0} 
run over all moduli, including the ``center of mass'' super-coordinates
$x$ and $\theta$. In presence of the $\epsilon$-deformations it is rather easy
to see 
that the integration over the super-space yields a volume factor growing as
$1/(\epsilon_1\epsilon_2)$ in the limit of small $\epsilon_{1,2}$.  
Therefore, to obtain the integral over the centered moduli this factor has to be
removed. 
In addition, we have to take into account the fact that the $k$-th order in the
$q$-expansion receives contributions not only from genuine $k$-instanton
configurations but also from disconnected ones, corresponding to copies of
instantons of lower numbers $k_i$ such that $\sum k_i = k$. To isolate
the connected components we have to take the logarithm of $Z^{(m)}(a,\epsilon)$.
Thus, we are led to
consider
\begin{equation}
\label{ipf}
\cF^{(m)}(a,\epsilon) =\epsilon_1 \epsilon_2  \log  Z^{(m)}(a,\epsilon) ~.
\end{equation}
The prepotential will be extracted from  $\cF^{(m)}(a,\epsilon)$ after taking
the appropriate $\epsilon_I \to 0$ limit and replacing $a$ with the complete
superfield $\Phi$. It is worth remarking 
that the function $\cF^{(m)}(a,\epsilon)$ contains also information about more
general 
interactions in the four-dimensional theory such as non-perturbative
gravitational couplings, flux induced mass terms, etc. For instance, the
gravitational terms can be extracted from 
$\cF^{(m)}(a,\epsilon)$ after promoting $\epsilon_{1,2}$ to dynamical
superfields describing a graviphoton multiplet, as done in
Ref.~\cite{Billo:2006jm} or, in the eight-dimensional context, in
Refs.~\cite{Billo:2009di,Fucito:2009rs}.
Similarly, the terms involving $\epsilon_{3,4}$ can be interpreted as mass
deformations for the antisymmetric matter that are induced by RR fluxes. In this
paper we focus only on corrections 
to gauge kinetic functions, and therefore higher order terms in $\epsilon$'s
will be systematically 
discarded.

\subsection{Localization formul\ae}
\label{subsec:locfor}
The localization procedure is based on the co-homological structure of the
instanton moduli action which is exact with respect to a suitable BRST
charge $Q$:
\begin{equation} 
\Smod= Q \Xi~.
\label{q}
\end{equation}
$Q$ can be obtained by choosing any component of the supersymmetry charges
preserved on the brane system. Supersymmetry charges
 are invariant under $\mathrm{U}(k)\times \mathrm{U}(m)\times \mathrm{U}(N)$
but transform as a spinor of $\mathrm{SO}(4)^2$, so that the choice of $Q$
breaks 
this symmetry to the $\mathrm{SU}(2)^3$ subgroup which preserves this spinor.
In our case we take%
\footnote{In the D(--1)/D7 system considered in Ref.~\cite{Billo:2009di}, the
subgroup 
of the SO(8) Lorentz symmetry preserving a fixed spinor is
SO(7), embedded in such a way that a vector of SO(8) becomes a spinor. In
our case, \eq{su23def} represents the subgroup of this SO(7) which is
compatible with the 4 $+$ 4 split of the eight-dimensional space induced by the
orbifold compactification on $\cT_4/\Z_2$.}
\begin{equation}
\label{su23def}
\mathrm{SU}(2)_1\times \mathrm{SU}(2)_2 \times \mathrm{SU}(2)_3
= \mathrm{SU}(2)_- \times \widehat{\mathrm{SU}}(2)_- \times 
\diag\left[\mathrm{SU}(2)_+ \times
\widehat{\mathrm{SU}}(2)_+\right]~.
\end{equation}
This reduction is achieved by identifying the spinor indices ``$\alpha$'' and
``$a$''
of the first and third $\mathrm{SU}(2)$'s in Tabs.~2, 3 and 4 of
Section~\ref{sec:Dmod}.
After this identification is made, the fermionic moduli $M_{\dot\alpha a}$ and
$M_{\alpha \dot a}$ 
can be renamed as $M_{\ell=\dot\alpha\alpha}$ and $M_{\dot \ell=a\dot a}$,
and paired with $B_\ell$ and $B_{\dot \ell}$ into BRST multiplets.
Similarly, the singlet component $\eta\equiv N_{\alpha a}\epsilon^{\alpha a}$
and
the $(-1)/3$ fermionic moduli $\mu_{\alpha=a}$ have the right transformation
properties 
to qualify for the BRST partners of $\bar\chi$ and  $w_\alpha$ respectively.

The remaining fields $N_m\equiv\sigma_m^{\alpha a}N_{\alpha a}$, $N_{\dot \alpha
\dot a}$, $\mu_{\dot a}$ and $\mu'$ are unpaired, and should be supplemented
with auxiliary fields
having identical transformation properties. We denote such fields
as $d_{m}$, $D_{\dot \alpha \dot a}$, $h_{\dot a}$ and $h'$, respectively. 
In ordinary cases the auxiliary fields collect the D- and F-terms of 
the gauge theory on the D(--1)'s, and the corresponding D- and F-flatness
conditions 
are the ADHM constraints on the instanton moduli space (see for example
Refs.~\cite{Bruzzo:2002xf,Fucito:2001ha,Bruzzo:2003rw} for details). In our case
we have 
an extension of this construction, defining a sort of generalized ``exotic''
instanton moduli space. 
More precisely, the seven auxiliary moduli $d_{m}$, $D_{\dot \alpha \dot a}$, of
dimension $L^2$,     linearize the quartic interactions among the scalars
$B_\ell$ and $B_{\dot\ell}$ and correspond to vertex operators%
\footnote{One can see that vertices associated to the moduli $D_{\dot \alpha
\dot a}$ transform under the discrete parities $\Omega'$ and $g$ in the same way
as the fermionic moduli $N_{\dot \alpha \dot a}$, while the triplet $d_{m}$
transforms like the
fermions $N_{m}$. Thus, the structure of their CP factors match that of their
BRST partners as expected.}
that are bilinear in the fermionic string coordinates
\cite{Billo:2002hm,Billo:2009di}.
In particular, the triplet $d_{m}$ disentangles the quartic
interactions of $B_\ell$ and $B_{\dot\ell}$ among themselves, while the quartet
$D_{\dot \alpha \dot a}$ decouples the quartic interactions
between $B_\ell$ and $B_{\dot\ell}$. Likewise,
the dimensionless auxiliary $(-1)/3$ moduli $h_a$ disentangle the 
quartic interactions between $B_{\dot\ell}$ and $w_\alpha$.
Finally, $h'$ completes the $(-1)/7$ BRST multiplet. In the end, $\chi$ remains 
unpaired and therefore $Q\chi=0$. 

In this way, all moduli but $\chi$ form
BRST doublets, which we will schematically denote as $\big(\phi,\psi\equiv
Q\phi\big)$ in the following and which are explicitly listed in the first column
of Tab.~5. 
Note that $\phi$ is a boson if the multiplet 
is built out of physical moduli, and is a fermion if instead it contains
auxiliary fields.
Indeed, the auxiliary fields, being related to D- and F-terms, can only appear
as highest
components in the BRST multiplets while the physical bosonic moduli enter as the
lowest
components of the pair. These statistical properties are listed in the second
column of
Tab.~5.  With all these ingredients at hand, one can show that the moduli action
$\Smod$ can be written in the form (\ref{q}). The details of the fermion $\Xi$
are irrelevant to the computation, 
since integrals are insensitive to $Q$-exact terms.  

Since the length dimension of the BRST charge is $L^{-1/2}$,
the length dimensions of the components $(\phi,\psi)$ of $Q$-multiplet are
$\big(\Delta,\Delta-\ft12\big)$. Thus, recalling that a fermionic variable and
its differential 
have opposite dimensions, we find that the measure on the instanton moduli space
\begin{equation}
d\cM_{k,m} \equiv  d\chi\,\prod_{(\phi,\psi)} d\phi ~d\psi 
\label{meas}
\end{equation}
has the following scaling dimensions
\begin{equation}
L^{-k^2 + \frac 12 (n_+-n_-)}~.
\label{dim} 
\end{equation}
Here, the first term in the exponent accounts for the unpaired $k^2$ bosonic
moduli $\chi$, of dimension $L^{-1}$, and $n_\pm$ denotes the number of
$Q$-multiplets where the statistics of
the lowest component is $(-)^{F_\phi}=\pm$. Using Tab.~5, we can explicitly
rewrite
\eq{dim} as  
\begin{equation}
L^{-k^2 +\frac 12(n_{B} + n_{\bar\chi} + n_w -
n_N - n_\mu - n_{\mu'})} = L^{\frac{4 - N}{2}\,k}
\label{ridim}
\end{equation}
where $n_\phi$ is the number of real component of a modulus of type $\phi$. The
measure is therefore dimensionless for $N=4$, {\it i.e.}  for the
$\mathrm{U}(4)$ D7-brane gauge theory in our model. Note that this result is
independent from the number $m$ of D3-branes at the fixed point where the
instantons sit, and therefore holds for both the two relevant cases $m=0,1$ in
our setup.

To localize the integral over moduli space, it is necessary to make the charge
$Q$ equivariant with respect to all symmetries, which in our case are the gauge
symmetry $\mathrm{U}(k)\times \mathrm{U}(N)\times \mathrm{U}(m)$, and the
residual Lorentz symmetry $\mathrm{SU}(2)^3$. For our purposes it is enough to
consider the Cartan directions of the various groups. 
We label the Cartan components of the U($k$) parameters of
$Q$ by $\vec \chi$, those of U($m$), those of $\mathrm{U}(m)$ by
$\vec{b}$ and those of $\mathrm{U}(N)$ by $\vec{a}$. {From} the string
perspective $\vec \chi$, $\vec b$ and $\vec a$ parametrize, respectively, the
positions of the D(--1), D3 and D7-branes along the overall transverse
two-dimensional  plane, and their appearance in the moduli action can be deduced
from disk amplitudes with (part of) their boundary on the D-instantons and with
insertion of (--1)/(--1), 3/3 or 7/7 fields. Thus, $\vec a$ can be interpreted
as the vacuum expectation value of the chiral superfield $\Phi$ of gauge theory
on the D7-branes, and $\vec b$ as the analogue for the gauge theory on the
D3-branes. Finally, the Cartan directions of the residual Lorentz group
$\mathrm{SU}(2)^3$ are parametrized by   $\epsilon_I$ ($I=1,\ldots,4$) subject
to the constraint
\begin{equation}
\epsilon_1+ \epsilon_2+ \epsilon_3 +\epsilon_4=0 ~.
\label{zero}
\end{equation}
Although only three out of the four $\epsilon$'s are independent variables, 
it is convenient during the computation to keep all of them as independent
variables and impose 
the relation (\ref{zero}) only at the very end.

After the equivariant deformation, the charge $Q$ becomes
nilpotent up to an element of the symmetry group. It is convenient to 
use the basis provided by the weights of this group, and thus we 
denote by $\phi_q$ and
$\psi_q$ the components of $\phi$ and $\psi$ along a weight
\begin{equation}
\label{wd}
\vec q \equiv \bigl({\vec q}_{\mathrm{U}(k)}, \vec q_{\mathrm{U}(N)},
{\vec q}_{\mathrm{U}(m)}, {\vec q}_{\mathrm{SU}(2)^3} \bigr)\in \cW(\phi)~,
\end{equation}
where $\cW(\phi)$ is the set of weights of the representation
under which $\phi$ transforms, which can be read from the third and fourth
columns 
of Tab.~5. Then, in this basis the charge $Q$ acts diagonally as follows
\begin{equation}
\label{brspair3}
Q\phi_q = \psi_q~,~~~
Q\psi_q = \Omega_q \phi_q~,
\end{equation}
where $\Omega_q$ parametrizes the equivariant deformation, i.e. the eigenvalues
of $Q^2$.
From the brane perspective, $\Omega_q$ specifies the distance 
in the overall two-dimensional transverse plane between the branes at the two
endpoints of 
the open string. Explicitly, we have
\begin{equation}
\label{omephis}
 \Omega_q = \vec\chi\cdot {\vec q}_{\mathrm{U}(k)} + \vec a \cdot 
 \vec q_{\mathrm{U}(N)} + \vec b \cdot {\vec q}_{\mathrm{U}(m)}
 + \vec \epsilon \cdot {\vec q}_{\mathrm{SU}(2)^3}~.
\end{equation}
The $\vec \epsilon \cdot {\vec q}_{\mathrm{SU}(2)^3}$ eigenvalues appearing
above
can be deduced from  $\vec \epsilon \cdot {\vec q}_{\mathrm{SO}(4)^2}$, 
where $\vec q_{\mathrm{SO}(4)^2}$ is the $\mathrm{SO}(4) \times
\widehat{\mathrm{SO}}(4)$ weight vectors of the physical moduli inside each
multiplet, using the relation (\ref{zero}) among the $\epsilon_I$ parameters. 
For example, the complex moduli $B_\ell$, transforming
as a vector of the first $\mathrm{SO}(4)$, have $ {\vec q}_{\mathrm{SO}(4)^2}$
weights $(\pm1,0,0,0)$ or $(0,\pm1,0,0)$, and thus
their contribution to $\Omega_q$ is $\pm\epsilon_{1}$ or $\pm\epsilon_2$.
Similarly, for $B_{\dot\ell}$ which is a vector of $\widehat{\mathrm{SO}}(4)$,
we find $\pm\epsilon_3$
or $\pm\epsilon_4$. The same results are found for the $M$-fermions which
transform as a right spinor
of $\mathrm{SO}(4) \times \widehat{\mathrm{SO}}(4)$ and have weights
$(\pm\ft12,\pm\ft12,\pm\ft12,\pm\ft12)$ with an odd number of plus signs.
For the $N$-fermions, transforming instead as a left spinor with an even number
of plus signs, 
after using \eq{zero} we find $0$, $\pm\ft12(\epsilon_1 + \epsilon_2)$,
$\pm\ft12(\epsilon_1 + \epsilon_3)$ and $\pm\ft12(\epsilon_2 + \epsilon_3)$.
Finally, $w_\alpha$ transforming
as a right spinor of $\mathrm{SO}(4)$ has eigenvalues $\pm\ft12(\epsilon_1 +
\epsilon_2)$, while
$\mu_{\dot a}$ transforming as a left spinor of $\widehat{\mathrm{SO}}(4)$
corresponds to $\pm\ft12(\epsilon_3 - \epsilon_4)$.
Alternatively, these eigenvalues can be read from the formula
\begin{equation}
\label{qsu23}
\vec \epsilon \cdot {\vec q}_{\mathrm{SU}(2)^3}
= q_1(\epsilon_1 - \epsilon_2)+ q_2( \epsilon_3  - \epsilon_4)+q_3(\epsilon_1 +
\epsilon_2)   
\end{equation}
with $q_i=0$ for states in the ${\mathbf 1}$,
$q_i= \pm\ft12$ for states in the ${\mathbf 2}$ and so on%
\footnote{To see this, associate to each modulus the $\mathrm{SU}(2)^4$ charges
$q_\pm, \hat q_\pm$ and the eigenvalue
$\vec \epsilon \cdot {\vec q}_{\mathrm{SU}(2)^4}=\epsilon_1
(q_++q_-)+\epsilon_2(q_+-q_-)+\epsilon_3 (\hat q_+ +\hat q_-)+\epsilon_4(\hat
q_+ -\hat q_-)$. Then, \eq{qsu23} follows
after the identification $q_1=q_-$, $q_2=\hat{q}_-$, $q_3=q_+-\hat{q}_+$ and the
use of (\ref{zero}).
For example, $B_{1,2} \in \repst 212$ have $\mathrm{SU}(2)^3$ weights
$(\pm\ft12,0,\pm\ft12)$
that once plugged into \eq{qsu23} lead to $\pm\epsilon_1$ and $\pm
\epsilon_2$.}.
All this is summarized in the last column of Tab.~5, where we have displayed 
the positive eigenvalues of $\vec \epsilon \cdot {\vec q}_{\mathrm{SU}(2)^3}$
(assuming 
$\epsilon_1>\epsilon_2>\epsilon_3>\epsilon_4$) corresponding to the holomorphic
components of the various fields.

\begin{table}[htb]
\begin{center}
\begin{tabular}{|c|c|c|c|c|}
\hline
\begin{small} $(\phi,\psi)$ \end{small}
&\begin{small} $(-)^{F_\phi}$ \end{small}
&\begin{small} $\mathrm{U}(k) \times \mathrm{U}(N)\times \mathrm{U}(m)$
\end{small} 
&\begin{small} $\mathrm{SU}(2)^3$ \end{small}
&\begin{small}  $\vec\epsilon\cdot \vec{q}_{SU(2)^3}$ \end{small}
\\
\hline\hline
& & & & \\
$(B_\ell,M_\ell)$ & + & $\bigl(\mbox{adj}, \bone, \bone\bigr)$ 
& $\repst 212$ &   $ \epsilon_1,\epsilon_2$\\
$(B_{\dot\ell},M_{\dot\ell})$ & + 
& $\bigl(\Ysymm, \bone, \bone\bigr) + \mbox{h.c.}$
& $\repst 122$   & $ \epsilon_3,\epsilon_4$\\
$(N_{\dot\alpha\dot a},D_{\dot\alpha\dot a})$ & $-$ 
& $\bigl(\Yasymm, \bone, \bone\bigr) + \mbox{h.c.}$
& $\repst 221$   & $ \epsilon_2+\epsilon_3,\epsilon_1+\epsilon_3$\\
$(N_{m},d_{m})$ & $-$ & $\bigl(\mbox{adj}, \bone, \bone\bigr)$ 
& $\repst 113$ &  $ 0_{\R},\epsilon_1+\epsilon_2$\\
$(\bar\chi,\eta)$ & $+$ & $\bigl(\mbox{adj}, \bone, \bone\bigr)$ 
& $\repst 111$ &   $ 0_{\R} $\\
& & & &  \\
$(\mu',h')$ & $-$ 
& $\bigl(\Yfund, \overline{\Yfund}, \bone\bigr) + \mbox{h.c.}$
& $\repst 111$    & $ 0$\\
& & & & \\
$(w_\alpha,\mu_\alpha)$ & $+$ 
& $\bigl(\Yfund, \bone, \overline{\Yfund}\bigr) + \mbox{h.c.}$
& $\repst 112$   & $ \ft12(\epsilon_1+\epsilon_2 )$  \\
$(\mu_{\dot a},h_{\dot a})$ & $-$ 
& $\bigl(\Yfund, \bone, \Yfund\bigr) + \mbox{h.c.}$
& $\repst 121$ & $ \ft12(\epsilon_3-\epsilon_4 )$ \\
& & & & \\
\hline
\end{tabular}
\caption{BRST structure and symmetry properties of the D(--1)/D3/D7 moduli. With
$(-)^{F_\phi}=\pm$ we denote the statistics, bosonic or fermionic, of the lower
component
of the doublet.
The  third and fourth columns report the transformation properties under the
symmetry groups.
The last column collects the eigenvalues  $\vec\epsilon\cdot \vec{q}_{SU(2)^3}$
for the positive weights $\vec{q}$'s specified in the third column.}
\end{center}
\label{tab:bs}
\end{table}

The complete localization of the integral around isolated fixed points implies
that the integral
is given by  the (super)-determinant of $Q^2$ evaluated at the fix points of
$Q$ 
\cite{Nekrasov:2002qd,Bruzzo:2002xf,Bruzzo:2003rw}.
As we already mentioned, the moduli $\chi$ and $\bar\chi$ appear very
asymmetrically
in the BRST formalism: $\chi$ parametrizes the $\mathrm{U}(k)$ gauge rotations,
while $\bar\chi$ falls into one of the doublets. Moreover, 
the contribution of the  $(\bar\chi,\eta)$ multiplet to the super-determinant
cancels against an identical contribution coming from a real component in
$(N_m,d_m)$ with identical
transformation properties and opposite statistics.
After discarding these contributions, the super-determinant of $Q^2$ takes 
a simple product form in terms of the $\Omega_q$-eigenvalues over complex
variables 
and can be restricted to holomorphic components of the latter, corresponding to
the
positive weights $\in \cW^+(\phi)$.
Thus, the $k$-instanton partition function $Z_k^{(m)}$ of \eq{ipf0}, now
deformed also with
the $\mathrm{U}(m)$ parameters $b$, is given by the localization formula
\begin{equation}
\begin{aligned}
 Z_k^{(m)}(a,b,\epsilon)  &= \int d\cM_{k,m} 
~\ee^{-\Smod^\epsilon(\cM_{k,m},a,b)}
=\int d{\chi} \, \prod_{(\phi,\psi)} d\phi \,d\psi~
\ee^{-Q\Xi}\\
& = 
\int \prod_{i=1}^k \frac{d{\chi_i}}{2\pi\ii}~ \prod_{i<j}^{k} (\chi_i
-\chi_j)^2\,
\prod_\phi\!
\prod_{q\in\cW^+(\phi)} \Omega_{q}^{-(-)^{F_\phi} }~.
\end{aligned}
\label{zgeneral}
\end{equation}
The factor $\prod_{i<j}(\chi_i -\chi_j)^2$, known as Vandermonde determinant, 
comes from the Jacobian resulting from bringing
$\chi$ into the diagonal form $\chi={\rm diag}(\chi_1,\chi_2,...\chi_k)$.  
The integral over  $\chi_i$ in the second line above has to be thought of as a
multiple contour integral, according to the prescription introduced in
Ref.~\cite{Moore:1998et}.

The explicit expression for the products appearing in \eq{zgeneral} can be
easily deduced from \eq{omephis} by considering in turn, for each
modulus $\phi$ in Tab.~5, the set of weights corresponding to its symmetry
representations.
Introducing for notational convenience
\begin{equation}
\label{stoepsi}
s_1 = \epsilon_2 +\epsilon_3~,~~~
s_2 =  \epsilon_1+ \epsilon_3~,~~~
s_3 = \epsilon_1 + \epsilon_2~,
\end{equation}
the products for the (--1)/(--1) moduli can be written as
\begin{equation}
\begin{aligned}
\prod_{\vec q \in {\cal W}^+(B_\ell)} \Omega^{-(-)^{F_\phi}}_q &=
\prod_{{\ell}=1}^2  \prod_{i\leq j}^k  \Big(
(\chi_i-\chi_j)^2-\epsilon_{\ell}^2\Big)^{-1}~,\\
\prod_{\vec q \in {\cal W}^+(B_{\dot \ell})} \Omega_q^{-(-)^{F_\phi}} &= 
\prod_{{\dot \ell}=3}^4\prod_{i<j}^{k} \Big((\chi_i+\chi_j)^2 -
\epsilon_{\dot\ell}^2\Big)^{-1} ~ \prod_{i=1}^{k} \Big(4\chi_i^2-\epsilon_{\dot
\ell}^2\Big)^{-1}~,\\
\prod_{\vec q \in {\cal W}^+(N_{\dot \alpha\dot a})} \Omega_q^{-(-)^{F_\phi}} &=
\prod_{\ell=1}^2 \prod_{i<j}^{k} \Big((\chi_i +\chi_j)^2-s_\ell^2\Big)~, \\
\prod_{\vec q \in {\cal W}^{+}(N_m,\bar\chi)} \Omega_q^{-(-)^{F_\phi}}  &=
\prod_{i\leq j}^{k} \Big((\chi_i -\chi_j)^2-s_{3}^2\Big) ~,
\end{aligned}
\label{prodsome}
\end{equation} 
while for the (--1)/7 moduli we have
\begin{equation}
\label{prodsome1}
\prod_{\vec q \in {\cal W}^+(\mu')} \Omega_q^{-(-)^{F_\phi}}  = 
\prod_{i=1}^k \prod_{u=1}^n \Big(\chi_i - a_u\Big)~,
\end{equation}
and for the (--1)/3 moduli we have
\begin{equation}
\begin{aligned}
\prod_{\vec q \in {\cal W}^+(\omega_{\alpha)}} \Omega_q^{-(-)^{F_\phi}}  &=
\prod_{i=1}^k \prod_{r=1}^m \Big((\chi_i - b_r)^2
-\frac{(\epsilon_1+\epsilon_2)^2}{4}\Big)^{-1}~,\\
\prod_{\vec q \in {\cal W}^+(\mu_{\dot a})} \Omega_q^{-(-)^{F_\phi}}  &=
\prod_{i=1}^k \prod_{r=1}^m \Big((\chi_i +
b_r)^2-\frac{(\epsilon_3-\epsilon_4)^2}{4}\Big)~.
\end{aligned}
\label{prodsome2}
\end{equation}
Putting everything together, the instanton partition function (\ref{zgeneral})
is then given by
\begin{eqnarray}
 Z_k^{(m)}(a,b,\epsilon)   &=& \left( \frac{s_3}{\epsilon_1 \epsilon_2 }
\right)^k\int \prod_{i=1}^k \!\frac{d{\chi_i}}{2\pi\ii} ~
\prod_{i<j}^{k} \big(\chi_i -\chi_j\big)^2\,\Big( (\chi_i -\chi_j)^2-s_{3}^2
\Big)\nonumber\\
&&\times  \prod_{i<j}^{k}\, \prod_{\ell=1}^{2} 
\frac{ \Big((\chi_i +\chi_j)^2-s^2_{\ell}\Big)}
{\Big((\chi_i-\chi_j)^2-\epsilon_{\ell}^2\Big)
\Big( (\chi_i+\chi_j)^2-\epsilon_{\ell+2}^2\Big)}\label{Z}\\
&&\times 
\prod_{i=1}^k\left[\,\prod_{\ell=1}^{2} \frac{1}{\Big(4\chi_i^2
-\epsilon^2_{\ell+2}\Big)} \,
\prod_{r=1}^{m}
\frac{\Big(( \chi_i
+b_r)^2-\frac{(\epsilon_3-\epsilon_4)^2}{4}\Big)}{\Big((\chi_i
-b_r)^2-\frac{(\epsilon_1+\epsilon_2)^2}{4}\Big)} 
\,\prod_{u=1}^{n} \Big(\chi_i-a_u\Big)\right]~.
\nonumber
\end{eqnarray}
This multiple integral should be supplemented by a pole prescription. Here,
inspired by the
prescription of Ref.~\cite{Moore:1998et}, we take 
$\mathrm{Im} b_r=0$ and $\mathrm{Im} \epsilon_1\gg \mathrm{Im} \epsilon_2\gg
\mathrm{Im} \epsilon_3
\gg \mathrm{Im} \epsilon_4 >0$,
and compute the integrals closing the contours in the upper half-plane
$\mathrm{Im} \chi_i>0$.
In Appendix~\ref{app:3inst}, we provide
explicit results of these integrals up to $k=3$ instantons.

It is interesting to observe that the exotic instantons we have considered here
can also be re-interpreted as standard gauge instantons from the D3-brane
perspective. Indeed, the D(--1)-instanton partition function (\ref{Z}) coincides
with that describing gauge instantons in a $\mathrm{U}(m)$ gauge theory, with
two antisymmetric hyper-multiplets with masses $-\epsilon_3$ and $-\epsilon_4$,
and four fundamentals with masses $a_u$. The cases we are interested in,
$m=0,1$, correspond however to a rather bizarre choice of the gauge theory where
the standard field theory notions tend to lose their meaning. In this sense, our
exotic instantons can be thought as a extrapolation of ordinary gauge instanton
effects to degenerated limits of quantum field theories. 

\subsection{Non-perturbative prepotential}
\label{subsec:prepot}
%
In order to obtain the non-perturbative prepotential for the D7-brane gauge
theory 
from the partition function $Z^{(m)}(a,b,\epsilon)$, we first set 
the vacuum expectation values $b_r=0$ of the 3/3 scalars
to zero, since in our string vacua the D3-branes are fixed at one of the
orbifold fixed-points.
Thus, from now on we will not consider any more the $b$-dependence of the
instanton partition
function.
As a second step, we take the limit $\epsilon_I\to 0$ to remove the Lorentz
breaking deformations. 
A simple inspection of the explicit results for $\log Z^{(m)}(a,\epsilon)$ given
in 
Eqs.~(\ref{z30}) and (\ref{z31}), shows that this expression diverges as 
$1/(\epsilon_1 \epsilon_2\epsilon_3 \epsilon_4)$ in this limit. Such a
divergence
is typical of interactions in eight dimensions where the $\cN=2$ super-space
volume grows
like $\int d^8x d^8 \theta \sim 1/(\epsilon_1 \epsilon_2\epsilon_3 \epsilon_4)$.
Indeed, although  the $\sigma_3$-trace components $\hat x$ and $\hat\theta$ of
the
moduli $B_{\dot \ell}$ and $M_{\dot \ell}$ do not in general decouple from the
moduli action, they do in some of the fixed points that contribute to the
completely localized integral. In these points, $\hat x$ and $\hat\theta$
effectively represent the
super-coordinates of the internal orbifold where the D7-branes are wrapped, and
together with the true super-space coordinates $x$ and $\theta$ reconstruct an
eight-dimensional volume factor. 
These contributions can then be thought of as coming from regular
D(--1)-instantons moving in the full eight-dimensional world-volume of the
D7-branes.
Moreover, from the explicit results presented in Appendix~\ref{app:3inst}, we
can see that the terms
proportional to $1/(\epsilon_1 \epsilon_2\epsilon_3 \epsilon_4)$ in  $\log
Z^{(m)}(a,\epsilon)$ are independent of $m$ (and also of $b$'s if we keep these
parameters switched on). 
Thus, they can be associated to a universal quartic prepotential defined as
\begin{equation}
\label{F4}
{\mathcal F}_{\mathrm{IV}}(a)
=  \lim_{\epsilon_I\to 0} \epsilon_1 \epsilon_2\epsilon_3 \epsilon_4  
\log Z^{(m)}(a,\epsilon)~.
\end{equation}
Explicitly we have
\begin{equation}
\label{F4b}
{\mathcal F}_{\mathrm{IV}}(a) =  \Big( 4 a_1 a_2 a_3 a_4\Big)\,q  
-\Big(\sum_{i<j} a_i^2 a_j^2\Big) \, q^2  + \Big( \frac{16}3 a_1 a_2 a_3
a_4\Big)\,q^3 +\ldots
\end{equation}
which has indeed quartic mass dimension.
Surprisingly, the result (\ref{F4b}) matches precisely (half of) the
non-perturbative 
quartic prepotential induced by D-instantons on the parent eight-dimensional
SO(8) 
gauge theory living on D7-branes of type I$^\prime$ \cite{Billo:2009di}. 
This is consistent with the interpretation of these contributions as coming from
bulk or regular
instantons, since such instantons  are insensitive to the $\Z_2$-orbifold
projection. 

More interestingly, we can extract a finite quadratic prepotential 
by subtracting the divergence coming from the eight-dimensional interactions.
This quadratic 
prepotential is defined as
\begin{equation}
\label{F2}
{\mathcal F}_{\mathrm{II}}^{(m)}(a)
=  \lim_{\epsilon_{I}\to 0} \Big(
\epsilon_1 \epsilon_2 \log Z^{(m)}(a,\epsilon) 
-\frac{1}{\epsilon_3 \epsilon_4} {\cal F}_{\mathrm{IV}} \Big)~.
\end{equation}
Since the moduli measure is dimensionless, as shown in \eq {ridim}, no
dynamically 
generated scale may appear and the contributions at\emph{all} instanton numbers 
must be constructed only out of the $a$'s once the $\epsilon$-deformations are
switched off.
This is what happens. In fact, for $m=0,1$ we find the following quadratic
prepotentials
\begin{equation}
\begin{aligned}
 \!\!{\mathcal F}_{\mathrm{II}}^{(m=0)}(a) 
&=\Big(\!\!-\sum_{i<j}a_ia_j \Big) \, q +
\Big(\sum_{i<j}a_ia_j-\frac14\,\sum_ia_i^2 
\Big)\,q^2 +\Big(\!\!-\frac{4}{3}\sum_{i<j}a_ia_j \Big)\,q^3+\cdots~,\\
\!\!{\mathcal F}_{\mathrm{II}}^{(m=1)}(a) 
&=\Big(3\sum_{i<j}a_ia_j \Big) \, q +
\Big(\sum_{i<j}a_ia_j+\frac74\,\sum_ia_i^2 
\Big)\,q^2 +\Big(4\sum_{i<j}a_ia_j \Big)\,q^3+\cdots~.
\end{aligned}
\label{F2b}
\end{equation}
The total quadratic prepotential ${\mathcal F}{\mathrm{n.p.}}(a)$, which takes
into account the contributions from the various $m=0,1$ configurations with
their appropriate multiplicity, 
is obtained inserting \eq{F2b} into \eq{prepsum}, and reads
\begin{equation}
\label{Ff}
{\mathcal F}_{\mathrm{n.p.}}(a) = 
4  \Big[ 2\big(\tr a\big)^2 - \tr a^2 \Big]\, q^2 + O(q^4)~,
\end{equation}
where we have rewritten in a basis-independent way the sums over the $a$'s. It
is important
to stress that the terms proportional to $q$ and $q^3$ cancel when we sum over
all possible
D-instanton configurations, and that the relative factor inside the square
brackets is a consequence
of the explicit numerical coefficients we have found in evaluating the instanton
integrals using
the localization technique. 

We can now promote the vacuum expectation values $a$'s to the corresponding
dynamical 
superfield $\Phi(x,\theta)$ and determine the quadratic non-perturbative
action according to \eq{StoF}. Performing the $\theta$-integration, we then
obtain
\begin{equation}
\label{snpq}
S_{\mathrm{n.p.}} \propto \int d^4x\, \Big[ 2\big(\tr F\big)^2 - \tr F^2 \Big]
\,q^2 + O(q^4)~~+~
\mathrm{c.c.}
\end{equation}
{From} this expression, we can say that the non-perturbative part of the
holomorphic couplings
$f$ and $f'$ of our $\cN=2$ theory is given by
\begin{equation}
f_{\mathrm{n.p.}} = \alpha\,q^2+ O(q^4)~,~~~f'_{\mathrm{n.p.}} = -2\alpha\,q^2+
O(q^4)
\label{ffnp}
\end{equation}
where $\alpha$ is an overall coefficient that accounts for the normalization of
the
instanton partition function and the numerical factors arising from the $\theta$
integrations.
We would like to stress again that the vanishing of the contributions at the one
and three instanton level is due to the non-trivial cancellations between
contributions coming from fixed points with one D3-brane or with none.  In
the next section we will test this result against a dual heterotic computation
that predicts the absence of these odd instanton number contributions to any
order and reproduces the relative factor of $-2$ between the two structures at
$k=2$. This will provide a robust test of our explicit calculations.

\section{Heterotic gauge couplings}
\label{sec:prephet}

In this section we exploit the heterotic/type I$^\prime$ duality to test the 
results we have found via localization of the integrals on the moduli space of
the
D(--1)/D3/D7-brane system. The heterotic dual model can be built from the 
U(16) compactification of the SO(32) heterotic string on $\cT_4/\mathbb Z_2$ 
(with standard embedding of the orbifold curvature into the gauge bundle)
\cite{Berkooz:1996iz,Aldazabal:1997wi}
and further reduced on $\cT_2$ with Wilson lines that break U(16) to
$\mathrm{U}(4)^4$.
The gauge kinetic terms in this heterotic set-up are corrected at 1-loop by an
infinite tower of world-sheet instantons wrapping $\cT_2$, which are dual to the
D-instantons of the type I$^\prime$ theory \cite{Bachas:1997xn,Bachas:1997mc}. 
In this section we will compute the 1-loop heterotic thresholds and, after
applying the duality map, show a perfect match against the stringy 
multi-instanton contributions found in Section~\ref{sec:loc}.

\subsection{The heterotic orbifold}
\label{subsec:hetsetup}

We first give some details on the heterotic model we will consider. 
We start from the SO(32) heterotic string with super-string coordinates
$X^M$ and $\psi^M$ ($M=0,\ldots,9$), and a left-moving SO(32) current algebra
realized in terms of 16 complex fermions $\Lambda^I$ ($I=1,\ldots,16$).
To find a four-dimensional ${\mathcal N}=2$ vacuum with gauge group
$\mathrm{U}(4)^4$ we
compactify the theory on $\cT_4/\mathbb Z_2\times \cT_2$ with a proper choice of
Wilson lines on $\cT_2$. More precisely, the $\mathbb Z_2$ orbifold group is
generated by
\begin{equation}
\hat g_0 ~: \qquad   X^i \to -X^i ~,~~~ \psi^i \to -\psi^i  ~,~~~\Lambda^I \to
\ii
\Lambda^I
\end{equation}
where $X^i$ and $\psi^i$ ($i=4,5,6,7$) are the string coordinates along
$\cT_4$. 
This action breaks the gauge group SO(32) down to U(16) corresponding to the 256
massless vectors
of the form $\psi_{-\frac{1}{2}}^{\mu} \Lambda_{-\frac{1}{2}}^I
\bar\Lambda_{-\frac{1}{2}}^{\bar J}|0\rangle$ which are even under $\hat g_0$. 
The further breaking to $\mathrm{U}(4)^4$ is achieved by turning on 
discrete Wilson lines on $\cT_2$. These can be realized in terms of a
$\Z_2\times \Z_2$ freely 
acting orbifold with each $\Z_2$ acting as a reflection in the U(16) lattice and
a 
half-shift along $\cT_2$. More precisely, if we denote by $X^8$ and $X^9$ the
bosonic 
coordinates of $\cT_2$, and for simplicity take the latter to be a square torus
with
radii $R_8$ and $R_9$, then the two generators $\hat g_1$ and $\hat g_2$ 
of $\Z_2\times \Z_2$ are defined as
\begin{equation}
\begin{aligned}
\hat g_1 &:\quad   X^8\to X^8+\pi R_8~,~~~ \Lambda^{I=1,\ldots, 8} \to -
\Lambda^{I=1\ldots 8}~,\\
\hat g_2 &:\quad   X^9\to X^9+\pi R_9~,~~~ \Lambda^{I=5,\ldots 12} \to -
\Lambda^{I=5,\ldots 12} 
\end{aligned}
\label{hetz2z2}
\end{equation}
This action splits the 16 complex fermions $\Lambda^I$ into four groups of four,
thus realizing 
the desired breaking from U(16) to $\mathrm{U}(4)^4$.

\subsubsection{Partition function and massless spectrum}
As a preliminary test, we check that the massless spectrum of this heterotic
orbifold
is the same as that of the dual type I$^\prime$ model.
To do so, we compute the heterotic partition function
\begin{equation}
\begin{aligned}
\mathcal{Z}_{\mathrm{het}} &= \int_{\cF}\frac{d^2\tau}{2\tau_2}\,
\Tr\Big(\frac{1+\hat g_0}{2}\,\frac{1+\hat g_1}{2}\,\frac{1+\hat g_2}{2}\,
\frac{1+(-1)^F}{2}\,
q^{L_0-\frac{c}{24}}\,\bar q^{\bar L_0-\frac{\bar c}{24}}\Big)
\end{aligned}
\label{z000}
\end{equation}
where $q=\ee^{2\pi\ii\tau}$, $d^2\tau=d\tau_1d\tau_2$, $\cF$ is the fundamental
domain of the torus 
and the GSO projection acts on the right-moving fields. Performing the
conformal field theory trace over all sectors, we obtain
\begin{equation}
\mathcal{Z}_{\mathrm{het}}
=  \frac{1}{2^4}\,\int_\cF \frac{d^2 \tau}{\tau_2} ~\frac{\nu_4}{\tau_2^2}
\sum_{g_i,h_i=0}^1\, \rho\sp{h_0}{g_0}\!(0;\tau) 
~\chi\sp{h_0\,h_1\,h_2}{g_0\,g_1\,g_2}\!(0;\bar \tau)
~ \Gamma\sp{h_0\,h_1\,h_2}{g_0\,g_1\,g_2}\!(\tau,\bar\tau)
~,
\label{z00}
\end{equation}
where the factor 
\begin{equation}
\frac{\nu_4}{\tau_2^2} \equiv \frac{V_4}{(4\pi^2\alpha')^2\,\tau_2^2}
\label{v4}
\end{equation}
arises from the integration over the bosonic zero-modes of the four non-compact
directions with a (regularized) volume $V_4$, and the functions
$\rho\sp{h_0}{g_0}\!(0;\tau)$ and
$\chi\sp{h_0\,h_1\,h_2}{g_0\,g_1\,g_2}\!(0;\bar \tau)$ account for the
contributions coming from the trace over the right- and left-moving oscillators,
respectively. Their explicit definition and properties can be found in
App.~\ref{app:dethet} (see in particular Eqs.~(\ref{char00})~-~(\ref{chiw})).
Finally, $\Gamma\sp{h_0\,h_1\,h_2}{g_0\,g_1\,g_2}\!(\tau,\bar\tau)$ represents
the contribution of the bosonic zero-modes in the internal compact directions
and is given explicitly in Eqs.~(\ref{zeromodes}-\ref{momwin}) (see 
also App.~\ref{app:notations}).

With this information, we can check the massless spectrum. To do so, we first
notice that the massless states contribute only 
to the untwisted amplitudes with $h_1=h_2=0$, since the $\Z_2\times
\Z_2$-actions (\ref{hetz2z2}) have 
no fixed points. Indeed, expanding the lattice sums for large values of $\tau$,
we have
\begin{equation}
\begin{aligned}
\Gamma\sp{h_0\,0\,\,\,0}{g_0\,g_1\,g_2} &=1+\ldots~,\\
\Gamma\sp{h_0\,h_1\,h_2}{g_0\,g_1\,g_2}&=0+\ldots~~\mbox{for}~
(h_1,h_2)\neq (0,0)~.
\end{aligned}
\end{equation}
Thus, only the left-moving contributions $\chi\sp{h_0\,0\,\,\,0}{g_0\,g_1\,g_2}$
need to be considered. 
Using the results derived in App.~\ref{app:dethet} (see in particular
\eq{chiapp}),
it is not difficult to obtain their asymptotic expansions for large values of
$\tau$, namely
\begin{equation}
\begin{aligned}
&\chi\sp{0\,\,0\,\,0}{0\,\,0\,\,0} =\frac{1}{\bar q}+(2_v+502)+\ldots ~,\\
&\chi\sp{0\,\,0\,\,\,0}{0\,g_1\,g_2}=\frac{1}{\bar q}+(2_v-10)+\ldots ~~~~
\mbox{for}~(g_1,g_2)\neq(0,0)~,
\\
&\chi\sp{0\,\,0\,\,\,0}{1\,g_1\,g_2}=\frac{1}{\bar q}+(2_v+14)+\ldots ~, 
\\
&\chi\sp{1\,\,0\,\,0}{0\,\,0\,\,0} =\chi\sp{1\,\,0\,\,0}{1\,\,0\,\,0} = 128
+\ldots
\phantom{\frac{1}{\bar q}},\\
&\chi\sp{1\,\,0\,\,\,0}{0\,g_1\,g_2}=\chi\sp{1\,\,0\,\,\,0}{1\,g_1\,g_2}=0~~~~
\mbox{for}~(g_1,g_2)\neq(0,0)~.
\end{aligned}
\label{chis}
\end{equation}
Here and below we use the notation $2_{v,c,s}$ to keep track of the 
transformation properties of the various states with respect to the
SO(2) Lorentz little group in four dimensions; thus $2_v$ stands
for two  degrees of freedom in the vector representation. Likewise, for the
right-moving contributions we find
\begin{equation}
\rho\sp{0}{0}=\mathbf{V}+\mathbf{H}  +\ldots  ~,~~~
\rho\sp{0}{1}=\mathbf{V}-\mathbf{H}+\ldots
~,~~~\rho\sp{1}{0}=\rho\sp{1}{1}= 2\mathbf{H} +\ldots~,
\label{rhos}
\end{equation}
where we have denoted by
\begin{equation}
\mathbf{V}= 2_v+2 -2_s -2_c  ~,~~~ \mathbf{H}=4  - 2\times 2_s -2 \times 2_c
\label{VH}
\end{equation}
the number of physical states in a vector and a hyper-multiplet.

Taking the product of the left- and right-moving contributions, we find that the
number of massless
states can be written as 
\begin{equation}
\Big[\mathbf{V}\times ( 2_v+2 )+ 4\mathbf{H}\Big] + 4\Big[\mathbf{V}\times
\mathbf{16}+
\mathbf{H}\times(\mathbf{6}+\overline{\mathbf{6}} +4\times\mathbf{4})\Big]
\label{massless}
\end{equation}
where the first square bracket corresponds to the gravity multiplet and four
hypers
containing the geometrical moduli of the $\cT_4/\Z_2$ orbifold, and the
remaining terms
build up four copies of a vector multiplet in the adjoint of U(4) plus one hyper
in the
$\Yasymm$, one in the $\overline\Yasymm$ and four in the $\Yfund$
representations, in perfect
agreement with the type I$^\prime$ dual spectrum of Tab.~1. 

\subsection{Threshold corrections}
\label{subsec:twist}

The moduli dependence of the gauge kinetics terms in the heterotic model at
1-loop can be
extracted from the 2-gluon scattering amplitude on the torus $\langle V_F V_F
\rangle$,
where
\begin{equation}
V_F=(2\pi\alpha')\,F^I_{\mu\nu} \int d^2 z \,\big(X^\mu\partial
X^\nu+\psi^\mu\psi^\nu\big)(z)
~J_{\mathrm{int}}^I(\bar z)
\label{VF}
\end{equation}
is the vertex operator for the emission of a gauge field along the Cartan
directions
of $\mathrm{U}(4)^4$ for which the corresponding current is
$J_{\mathrm{int}}^I=\lambda^I \bar\lambda^{\bar I}$.
In a theory with eight supercharges, like ours, this is a BPS saturated
amplitude
and therefore non-trivial contributions come only from the fermionic zero mode
part of the string vertices, namely
\begin{equation}
V_F = (2\pi\alpha')\, 2\tau_2\, F_{s}^I \,J^s_{\mathrm{Lor}} \,
J^I_{\mathrm{int}}  
\label{vertices}
\end{equation}
with $J^s_{\mathrm{Lor}}=\psi^s \bar \psi^s$, $s=1,2$, being the right-moving 
fermionic currents along the two Cartan directions of the SO(4) Lorentz
group. In the above expression 
the factor of $\tau_2$ comes from the integration over the position of the  
vertices over the world-sheet torus.

To compute the 2-point function $\langle V_F V_F \rangle$ it is
convenient to exponentiate the string vertices and rewrite the threshold
amplitudes
as a second derivative of a generating function. Since both the left- 
and the right-moving parts of the vertices (\ref{vertices}) are quadratic in
free 
fermions, this generating function is nothing but the weighted partition
function%
\footnote{For convenience, in the exponentiation we have attached the factor of
$2\tau_2$ of the
gluon vertex (\ref{vertices}) to the Lorentz current. We will instead keep track
explicitly of the
dimensional factors of $(2\pi\alpha')$.}
\begin{equation}
\mathcal{Z}_{\mathrm{het}}(w, \vec v) = \!\!\int_{\cF}\frac{d^2\tau}{2\tau_2}
\Tr\!\Big(\frac{1+\hat g_0}{2}\,\frac{1+\hat g_1}{2}\,\frac{1+\hat
g_2}{2}\,\frac{1+(-1)^F}{2}\,
q^{L_0-\frac{c}{24}}\,\bar q^{\bar L_0-\frac{\bar c}{24}}
\,\ee^{2\pi \ii [2\tau_2 w_s
J^s_{\mathrm{Lor}}  + v_I  J^I_{\mathrm{int}}]}\Big)
\label{weightz}
\end{equation}
As we know, there two possible tensor structures for the gauge kinetic terms:
the single trace $\tr F^2$ term, and the double trace term $(\tr F)^2$.
The 1-loop thresholds for each structure can be read from appropriate
combinations of $v_I$-derivatives and will be denoted again as $\Delta$ and
$\Delta'$ respectively.
To find them, it is enough to specify one Cartan direction along SO(4) and two
Cartan directions along one of the four U(4)'s. Thus, we can fix for example
$s=1$ and $I=1,2$, and define
\begin{equation}
\Delta_{IJ} = \frac{1}{4!(2\pi)^4}\,
\partial_w^2 \partial_{v_I}\partial_{v_J}
\mathcal{Z}_{\mathrm{het}}(w, \vec v) \Big|_{w=\vec v=0} 
\label{DeltaIJ}
\end{equation}
with $w=w_1$. 
Then the quadratic gauge thresholds $\Delta$ and $\Delta'$ for the two tensor
structures can
be obtained from the following relations
\begin{equation}
\begin{aligned}
& \frac{V_4}{16\pi^2}\,\Delta= (2\pi\alpha')^2\,\big(\Delta_{11} -
\Delta_{12}\big)~,\\
& \frac{V_4}{16\pi^2}\,\Delta'=(2\pi\alpha')^2\,\Delta_{12}~.
\end{aligned}
\label{d2z}
\end{equation}
We now give some details on the calculation of such quantities.

\subsubsection{Calculation}

Just like $\mathcal{Z}_{\mathrm{het}}$, also the weighted partition function
(\ref{weightz}) 
can be written as a sum over all projected and twisted sectors. Indeed, using
the functions
defined in App.~\ref{app:dethet}, we have
\begin{equation}
\mathcal{Z}_{\mathrm{het}}(w, \vec v) 
= \frac{\nu_4}{2^4} \, \int_\cF \frac{d^2 \tau}{\tau_2^3} 
\sum_{g_i,h_i=0}^1\, \rho\sp{h_0}{g_0}\!(w,\tau)
~\chi\sp{h_0\,h_1\,h_2}{g_0\,g_1\,g_2}\!(\vec v,\bar \tau)
~ \Gamma\sp{h_0\,h_1\,h_2}{g_0\,g_1\,g_2}\!(\tau,\bar\tau)
~,
\label{weightz2}
\end{equation}
which is a rather obvious generalization of the partition function (\ref{z00}).
To proceed, it is convenient to organize the various contributions
according to the orbits of the modular group. More precisely, exploiting the
modular transformation
properties of the various building blocks given in App.~\ref{app:dethet}, one
can show
that the contributions coming from the sectors $\sp{1\, h_1 h_2}{0\, g_1 g_2}$
and 
$\sp{1\, h_1 h_2}{1\, g_1 g_2}$ can be obtained by applying, respectively, the
$S$ and $TS$
transformations of the modular group to the amplitudes in the sectors $\sp{0\,
h_1 h_2}{1\, g_1 g_2}$.
The latter can therefore be taken as representatives of a modular orbit, and the
weighted
partition function (\ref{weightz2}) can be rewritten as
\begin{equation}
\begin{aligned}
\mathcal{Z}_{\mathrm{het}}(w, \vec v) 
&= \frac{\nu_4}{2^4} \, \int_\cF \frac{d^2 \tau}{\tau_2^3}
\sum_{g_i,h_i=0}^1\,\Bigg[  \rho\sp{0}{0}\!(w,\tau)
~\chi\sp{0\,h_1\,h_2}{0\,g_1\,g_2}\!(\vec v,\bar \tau)
~ \Gamma\sp{0\,h_1\,h_2}{0\,g_1\,g_2}\!(\tau,\bar\tau)\\
&\hspace{25pt}+ \Big(\rho\sp{0}{1}\!(w,\tau)
~\chi\sp{0\,h_1\,h_2}{1\,g_1\,g_2}\!(\vec v,\bar \tau)
~ \Gamma\sp{0\,h_1\,h_2}{1\,g_1\,g_2}\!(\tau,\bar\tau)\,+\,
\mbox{orb}~\Big)\Bigg]~.
\end{aligned}
\label{weightz22}
\end{equation}

Now we are ready to evaluate \eq{DeltaIJ}. Using the Riemann identity for 
the $\vartheta$-functions, we first rewrite the right-moving contributions as
follows
\begin{equation}
\begin{aligned}
\rho\sp{0}{0}\!(w,\tau) &=\frac{\vartheta_1(\tau_2w)^4}{\eta^{12}}~,\\
\rho\sp{h_0}{g_0}\!(w,\tau) &=-4\,\ee^{\ii\pi h_0}\,
\frac{\vartheta_1(\tau_2w)^2\,\vartheta\sp{1+h_0}{1+g_0}\!(\tau_2w)^2}
{\eta^{6}\,\vartheta\sp{1+h_0}{1+g_0}^2}\qquad(g_0,h_0)\neq (0,0)~,
\end{aligned}
\label{rhow1}
\end{equation}
from which we easily find
\begin{equation}
\begin{aligned}
\partial_w^2\rho\sp{0}{0}\!(w,\tau) \Big|_{w=0} &=0~,\\
\partial_w^2\rho\sp{h_0}{g_0}\!(w,\tau)\Big|_{w=0} &= -8\,(2\pi)^2\,\ee^{\ii\pi
h_0}\,\tau_2^2\qquad(g_0,h_0)\neq (0,0)~.
\end{aligned}
\label{rho0}
\end{equation}
Thus, only the sectors with $(g_0,h_0)\neq (0,0)$ contribute to the quadratic
thresholds.
Inserting these results in \eq{DeltaIJ} and using \eq{zeromodes}, we have
\begin{eqnarray}
\Delta_{IJ} &=& -\frac{\nu_4}{4!\,8\pi^2}\int_\cF \frac{d^2 \tau}{\tau_2^2}\!
\sum_{g_i,h_i=0}^1\!\Bigg[\partial_{v_I}\partial_{v_J}\chi\sp{0\,h_1\,h_2}{1\,
g_1\,g_2}\!(\vec v,\bar \tau)\Big|_{\vec v=0}
\,\Gamma_{2,2}\sp{h_1\,h_2}{g_1\,g_2}\!(\tau,\bar\tau;T,U)\,+\,
\mbox{orb}\Bigg]\nonumber\\
&=&-\frac{\nu_4}{4!\,8\pi^2}\int_\cF \frac{d^2 \tau}{\tau_2^2}\!
\sum_{h_i=0}^1\!\Bigg[\partial_{v_I}\partial_{v_J}\chi\sp{0\,h_1\,h_2}{1\,\,0\,\
,\,0}\!(\vec v,\bar \tau)\Big|_{\vec v=0}
\sum_{g_i=0}^1\Gamma_{2,2}\sp{h_1\,h_2}{g_1\,g_2}\!(\tau,\bar\tau;T,U)\,+\,
\mbox{orb}\Bigg]\nonumber\\
&\equiv&\sum_{h_i=0}^1\Delta_{IJ}^{~(h_1\,h_2)}
\label{DeltaIJ1}
\end{eqnarray}
where the second line follows from the fact that the double derivatives of the
$\chi$-functions are independent of $g_1$ and $g_2$. We would like to stress
that, unlike in the case
of the partition function where only the untwisted 
sectors with $(h_1,h_2)=(0,0)$ were relevant to derive the
massless spectrum, in the threshold calculation all sectors, including the
twisted
ones, contribute. In the following
we will in turn analyse the two types of contributions, the untwisted one
arising from sectors with $(h_1,h_2)=(0,0)$ and
the twisted ones arising from sectors with $(h_1,h_2)\neq(0,0)$.

\subsubsection*{Orbits of $\chi\sp{0\,0\,0}{1\,0\,0}$}

In this case we have
\begin{equation}
\Delta_{IJ}^{~(0\,0)} = -\frac{\nu_4}{4!}\int_\cF \frac{d^2 \tau}{\tau_2^2}
\,\Bigg[\chi_{IJ}(\bar \tau)
\sum_{g_i=0}^1\Gamma_{2,2}\sp{0\,\,\,0}{g_1\,g_2}\!(\tau,\bar\tau;T,U)\,+\,
\mbox{orb}\Bigg]
\label{DeltaIJ0}
\end{equation}
where we have introduced the notation
\begin{equation}
\chi_{IJ}(\bar \tau)=\frac{1}{8\pi^2}\,\partial_{v_I}\partial_{v_J}
\chi\sp{0\,0\,0}{1\,0\,0}\!(\vec v,\bar \tau)\Big|_{\vec v=0}~.
\label{chi1112}
\end{equation}
Using the summation identity
(see also \eq{sumg}) 
\begin{equation}
\sum_{g_i=0}^1\Gamma_{2,2}\sp{0\,\,\,0}{g_1\,g_2}\!(\tau,\bar\tau;T,U) = 
2\,\Gamma_{2,2}(2\tau,2\bar\tau;\ft{T}{2},U)~,
\label{sumGamma}
\end{equation}
and exploiting the modular properties of the right hand side, we have
\begin{equation}
\begin{aligned}
 \Delta_{IJ}^{~(0\,0)} &= -\frac{2\nu_4}{4!}\int_\cF \frac{d^2 \tau}{\tau_2^2}
\,\Big[\chi_{IJ}(\bar \tau)\,+\,
\mbox{orb}\Big]\,\Gamma_{2,2}(2\tau,2\bar\tau;\ft{T}{2},U)\\
&= - \frac{2\nu_4}{4!}\int_\cF \frac{d^2 \tau}{\tau_2^2}
\,c_{IJ}(\bar \tau)\,\Gamma_{2,2}(\tau,\bar\tau;\ft{T}{2},U)
\end{aligned}
\label{DeltaIJ00}
\end{equation}
where
\begin{equation}
 c_{IJ}(\bar\tau)=
 \left[\chi_{IJ}\Big(\ft{\bar\tau}{2} \Big)+\chi_{IJ}\Big(\ft{\bar\tau+1}{2}
 \Big)+\chi_{IJ}\Big(-\ft{1}{2\bar\tau} \Big)\right] ~
\label{cij}
\end{equation}
is simply twice the Hecke operator $H_{\Gamma^-}(\chi_{IJ})$.   
It is important to note that $c_{IJ}$ are modular forms of weight zero with no
poles and therefore are constants. Indeed, as shown in App.~\ref{app:dethet}
(see
\eq{c1112}), it turns out that
\begin{equation}
c_{IJ}(\bar\tau)= 6~,
\label{cijres}
\end{equation}
so that to obtain $\Delta_{IJ}^{~(0\,0)}$ we may simply use the general
integration formula
\cite{Dixon:1990pc}
\begin{equation}
\int_\cF \frac{d^2 \tau}{\tau_2^2}\,\Gamma_{2,2}(\tau,\bar\tau;{T},U)
= -\log\Big( c \,T_2\,U_2\,\big| \eta \left(T \right) 
\eta( U)\big|^4\Big)
\label{integral}
\end{equation}
with $c=\frac{8\pi \ee^{1-\gamma_E}}{3\sqrt{3}}$.
Putting everything together and discarding all moduli independent
terms, we finally obtain
\begin{equation}
\Delta_{IJ}^{~(0\,0)}= \frac{\nu_4}{2}\,  \log\Big(T_2\,U_2\Big| \eta\left(
\ft{T}{2} \right)\eta
\left( U \right)\Big|^4\Big) ~. 
\label{resint}
\end{equation}

\subsubsection*{Orbits of $\chi\sp{0\,h_1\,h_2}{1\,\,0\,\,\,0}$ with
$(h_1,h_2)\neq (0,0)$}

In this case, from \eq{DeltaIJ1} we have
\begin{equation}
\Delta_{IJ}^{~(h_1\,h_2)} = -\frac{\nu_4}{4!}\int_\cF \frac{d^2 \tau}{\tau_2^2}
\,\Bigg[\chi_{IJ}^{~(h_1\,h_2)}(\bar \tau)
\sum_{g_i=0}^1\Gamma_{2,2}\sp{h_1\,h_2}{g_1\,g_2}\!(\tau,\bar\tau;T,U)\,+\,
\mbox{orb}\Bigg]
\label{DeltaIJh}
\end{equation}
where
\begin{equation}
\chi_{IJ}^{~(h_1\,h_2)}(\bar \tau)
=\frac{1}{8\pi^2}\,\partial_{v_I}\partial_{v_J}\chi\sp{0\,h_1\,h_2}{1\,\,0\,\,\,
0}\!(\vec v,\bar \tau)\Big|_{\vec v=0}
\label{chih}
\end{equation}
for any $(h_1,h_2)\neq(0,0)$. Actually, as shown in App.~\ref{app:dethet}, it
turns out that
these functions are constants, namely
\begin{equation}
\chi_{11}^{~(h_1\,h_2)}(\bar \tau)=-\chi_{12}^{~(h_1\,h_2)}(\bar \tau)=2~,
\label{chihres}
\end{equation}
so that the calculation of $\Delta_{IJ}^{~(h_1\,h_2)}$ drastically simplifies.
Furthermore,
if we use the summation identity (see also \eq{sumu})
\begin{equation}
\sum^1_{h_i,g_i=0}\!\!'~\Gamma_{2,2}\sp{h_1\,h_2}{g_1\,g_2}(\tau,\bar\tau;T,U)=
4\Gamma_{2,2}(\tau,\bar\tau;\ft{T}{4},U) - 2
\Gamma_{2,2}(2\tau,2\bar\tau;\ft{T}{2},U)
\label{sumu1}
\end{equation}
where the $'$ means that $(h_1,h_2)=(0,0)$ is excluded from the sum, we obtain
\begin{equation}
\begin{aligned}
\sum^1_{h_i,=0}\!\!'~\Delta_{11}^{~(h_1\,h_2)} &=
-\sum^1_{h_i,=0}\!\!'~\Delta_{12}^{~(h_1\,h_2)}
\\&=- \frac{\nu_4}{2}\int_\cF \frac{d^2 \tau}{\tau_2^2} 
\Bigg[2 \Gamma_{2,2}(\tau,\bar\tau;\ft{T}{4},U) - 
\Gamma_{2,2}(2\tau,2\bar\tau;\ft{T}{2},U)\Bigg]\\
&=
\frac{\nu_4}{2}\,\log\Bigg(T_2 \,U_2\,|\eta(U)|^4 \,
\frac{|\eta(\ft{T}{4})|^8}{|\eta (\ft{T}{2})|^4}
\Bigg)
\end{aligned}
\label{resint2}
\end{equation}
up to moduli independent terms.

\subsubsection{Results}

Now we can collect all results and obtain the final expressions for the
quadratic gauge thresholds
in our heterotic model.
Plugging the contributions of the two types of orbits (\ref{resint}) and
(\ref{resint2}) into 
\eq{DeltaIJ1}, using \eq{d2z} and recalling that $\nu_4=V_4/(4\pi^2\alpha')^2$,
we find
\begin{equation}
\begin{aligned}
&\Delta =4\log\Bigg(T_2 \,U_2\,|\eta(U)|^4 \, \frac{|\eta(\ft{T}{4})|^8}{|\eta
(\ft{T}{2})|^4}
\Bigg)~,\\
&\Delta'= 4\log\Bigg(\frac{|\eta(\ft{T}{2})|^4}{|\eta (\ft{T}{4})|^4}
\Bigg)~.
\end{aligned}
\label{resdelta}
\end{equation}
It is interesting to observe that these expressions are invariant under the
following target-space modular transformations 
\begin{equation}
\Gamma^0(4)_T \otimes \Gamma_U
\label{modular}
\end{equation}
where $\Gamma_U$ is the standard modular group acting on $U$, while
$\Gamma^0(4)_T$ is the subgroup of the modular transformations on $T$ of the
form 
\begin{equation}
T~\rightarrow ~\frac{a T + b }{c T + d}~~~~~\mbox{with}~~ a,b,c,d \in \mathbb
Z~,~~
ad-bc=1 ~~\mbox{and}~~b=0~\mbox{mod}\,4~.
\label{gamma4}
\end{equation}
The target-space modular transformations (\ref{modular}) are those that are
consistent with the Wilson lines which break U(16) to $\mathrm{U}(4)^4$. Any
meaningful string amplitude should therefore be invariant under such
transformations.

\subsection{Holomorphic gauge couplings and duality check}
\label{DKLhet}

To obtain from the above thresholds the holomorphic gauge couplings of the
heterotic
model, we can follow the same reasoning described in Sec.~\ref{DKLI} for the
dual type I$^\prime$
theory. We only have to remember that in the heterotic set-up
the bulk K\"ahler potential reads as
\begin{equation}
\label{ktoth}
{K} = -\log S_2  -\sum_{i=1}^3\log\big(T_2^{(i)}\,U_2^{(i)}\big) ~,
\end{equation}
where $S_2$, related to the four-dimensional dilaton $\phi_4$ by $S_2 =
\ee^{-2\phi_4}$, is 
the imaginary part of the chiral superfield $S$ 
which determines the holomorphic coupling function at tree level and plays
the same r\^ole as the $t$ superfield of the type I$^\prime$ theory,
and that 
\begin{equation}
\widehat {K} =-\log\big(T_2\,U_2\big)~.
\label{kalerh} 
\end{equation}
Then, as shown in App.~\ref{app:hol}, the 1-loop contributions
$f_{(1)}$ and $f'_{(1)}$ are given by the same relations (\ref{f1a}), now
expressed in terms 
of the heterotic variables \cite{Dixon:1990pc,Kaplunovsky:1995jw}.
In particular, for the single-trace Yang-Mills term, using \eq{resdelta} and
recalling that $b=0$, we have
\begin{equation}
\re f_{(1)} 
= \Delta+\Delta_{\mathrm{univ}} + b\,\widehat K =
4\log\Bigg(T_2 \,U_2\,|\eta(U)|^4 \, \frac{|\eta(\ft{T}{4})|^8}{|\eta
(\ft{T}{2})|^4}
\Bigg)+
\Delta_{\mathrm{univ}}~.
\label{fsu4}
\end{equation}
It is important to stress that the universal term $\Delta_{\mathrm{univ}}$
is related to the 1-loop corrections of the K\"ahler potential, which in the
heterotic
setup is of order $(g_{\mathrm{s}})^0$ \cite{Dixon:1990pc,Antoniadis:1996vw},
and, like any meaningful amplitude, it must respect all symmetries of the
compactification manifold including the target-space modular invariance
(\ref{modular}). From these considerations we are then led to write
\begin{equation}
\re f_{(1)} =  4\log\Bigg(\frac{|\eta(\ft{T}{4})|^4}{|\eta (\ft{T}{2})|^4}
\Bigg)~,
\label{fa11}
\end{equation}
and
\begin{equation}
\Delta_{\mathrm{univ}} =-4\log\Big(T_2 \,U_2\,|\eta(U)|^4
\,|\eta(\ft{T}{4})|^4\Big)~.
\label{deltauniv}
\end{equation}
Likewise, for the double-trace coupling we have
\begin{equation}
\re f'_{(1)} = \Delta'+\Delta_{\mathrm{univ}} +b'\,\widehat K
=- 4\log|\eta(U)|^4
+4\log\Bigg(\frac{|\eta(\ft{T}{2})|^4}{|\eta(\ft{T}{4})|^8}\Bigg)
\label{hetdoublefinal}
\end{equation}
where the second equality follows upon using $b'=-4$ and Eqs.~(\ref{resdelta})
and (\ref{kalerh}).
Notice that all non-holomorphic terms correctly compensate each other and yield
a holomorphic result
for $f'_{(1)}$. This is an {\it a posteriori} confirmation of the universal term
(\ref{deltauniv}).

The heterotic holomorphic couplings (\ref{fa11}) and (\ref{hetdoublefinal}) are
exact and
do not receive any kind of corrections beyond 1-loop. Therefore, when translated
with the duality map to the type I$^\prime$ theory, they must contain all
information, both
perturbative and non-perturbative, on the corresponding type I$^\prime$
couplings, including the (exotic) instanton corrections computed in
Section~\ref{sec:loc}.
We now show that this is indeed what happens.

Under the heterotic/type I$^\prime$ duality map, the K\"ahler modulus of
the heterotic theory $T$ is mapped into the axio-dilaton $\lambda$ of the type
I$^\prime$ model according to (see also Ref.~\cite{Lerche:1998nx})
\begin{equation}
\frac{T}{4} \longleftrightarrow \lambda~.
\label{dualmap}
\end{equation}
Thus, the weak coupling regime $g_s \sim 1/\lambda_2 \to 0$ 
in type I$^\prime$ can be recovered from the large volume expansion $T_2 \to
\infty$ 
of the heterotic theory and {\it viceversa}. 
Expanding Eqs.~(\ref{fa11}) and (\ref{hetdoublefinal}) for large $T_2$, we find
\begin{equation}
\begin{aligned}
&\re f_{(1)} = \left.\frac{\pi}{3}\,T_2 + 8\sum_{k=1}^\infty \Bigg[
\sum_{d|k} \frac{1}{d}\,\Big(\ee^{2\pi \ii  k\ft{T}{4}} - \ee^{2\pi \ii 
k\ft{T}{2}} \Big) +\mathrm{h.c.}\Bigg]\right. ~,\\
&\re f'_{(1)} = \left.-4\log |\eta(U)|^4+ 8\sum_{k=1}^\infty \Bigg[
\sum_{d|k} \frac{1}{d}\,\Big(\ee^{2\pi \ii  k\ft{T}{2}} -2 \ee^{2\pi \ii 
k\ft{T}{4}} \Big) +\mathrm{h.c.}\Bigg]\right.~.
\end{aligned}
\label{expansions}
\end{equation}
After translating into the type I$^\prime$ variables $q=\ee^{\pi \ii \lambda}=
\ee^{\pi \ii \frac{T}4}$, these formulas predict a tree-level term proportional
to
\begin{equation}
\lambda_2 \,\tr F^2~,
\label{tree}
\end{equation}
a 1-loop contribution
\begin{equation}
-4\big(\log|\eta(U)|^4\big)\,\big(\tr F\big)^2~,
\label{1loop}
\end{equation}
which agrees with the perturbative type I$^\prime$ result (\ref{fIaa}), 
as well as a series of instanton-like contributions with even instanton numbers,
correcting both the single and the 
double trace gauge kinetic functions, with a leading term proportional to
\begin{equation}
q^2\Big(\tr F^2 - 2 \big(\tr F\big)^2\Big)~.
\label{1inst}
\end{equation}
The relative coefficient between the two trace structures is in perfect
agreement 
with the multi-instanton calculus of the type I$^\prime$ theory, as one can see
from \eq{snpq}.
Also the absence of odd instanton corrections is in agreement with the results
found in 
Section~\ref{sec:loc}.

The presence of the tree-level term (\ref{tree}) proportional to $\lambda_2$ may
seem
puzzling at first sight, since $\lambda$ is the chiral field accounting 
for the gauge coupling on a D3-brane and not on a D7-brane. 
The same type of contribution was found also for the T-dual version of our
model in Ref.~\cite{Camara:2008zk} where a convincing explanation for its
presence was given. Indeed,
it was argued that since the gauge branes are entirely wrapped over the orbifold
$\cT_4/\Z_2$, 
their gauge kinetic function, besides the usual ``untwisted'' contribution,
should receive also ``twisted'' contributions from the exceptional 2-cycles at
the orbifold fixed
points, where a hidden non-trivial U(1) gauge bundle is localized. 
In the case of fractional D7-branes this mechanism is responsible for gauge
coupling corrections proportional to $\lambda_2$. It would be very interesting
to explicitly derive this result from
disk scattering amplitudes involving twisted fields at the orbifold fixed
points.

\section{Conclusions}
\label{sec:concl}
It is by now clear that ``exotic'' instanton corrections to the effective 
actions of D-brane worlds can have relevant consequences. Our work has been
motivated by the importance of putting on firm grounds the techniques to compute
such effects in a four-dimensional context. Let us summarize here our results.

We considered a type I$^\prime$/heterotic dual pair realizing a ${\mathcal N}=2$
super-conformal gauge theory in four dimensions with gauge group U(4) and a
matter content made of four fundamentals plus one antisymmetric hyper-multiplet
and its conjugate. The type I$^\prime$ model is built with D7- and D3-branes in
a
$\cT_4/\Z_2 \times \cT_2$ background with O7- and O3-planes. In this setup, the
U(4)
gauge theory lives in the uncompactified part of the world-volume of the
D7-branes
on top of one of the O7-planes. On the heterotic side this $\mathcal N=2$ vacuum
is realized
starting from the U(16) heterotic model on $\cT_4/\Z_2$ after a further
compactification on $ \cT_2$ with non-trivial Wilson lines 
breaking U(16) down to  $\mathrm{U}(4)^4$. 

In both settings we studied the terms of the low-energy effective action
quadratic in the gauge field strength plus their supersymmetric completion,
namely
\begin{equation}
S=\frac{1}{8\pi}\,\int d^4x\,\Big[{\mathrm{Re}}f~ {\tr} F^2 +
{\mathrm{Re}}f' \, \big({\tr} F\big)^2  \Big] + \cdots
\label{concl_action}
\end{equation}
where $f$ and $f'$ are the Wilsonian couplings. These are \emph{holomorphic}
functions 
of the bulk moduli, which depend on the following set of variables
\begin{equation}
\begin{aligned}
\mbox{type~I$^\prime$}~~&:& ~ \big(t, \lambda, U\big) ~,\\
\mbox{heterotic}~~&:& ~~ \big(S,T,U\big)~.
\end{aligned}
\label{concl_mod}
\end{equation}
Here $t$ and $S$ represent the tree-level Yang-Mills coupling in the type
I$^\prime$
and heterotic setups, respectively, $\lambda$ is the axio-dilaton of type
I$^\prime$, 
and $T$ and $U$ are the K\"ahler and complex structures of $\cT_2$.

In the type I$^\prime$ side, the holomorphic functions $f$ and $f'$ get
contributions 
at the tree-level, at 1-loop and also from D(--1)-branes, which represent 
exotic instantons for this system. We
computed such non-perturbative corrections by means of localization techniques
for the integration over the exotic moduli space up to $k=3$ instantons. 
Altogether we obtained
\begin{equation}
\label{concl_ffpI}
\begin{aligned}
 f & = -\ii t + \alpha\,q^2 + O(q^4)\phantom{\vdots}~,\\
 f'& = -8\log \eta(U)^2 -2\alpha\,q^2 + O(q^4)\phantom{\vdots}~.  
\end{aligned}
\end{equation}
where $q= \exp(\pi\ii\lambda)$ and $\alpha$ is a coefficient related to the
overall
normalization of the measure of the exotic instanton moduli space. 

In the heterotic side, instead, the holomorphic couplings $f$ and $f'$ are exact
at 1-loop due to their BPS nature. We determined them by computing the 1-loop
threshold corrections
finding
\begin{equation}
\label{concl_ffphet}
\begin{aligned}
 f & = -\ii S + 8 \log\Bigg(\frac{\eta(\ft{T}{4})^2}{\eta
(\ft{T}{2})^2}\Bigg)~,\\
 f'& = -8\log \eta(U)^2 + 8 \log\Bigg(\frac{\eta(\ft{T}{2})^2}{\eta
(\ft{T}{4})^4}\Bigg)~. 
\end{aligned}
\end{equation}
Expanding for large values of $T$ and using the duality map (\ref{dualmap}), 
these heterotic formulas predict no instanton corrections at $k=1$ and $k=3$, 
and a relative coefficient $-2$ between the $k=2$ corrections to $f$ and $f'$, 
in perfect agreement with the results obtained in the type I$^\prime$ setting.
Moreover, the precise match of the 1-loop terms of $f'$ between
(\ref{concl_ffpI}) and (\ref{concl_ffphet}) can be taken as a strong evidence
that the overall normalization of our coupling functions is the same in the two
settings, thus providing an indirect way to fix the numerical factor $\alpha$.
We regard these results as a nice and non-trivial confirmation of the validity
of the 
exotic instanton calculus, which can then be applied with confidence also to
four-dimensional theories and to models for which the heterotic dual is not
known or does not exist.

We think there are several lessons to be learned from our computations and
several new directions 
which deserve to be explored in the light of our results. 
In first place, the presence of branes with different world-volume dimensions
implies that the standard 
prescription of localization in four and eight dimensions needs to be changed in
a non trivial way in 
order to extract the corrections for the gauge couplings.
In second place, the heterotic computation gives a result for arbitrary 
instanton numbers, while the explicit integration over the instanton moduli
space 
could be computed only to the order $k=3$. More extended checks of this duality
would be desirable. 
This could be achievable by noticing that if one considers the D7-branes as non
dynamical, 
then the exotic instanton partition function can be reinterpreted as 
the ordinary gauge instanton partition function for the four-dimensional theory 
living on the D3-branes with fundamental matter hyper-multiplets having 
masses given by the positions of the D7-branes.   
An analysis to extract some sort of Seiberg-Witten curve in this case 
seems to be possible even if, as we have discussed, this is a rather
unconventional
extrapolation of the standard field theory notions. We leave this kind of
analysis for future work.

\vskip 1cm
\noindent {\large {\bf Acknowledgments}}
\vskip 0.2cm
We thank P. Camara, E. Dudas, L. Gallot and I. Pesando for several useful
discussions.
The work of R.P. was partially supported by the European
Commission FP7 Programme Marie Curie Grant Agreement PIIF-GA-2008-221571,
and the Institutional Partnership grant of the Humboldt Foundation of Germany.
The work of
F.F. and J.F.M. was partially supported by the ERC Advanced Grant n.226455, 
``Supersymmetry, Quantum Gravity and Gauge Fields'' (SUPERFIELDS) 
and by the Italian MIUR-PRIN contract 20075ATT78.

\vskip 1cm
\appendix

\section{Zero-mode traces and lattice sums}
\label{app:notations}
In this appendix we collect some formulas on the traces over the bosonic
zero-modes and the lattice
sums that are useful for the calculations of the 1-loop threshold corrections,
both in the type I$^\prime$ set-up and in the heterotic model.

\subsection*{Closed strings}

In untwisted sectors the bosonic string coordinates
have zero-modes $x_{\mathrm R}$, $x_{\mathrm L}$, $p_{\mathrm R}$ and
$p_{\mathrm L}$ that
contribute to the Virasoro characters since
\begin{equation}
\label{l0l0b}
 L_0 = \frac12 \,p_{\mathrm R}^2 + \mathrm{osc.}~,~~~
 {\bar L_0} = \frac12 \,p_{\mathrm L}^2 + \mathrm{osc.}~.
\end{equation}
For $d$ real non-compact directions, the right- and left-moving momenta are
$ (p_L)_\mu = (p_R)_\mu = \sqrt{\frac{\alpha'}{2}} k_\mu$,
with $k_\mu$ a continuous variable, 
and thus the zero-mode contribution to $\Tr q^{L_0} {\bar q}^{{\bar L}_0}$ is
simply
\begin{equation}
\label{0mt}
\int d^d x \int \frac{d^dk}{(2\pi)^d}~ \ee^{-\pi\alpha' \tau_2\, k^2} = 
\frac{V_d}{(4\pi^2 \alpha' \tau_2)^{\frac d2}}
\end{equation}
where we have set $q=\ee^{2\pi\ii\tau}$ with $\tau=\tau_1+\ii\tau_2$.

Now let us consider two directions compactified on a 2-torus $\cT_2$. They can
be
described by two periodic coordinates $x^i\in[0,\sqrt{\alpha'}]$, a complex
structure $U$ and a (complexified) K\"ahler parameter $T$ which
are encoded in the metric $G_{ij}$ and in the Kalb-Ramond field $B_{ij}$,
according to
\begin{equation}
\label{tormet}
G = \frac{T_2}{U_2} 
 \begin{pmatrix}
  1 & U_1 \\
  U_1 & |U|^2
 \end{pmatrix}~,~~~
B =   \begin{pmatrix}
  0 & - T_1 \\
  T_1 & 0
 \end{pmatrix}~.
\end{equation}
In the following we will denote as $G^{ij}$ the components of the inverse metric
\begin{equation}
\label{invmettor}
G^{-1} = \frac{1}{T_2 U_2}
 \begin{pmatrix}
  |U|^2 & - U_1 \\
  - U_1 & 1
 \end{pmatrix}~.
\end{equation}
In this case, the right and left bosonic zero-modes are given by
\begin{equation}
\label{pwc}
(p_L)_i  = \frac{1}{\sqrt{2}} \big(n_i - (G-B)_{ij} w^j\big)~,~~~  
(p_R)_i  = \frac{1}{\sqrt{2}} \big(n_i + (G+B)_{ij} w^j\big)
\end{equation}
with $n_i,w^i\in\Z$, and \eq{l0l0b} should actually read
\begin{equation}
\label{l0l0bc}
 L_0 = \frac 12 (p_R)_i G^{ij} (p_R)_j + \mathrm{osc.}~,~~~
 {\bar L_0} = \frac 12 (p_L)_i G^{ij} (p_L)_j + \mathrm{osc.}
\end{equation}
Thus, the zero-mode contribution to the Virasoro character $\Tr q^{L_0} {\bar
q}^{{\bar L}_0}$ becomes
\begin{equation}
\label{zmt1}
\sum_{(\vec n,\vec w)\in\Z^4} \ee^{-\pi\tau_2 n_i G^{ij} n_j + 2\pi\ii\tau_1
w^i n_i + 2\pi\tau_2 w^i (B G^{-1})_{i}^{~j} n_j - 
\pi\tau_2 w^i (G - B G^{-1} B)_{ij} w^j}~.
\end{equation}
Utilizing the Poisson resummation formula
\begin{equation}
 \label{poisson}
 \sum_{\vec n\in\Z^2} \ee^{-\pi\, n^T  X  n + 2\pi\ii Y^T n} =
\frac{1}{\sqrt{\det X}}
 \sum_{\vec m\in\Z^2} \ee^{-\pi(m - Y)^T\cdot X^{-1} (m-Y)}
\end{equation}
and the explicit form of the torus metric and $B$-field, we can rewrite
\eq{zmt1} as
\begin{equation}
\label{zmt}
 \frac{1}{\tau_2}\,\Gamma_{2,2}(\tau,\bar\tau;T,U)~,
\end{equation}
where we have defined
\begin{equation}
\begin{aligned}
\Gamma_{2,2}(\tau,\bar\tau;T,U) &= 
T_2 \sum_{(\vec m, \vec w)\in\Z^4} \ee^{-\frac{\pi}{\tau_2}(\tau w^i - m^i) 
(G -B)_{ij} (\bar\tau w^j -m^j)}\\
&=T_2~\sum_{ M } \ee^{2\pi \ii T \,\mathrm{det} M}\, \ee^{-
\frac{\pi T_2}{\tau_2 U_2}\,|( 1  \, U)\, M \,(_{-1}^{~\tau}) |^2}
\end{aligned}
\label{defW}
\end{equation}
with
\begin{equation}
\label{defM}
  M = \begin{pmatrix}
       w^1 & m^1 \\
       w^2 & m^2
      \end{pmatrix}~.
\end{equation}
In the right-hand sides of \eq{defW} the prefactor is just the volume of the
torus (in units of $\alpha'$), since 
$\int d^2x \sqrt{\det G} = \alpha'\, T_2$.

By suitably reshuffling the summation variables, it is easy to check that
$\Gamma_{2,2}$
is invariant under the modular group acting on the world-sheet parameter
$\tau$; indeed 
\begin{equation}
\label{modinvW}
\begin{aligned}
&\Gamma_{2,2}(\tau + 1, \bar\tau + 1;T,U) = \Gamma_{2,2}(\tau,\bar\tau;T,U)~,\\
& \Gamma_{2,2}(-1/\tau,-1/\bar\tau;T,U) = \Gamma_{2,2}(\tau,\bar\tau;T,U)~.
\end{aligned}
\end{equation}

In presence of Wilson lines and/or insertions of projection operators, the
lattice
sum corresponding to the trace over bosonic zero-modes is formally identical to
\eq{defW}, 
but with
\begin{equation}
\label{defMsh}
 M = \begin{pmatrix}
       w^1 + \frac{h_1}{2} & m^1 + \frac{g_1}{2} \\
       w^2 + \frac{h_2}{2} & m^2 + \frac{g_2}{2}
      \end{pmatrix}~,~~~(m^i,w^i)\in\Z^2
\end{equation}
where the parameters $g_i$ and $h_i$ depend on the type of Wilson lines or
projection
operators. To explicitly exhibit such a dependence we introduce the notation
\begin{equation}
\Gamma_{2,2}\sp{h_1\,h_2}{g_1\,g_2}(\tau,\bar\tau;T,U)
\label{defWhg}
\end{equation}
to denote the lattice sum on the torus. Of course, we have 
$\Gamma_{2,2}\sp{0\,0}{0\,0}\equiv\Gamma_{2,2}$.

Under the world-sheet modular group, the lattice sum (\ref{defWhg}) has the
following 
transformation properties
\begin{equation}
\label{modWsh}
\begin{aligned}
&\Gamma_{2,2}\sp{h_1\,h_2}{g_1\,g_2}(\tau + 1, \bar\tau + 1;T,U) 
= \Gamma_{2,2}\sp{~~h_1\,~~~~h_2}{g_1+h_1\,g_2+h_2}(\tau,\bar\tau;T,U)~,\\
& \Gamma_{2,2}\sp{h_1\,h_2}{g_1\,g_2}(-1/\tau,-1/\bar\tau;T,U) 
= \Gamma_{2,2}\sp{g_1\,g_2}{h_1\,h_2}(\tau,\bar\tau;T,U)~,
\end{aligned}
\end{equation}
which are a generalization of those in \eq{modinvW}.

In the heterotic threshold computation of Section~\ref{sec:prephet}, the shifts
$h_i$ and $g_i$ take the values 0 or 1 only, as they arise from a $\Z_2$
(freely-acting) 
orbifold procedure. In particular, $h_i=0$ and $h_i=1$ correspond, respectively,
to untwisted and 
twisted sectors, while $g_i=0$ and $g_i=1$ indicate to the absence or the
presence of the projection operator. In this case there are some useful
summation identities; in particular we have
\begin{equation}
\label{sumg}
\begin{aligned}
\sum_{g_i=0}^1 \Gamma_{2,2}\sp{h_1\,h_2}{g_1\,g_2}(\tau, \bar\tau;T,U) & = 2\,
\Gamma_{2,2}\sp{h_1\,h_2}{0\,\,\,0}(2\tau,2\bar\tau;\ft{T}{2},U)~, \\
\sum_{h_i=0}^1 \Gamma_{2,2}\sp{h_1\,h_2}{g_1\,g_2}\!(\tau, \bar\tau;T,U) & = 2\,
\Gamma_{2,2}\sp{0\,\,\,0}{g_1\,g_2}(\ft{\tau}{2},\ft{\bar\tau}{2};\ft{T}{2},U)~.
\end{aligned}
\end{equation}
Using these identities, we also find 
\begin{equation}
\label{sumu}
\begin{aligned}
\sum_{g_i=0}^1 &\Big(\Gamma_{2,2}\sp{0\,\,\,1}{g_1\,g_2}(\tau,\bar\tau;T,U)
+\Gamma_{2,2}\sp{1\,\,\,0}{g_1\,g_2}(\tau,\bar\tau;T,U)
+\Gamma_{2,2}\sp{1\,\,\,1}{g_1\,g_2}(\tau,\bar\tau;T,U)\Big)\\
&~~~~~=\,2\, \sum_{h_i=0}^1
\Gamma_{2,2}\sp{h_1\,h_2}{0\,\,\,\,0}(2\tau,2\bar\tau;\ft{T}{2},U)
-2\,\Gamma_{2,2}\sp{0\,0}{0\,0}(2\tau,2\bar\tau;\ft{T}{2},U)\\
&~~~~~=\,4\,\Gamma_{2,2}(\tau,\bar\tau;\ft{T}{4},U) - 2\,
\Gamma_{2,2}(2\tau,2\bar\tau;\ft{T}{2},U)~.
\end{aligned}
\end{equation} 

\subsection*{Open strings}
Also the bosonic zero-modes $x$ and $p$ of the open string coordinates 
contribute to the Virasoro characters since
\begin{equation}
\label{l0op}
 L_0 = \frac12 \,p^2 + \mathrm{osc.}~.
\end{equation}

In the case of $d$ real non-compact directions with Neumann-Neumann (NN)
boundary conditions,
we have $p_\mu=\sqrt{2\alpha'}k_\mu$, with $k_\mu$ a continuous variable, 
and thus the zero-mode contribution to $\Tr q^{L_0}$ is simply
\begin{equation}
\label{0open}
\int d^d x \int \frac{d^dk}{(2\pi)^d}~ \ee^{-\pi\alpha' \tau_2\, k^2} = 
\frac{V_d}{(4\pi^2 \alpha' \tau_2)^{\frac d2}} \equiv
\frac{\nu_d}{(\tau_2)^{\frac d2}}
\end{equation}
where we have set $q=\ee^{-\pi\tau_2}$ with $\tau_2$ being the real modular
parameter of an annulus.
This expression is formally identical to the closed string one in \eq{0mt}.

Now consider a pair of directions compactified on a 2-torus $\cT_2$.
If these directions have NN boundary conditions, the open strings carry a
quantized momentum given by $p_i=\sqrt2\,n_i$, with $n_i\in\Z$, and the Virasoro
generator becomes
\begin{equation}
L_0 = n_i\,{\mathcal G}^{ij}\,n_j+ 
\mathrm{osc.}~,
\label{L0NN}
\end{equation}
where ${\mathcal G}^{ij}$ is the inverse of the open string
metric~\cite{Seiberg:1999vs}
\begin{equation}
\cG_{ij} = \big(G_{ik}+B_{ik}\big)G^{kl}\big(G_{lj}-B_{lj}\big)~.
\label{openmetr} 
\end{equation}
In matrix form we have
\begin{equation}
\label{openmetr1}
\cG = \frac{|T|^2}{T_2U_2} 
 \begin{pmatrix}
  1 & U_1 \\
  U_1 & |U|^2
 \end{pmatrix}~,~~~
\cG^{-1}= \frac{T_2}{|T|^2U_2} 
 \begin{pmatrix}
   |U|^2& -U_1 \\
  -U_1 & 1
\end{pmatrix}~.
\end{equation}
Then, 
the zero-mode contribution to the partition function reads
\begin{equation}
\label{defPtr}
P(\tau_2;T,U) =
\sum_{\vec n\in\Z^2}\ee^{-\pi \tau_2\, n_i \cG^{ij} n_j} = 
\sum_{\vec n\in\Z^2}\ee^{-\pi\tau_2 \frac{|n_1 U-n_2|^2T_2}{|T|^2U_2}}~.
\end{equation}

In the case of open strings with Dirichlet-Dirichlet
(DD) boundary conditions, the bosonic zero-modes account for the integer
windings $w^i$ around
the torus, and the Virasoro operator becomes
\begin{equation}
L_0 = w^iG_{ij}w^j+ \mathrm{osc.}~.
\label{L0DD}
\end{equation}
Thus, the zero-mode trace for two compact DD directions is
\begin{equation}
\label{defYtr}
W(\tau_2;T_2,U) = 
\sum_{\vec w\in\Z^2} \ee^{-\pi \tau_2\, w^i G_{ij} w^j} = 
\sum_{\vec w\in\Z^2}\ee^{-\pi \tau_2\,\frac{|w^1+w^2 U|^2T_2}{U_2}}~. 
\end{equation}
Notice that $W$ does not depend on the $B$ field and hence on $T_1$, as opposed
to what happens
for $P$. Notice also that the two functions $P$ and $W$ are related to each
other by T-duality. 
Indeed, performing a T-duality along the directions of $\cT_2$, one exchanges NN
with DD boundary conditions and makes the following replacements 
\begin{equation}
U~\longrightarrow~ -\frac{1}{U}~,~~~T~\longrightarrow~-\frac{1}{T}~,
\end{equation}
under which $P$ and $W$ are mapped to each other as one can easily check from
the 
explicit expressions given above.

If the DD string endpoints are separated by a distance $\vec v$ along $\cT_2$,
the trace
(\ref{defYtr}) generalizes to
\begin{equation}
\label{defYtrv}
W_{\vec v}(\tau_2;T_2,U) = \sum_{\vec w\in\Z^2}\ee^{-\pi \tau_2\, (w^i - v^i)
G_{ij} (w^j - v^j)}~. 
\end{equation}
Furthermore, all these formulas can be easily generalized 
to higher dimensional factorized tori. For example,
for a 4-torus $\cT_2^{(1)}\times \cT_2^{(2)}$ the lattice sums over momentum and
winding
modes become
\begin{equation}
P_4 (\tau_2)=\prod_{i=1}^2  P(\tau_2;T^{(i)},U^{(i)})~~~\mbox{and}~~~
W_4 (\tau_2)=\prod_{i=1}^2  W(\tau_2;T_2^{(i)},U^{(i)})~.
\label{p4w4}
\end{equation}

\section{Details on the type I$^\prime$ computations}
\label{app:detI}
In this appendix we provide some details on the perturbative computations
performed in
the type I$^\prime$ model to recover the tadpole cancellation conditions and the
1-loop corrections to the D7-brane effective action.

Preliminarly, we note that the 1-loop amplitudes can be conveniently written in
terms 
of the SO(4) level-one characters defined by
\begin{equation}
O_{4} =\frac{\vartheta_3^2+\vartheta_4^2}{2\eta^2} ~,~~~
V_{4} =\frac{\vartheta_3^2-\vartheta_4^2}{2\eta^2} ~,~~~
S_{4} =\frac{\vartheta_2^2- \vartheta_1^2}{2\eta^2}~,~~~
C_{4} =\frac{\vartheta_2^2+\vartheta_1^2}{2\eta^2} ~,
\end{equation}
where the $\vartheta$-functions and their properties are collected in
App.~\ref{app:theta}.
On these characters the generators $T$ and $S$ of the modular group are
represented by the
following matrices
\begin{equation}
T=\ee^{-\frac{\pi\ii}{6}}\,
\left(\begin{array}{rrrr} 1&0&~0&~0\\
                     0&-1&0&0\\
                     0&0& \ii & 0\\
                     0&0& 0 & \ii  \end{array}\right)
~~,~~~
S=\frac{1}{2}\,
\left(\begin{array}{rrrr} 1&1&1&1\\
                     1&1&-1&-1\\
                     1&-1& -1 & 1\\
                     1&-1& 1 & -1  \end{array}\right) ~.
\label{modularST}
\end{equation}
Consequently, the transformation $P=T S T^2 S$, which relates the M\"obius
amplitudes 
in the direct and transverse channels according to
\begin{equation}
\frac{\ii \tau_2}{2}+\frac{1}{2} =\frac{i}{4 \ell}+\frac{1}{2}=P \left(i
\ell+\frac{1}{2}\right)~,
\label{ptransf}
\end{equation}
is represented on the characters by the matrix
\begin{equation}
P =\left(\begin{array}{rrrr} \,0\,&\,1\,&\,0\,&\,0\,\\
                     1&0&0&0\\
                     0&0& 0 & 1\\
                     0&0& 1 & 0  \end{array}\right)~.
                      \label{modularP} 
\end{equation}
In addition we introduce the supersymmetric combinations 
of characters
\begin{equation}
\begin{aligned}
Q_o &= V_4 O_4-C_4 C_4 \quad ,\quad
Q_v = O_4 V_4-S_4 S_4~,\\
Q_s &= O_4 C_4-S_4 O_4 \quad ,\quad
Q_c = V_4 S_4-C_4 V_4~,
\end{aligned}
\label{susychar}
\end{equation}
which have a definite parity under the $\Z_2$-orbifold generators. These
characters 
transform under $S$ and $P$ in the same way as $O_4$, $V_4$, $S_4$ and $C_4$
respectively.
Their expression in terms of $\vartheta$-functions is
\begin{equation}
\begin{aligned}
Q_o+Q_v &= \frac{\vartheta_3^4-\vartheta_4^4-\vartheta_2^4}{2 \eta^4}=-
\frac{\vartheta_1^4}{\eta^4}
~,\\
Q_o-Q_v &=\frac{\vartheta_3^2\vartheta_4^2-\vartheta_4^2 \vartheta_3^2}{2
\eta^4}=
-\frac{\vartheta_1^2 \,\vartheta_2^2}{\eta^4 }~,\\
Q_s+Q_c &=\frac{\vartheta_3^2\vartheta_2^2-\vartheta_2^2 \vartheta_3^2}{2
\eta^4}= 
- \frac{\vartheta_1^2 \,\vartheta_4^2}{\eta^4 }~,\\
Q_s-Q_c &=\frac{\vartheta_4^2\vartheta_2^2-\vartheta_2^2 \vartheta_4^2}{2
\eta^4}= 
- \frac{\vartheta_1^2 \,\vartheta_3^2}{\eta^4 } 
\end{aligned}
\label{qoqv}
\end{equation}
where the second equalities in the right-hand sides follow from the Riemann
identity (\ref{riemann}).
Finally, it is also useful to recall the modular transformations of the Dedekin
$\eta$-function
\begin{equation}
\begin{aligned}
&\eta(\tau+1) = \ee^{\frac{\ii\pi}{12}}\,\eta(\tau)~,\\
&\eta\left(-{1}/{\tau} \right) =  \sqrt{- \ii \tau}~ \eta(\tau)~,\\
&\eta(\ft{\ii \tau_2}{2}+\ft12) = \ee^{\frac{\pi \ii}{4}}\, \sqrt{ \ii/\tau_2}~
\eta(\ft{\ii}{ 2\tau_2}+\ft12)~.
\end{aligned}
\label{etamod}
\end{equation}

\subsection{Tadpole cancellation}
\label{subapp:tadpoles}
The tadpole constraints arise from the analysis of the  Klein bottle amplitude
and of the annuli and M\"obius diagrams with boundaries on D7- and/or on
D3-branes. For each boundary these 1-loop diagrams contain a trace over the
corresponding CP indices and hence, when the orbifold/orientifold generators are
inserted, also a trace on the corresponding $\gamma$ matrices. 
All such amplitudes can be constructed in a rather straightforward manner by
collecting such
CP factors and the traces over zero- and non-zero-modes. Here we give some
details for the various
amplitudes in the direct (open string) channel, and then perform a modular
transformation to obtain their expression in the transverse (closed string)
channel and determine the massless tadpoles. 
To simplify a bit the calculation, but without any loss of generality, we switch
off the $B$ field
in the internal space so that the momentum sum and the winding sum
become related to each other under the world-sheet modular transformations in a
simple way. 
Indeed, applying the Poisson resummation formula (\ref{poisson}) to
\eq{defYtrv}, we find
\begin{equation}
 \label{YtoQ}
 W_{\vec v}(\tau_2;T_2,U) = \frac{1}{\tau_2 T_2} \,
P_{\vec v}(\ft{1}{\tau_2};T_2,U)~,
\end{equation}
where we have defined
\begin{equation}
 \label{defQ}
 P_{\vec v}(\tau_2;T_2,U) = \sum_{\vec n\in\Z^2} \ee^{-\pi\tau_2 \, n_i
 G^{ij} n_j -2\pi\ii n_i v^i}~. 
\end{equation}
Note that for $v^i=0$, we have $P_{\vec 0}(\tau_2;T_2,U)=P(\tau_2;T_2,U)$.
Similar transformation
properties can be obtained also for the lattice sums $P_4$ and $W_4$ on a
4-torus. In the following
we will understand the dependence on the K\"ahler and complex structures of
these functions
to simplify the notation.

\subsubsection*{1-loop amplitudes in the direct channel} 
In our model the annulus amplitude is defined by 
\begin{equation}
\mathcal A = \int_0^\infty\frac{d\tau_2}{2\tau_2}~
\Tr\Big[\frac{1}{2}\,\frac{1+\hat g}{2}\,\frac{1+(-1)^F}{2}\,\ee^{-\pi
\tau_2(L_0-\frac{c}{24})}\,\Big]
\label{an}
\end{equation}
where the trace is taken over all types of open strings, {\it i.e.} 7/7, 3/3,
7/3 and 3/7, as well as
over their CP indices and sectors.
For the 7/7 strings we find 
\begin{equation}
\begin{aligned}
\mathcal A_{7/7} &= \frac{\nu_4}{4}\,
\int_0^\infty\frac{d\tau_2}{\tau_2^3}\,\sum_{\vec\alpha,\vec\alpha'} \Bigg\{
N_{\vec\alpha }\,N_{\vec\alpha'}\,\frac{Q_o+Q_v}{\eta^8} 
\big( \ft{\ii \tau_2}{2}\big) \, P_4(\tau_2)\\
& ~~~~~~~~~~+ 4\,
\tr\gamma_{\vec\alpha}(\hat g)\,\tr\gamma_{\vec\alpha'}(\hat g)\, 
\frac{Q_o-Q_v}{\eta^2 \theta_2^2}\big(\ft{\ii \tau_2}{2}\big)
\,\Bigg\}\, W_{\vec\alpha-\vec\alpha'}( \tau_2 )
\end{aligned}
\label{an77tot}
\end{equation}
where $\nu_4$ is the dimensionless volume introduced in \eq{0open}.
The two lines in (\ref{an77tot}) correspond, respectively,
to the two contributions with $1$ and $\hat g$ inserted in the traces, and the
sum is over
all pairs of D7-brane fixed points. 

Let us now consider the 3/3 annuli. The location of the D3-branes is identified
by a
6-vector $\vec \xi$ that we write as $\vec\xi=(\vec\xi_4,\vec\xi_2)$ to
exhibit the position along $\cT_4$ and $\cT_2$. Then, proceeding
similarly as for the 7/7 annuli, we find
\begin{equation}
\begin{aligned}
\mathcal A_{3/3} &= \frac{\nu_4}{4}\,
\int_0^\infty\frac{d\tau_2}{\tau_2^3}\, \Bigg\{
\sum_{\vec\xi,\vec\xi'} M_{\vec\xi}\,M_{\vec\xi'}
~\frac{Q_o+Q_v}{\eta^8}\big(\ft{\ii \tau_2}{2}\big)\,
W_{\vec\xi-\vec\xi'}(\tau_2) \\
&~~~~~~~~~~+4\sum_{\vec\xi,\vec\xi'} {}'
\tr\gamma_{\vec\xi }(\hat g)\,\tr\gamma_{\vec\xi' }(\hat g)~
\frac{Q_o-Q_v}{\eta^2 \vartheta_2^2}\big(\ft{\ii \tau_2}{2}\big)\,
W_{\vec\xi_2-\vec\xi_2'}(\tau_2)\Bigg\} 
\end{aligned}
\label{an33tot}
\end{equation}
where $W_{\vec\xi-\vec\xi'}$ is the obvious generalization to $\cT_4\times
\cT_2$ of the winding sum
(\ref{defYtrv}), and in the second line the $^\prime$ means that the sum is
restricted to
couples of fixed points $\vec\xi$ and $\vec\xi'$ lying on top of each other on
$\cT_4$ and separated
by a distance $\vec\xi_2-\vec\xi'_2$ on $\cT_2$. More precisely this sum is over
fixed-point
pairs satisfying  $\vec\xi_4=\vec\xi'_4$. This constraint arises because only
states with zero winding number along $\cT_4$ can contribute to the trace when
$\hat g$ is inserted.

Finally, the annulus contribution from $7/3$ and $3/7$ strings turns out to be 
\begin{equation}
\begin{aligned}
 \mathcal A_{7/3}+\mathcal A_{3/7} & =
-\frac{\nu_4}{4}\int_0^\infty\frac{d\tau_2}{\tau_2^3}\,
\sum_{\vec\alpha,\vec \xi} \Bigg\{2\,  N_{\vec \alpha}\,M_{\vec\xi}\,
\frac{Q_s+Q_c}{\eta^2 \vartheta_4^2}\big(\ft{\ii \tau_2}{2}\big)\\
&~~~~~~~~~~+2\,\tr\gamma_{\vec\alpha}(\hat g)\,\tr\gamma_{\vec \xi}(\hat g)~
\frac{Q_s-Q_c}{\eta^2 \vartheta_3^2} \big(\ft{\ii \tau_2}{2}\big)\Bigg\}
W_{\vec\alpha-\vec\xi_2}(\tau_2)~.
\end{aligned}
\label{a73totale}
\end{equation}

Let us now turn to the M\"obius amplitudes, which in our model are given by
\begin{equation}
\mathcal M = \int_0^\infty\frac{d\tau_2}{2\tau_2}~
\Tr\Big[\frac{\Omega'}{2}\,\frac{1+\hat g}{2}\,\frac{1+(-1)^F}{2}\,\ee^{-\pi
\tau_2(L_0-\frac{c}{24})}\,\Big]
\label{moeb}
\end{equation}
with the trace computed over open strings of type 7/7 and 3/3, and their CP
indices.
Of course the 7/3 and 3/7 strings do not contribute to the M\"obius amplitudes.

For the 7/7 strings we have
\begin{equation}
\begin{aligned}
 \mathcal M_{7/7}&=-\frac{\nu_4}{4}\int_0^\infty\frac{d\tau_2}{\tau_2^3}\,
\sum_{\vec \alpha}  \Bigg\{\tr\big(\gamma_{ \vec\alpha}^{-1}(\Omega')\,
\gamma_{\vec\alpha}^T(\Omega')\big)\,\frac{Q_o+Q_v}{\eta^8}
\big(\ft{\ii \tau_2}{2}+\ft12\big)\,P_4(\tau_2)\\
&~~~~~~~~~~+4 \,\tr\big(\gamma_{ \vec\alpha}^{-1}(\Omega' \hat g)\,
\gamma_{\vec\alpha}^T(\Omega' \hat g)\big)\, \frac{Q_o-Q_v}{\eta^2
\vartheta_2^2}
\big(\ft{\ii \tau_2}{2}+\ft12\big) \Bigg\}\, W(\tau_2)~,
\end{aligned}
\label{moeb77fin}
\end{equation}
while for M\"obius diagrams with their boundary on D3-branes we find
\begin{equation}
\begin{aligned}
 \mathcal M_{3/3}&=-\frac{\nu_4}{4}\int_0^\infty\frac{d\tau_2}{\tau_2^3}\,
\sum_{\vec \xi}  \Bigg\{4 \,\tr\big(\gamma_{ \vec\xi}^{-1}(\Omega')\,
\gamma_{\vec\xi}^T(\Omega')\big)\, \frac{Q_o-Q_v}{\eta^2 \vartheta_2^2}
\big(\ft{\ii \tau_2}{2}+\ft12\big)\\
&~~~~~~~~~~+\tr\big(\gamma_{ \vec\xi}^{-1}(\Omega'\hat g)\,
\gamma_{\vec\xi}^T(\Omega'\hat g)\big)\,\frac{Q_o+Q_v}{\eta^8}
\big(\ft{\ii \tau_2}{2}+\ft12\big)\,W_4(\tau_2)\Bigg\}\, W(\tau_2)~.
\end{aligned}
\label{moeb33fin}
\end{equation}

The last type of contribution which is relevant is that corresponding to a Klein
bottle. This is
a closed string amplitude which in our model is given by
\begin{equation}
\begin{aligned}
\mathcal K & = \int_0^\infty\frac{d\tau_2}{2\tau_2}~
\Tr\Big[\,\frac{\Omega'}{2}\,\frac{1+\hat
g}{2}\,\frac{1+(-1)^F}{2}\,\frac{1+(-1)^{\bar F}}{2}\,
q^{L_0-\frac{c}{24}}\,\bar q^{\bar L_0-\frac{c}{24}}\,
\Big]\\
&=\int_0^\infty\frac{d\tau_2}{2\tau_2}~
\Tr\Big[\,\frac{\Omega'}{2}\,\frac{1+\hat
g}{2}\,\frac{1+(-1)^F}{2}\,\ee^{-4\pi\tau_2(L_0-\frac{c}{24})}\,
\Big]
\end{aligned}
\label{kb}
\end{equation}
where $q=\ee^{2\pi\ii\tau}$ with $\tau=\tau_1+\ii\tau_2$ being the modular
paramenter, and the
trace taken over both the untwisted and twisted closed string spectra. Since
$\Omega'$
exchanges left- and right-movers, only those states with $L_0=\bar L_0$ and
$F=\bar F$
contribute, thus explaining
the espression in the second line. Evaluating the traces, we find
\begin{equation}
\cK  = \frac{\nu_4}{4}
\int_0^\infty\frac{d\tau_2}{\tau_2^3} 
\Bigg\{\frac{Q_o+Q_v}{\eta^8}\big(2\ii \tau_2\big)\,
\Big[P_4\big(\tau_2\big)+W_4\big(\tau_2\big)\Big] -
32\,\frac{Q_s+Q_c}{\eta^2 \vartheta_4^2}\big(2\ii \tau_2\big) \Bigg\}\,
W\big(\tau_2\big)~.
\label{untwklein1}
\end{equation}
The terms proportional to $(Q_o+Q_v)$ come from $\Tr 1$ and $\Tr \hat g$ in the
untwisted closed
string sectors, while the terms proportional to $(Q_s+Q_c)$ account for two
identical 
contributions from the twisted sectors.

\subsection*{1-loop amplitudes in the transverse channel}
In order to obtain the tadpole condition, we have to rewrite the above
amplitudes in the transverse channel. For the annulus this is achieved by
writing $\tau_2=\frac{2}{\ell}$~, 
and then using the modular transformation properties of the lattice sums and the
supersymmetric
characters. In particular, with the help of \eq{YtoQ} we have
\begin{equation}
\begin{aligned}
W_{\vec v}\big(\tau_2;T_2^{(i)},U^{(i)}\big) &= \frac{\ell}{2T_2^{(i)}}\,
P_{\vec v}\left(\ft{\ell}{2};T_2^{(i)},U^{(i)}\right)
~,\\
P_{\vec v}\big(\tau_2;T_2^{(i)},U^{(i)}\big) &= \frac{T_2^{(i)}\,\ell}{2}
W_{\vec v}\left(\ft{\ell}{2};T_2^{(i)},U^{(i)}\right)~.
\end{aligned}
\label{PW}
\end{equation}
Using these relations, we find that in the transverse channel the amplitudes
(\ref{an77tot}), (\ref{an33tot}) and (\ref{a73totale}) become
\begin{equation}
\begin{aligned}
\mathcal A_{7/7} &= \frac{\nu_4}{4}\,
\int_0^\infty \!d\ell~\sum_{\vec\alpha,\vec\alpha'} \Bigg\{
\frac{T_2^{(1)}T_2^{(2)}}{32T_2}\,
N_{\vec\alpha }\,N_{\vec\alpha'}\,\frac{Q_o+Q_v}{\eta^8}(\ii\ell) 
\, W_4\big(\ft{\ell}{2}\big)~~~~~~~~~~\\
& ~~~~~~~~~~- \frac{1}{2T_2}
\tr\gamma_{\vec\alpha}(\hat g)\,\tr\gamma_{\vec\alpha'}(\hat g)\, 
\frac{Q_s+Q_c}{\eta^2 \theta_4^2}(\ii\ell)
\,\Bigg\}\, P_{\vec\alpha-\vec\alpha'}\big(\ft{\ell}{2}\big)~,
\end{aligned}
\label{A77ex}
\end{equation}
\begin{equation}
\begin{aligned}
\mathcal A_{3/3} &= \frac{\nu_4}{4}\,
\int_0^\infty \!d\ell ~\Bigg\{\frac{1}{32 T_2^{(1)}T_2^{(2)}T_2}
\sum_{\vec\xi,\vec\xi'} M_{\vec\xi}\,M_{\vec\xi'}
~\frac{Q_o+Q_v}{\eta^8}(\ii\ell)\,
P_{\vec\xi-\vec\xi'}\big(\ft{\ell}{2}\big) \\
&~~~~~~~~~~-\frac{1}{2T_2}\sum_{\vec\xi,\vec\xi'} {}'
\tr\gamma_{\vec\xi }(\hat g)\,\tr\gamma_{\vec\xi' }(\hat g)~
\frac{Q_s+Q_c}{\eta^2 \vartheta_4^2}(\ii\ell)\,
P_{\vec\xi_2-\vec\xi_2'}\big(\ft{\ell}{2}\big)
\Bigg\} ~,
\end{aligned}
\label{A33ex}
\end{equation}
and
\begin{equation}
\begin{aligned}
 \mathcal A_{7/3}+\mathcal A_{3/7} & = \frac{\nu_4}{4}\int_0^\infty\! d\ell~
\sum_{\vec\alpha,\vec \xi} \Bigg\{\frac{1}{4T_2}\,  N_{\vec \alpha}\,M_{\vec\xi}
\,\frac{Q_o-Q_v}{\eta^2 \vartheta_2^2}(\ii\ell)\\
&~~~~~~~~~~-\frac{1}{4T_2}\,\tr\gamma_{\vec\alpha}(\hat g)\,\tr\gamma_{\vec
\xi}(\hat g)~
\frac{Q_s-Q_c}{\eta^2 \vartheta_3^2}(\ii\ell)\Bigg\}
P_{\vec\alpha-\vec\xi_2}\big(\ft{\ell}{2}\big)~.
\end{aligned}
\label{A73ex}
\end{equation}

The transverse channel for the M\"obius diagrams is reached by means of the
transformation
(\ref{ptransf}) which on the lattice sums implies in particular the following
relations
\begin{equation}
\begin{aligned}
P\big(\tau_2;T_2^{(i)},U^{(i)}\big)& =2\ell\,T_2^{(i)}\, 
W\big(2\ell;T_2^{(i)},U^{(i)}\big)~,\\
W\big(\tau_2;T_2^{(i)},U^{(i)}\big)& =\frac{2\ell}{T_2^{(i)}}\, 
P\big(2\ell;T_2^{(i)},U^{(i)}\big)~.
\end{aligned}
\label{PtoYMob}
\end{equation}
Then, in the transverse channel the M\"obius amplitudes (\ref{moeb77fin}) and
(\ref{moeb33fin}) become
\begin{equation}
 \begin{aligned}
  \mathcal M_{7/7}&=-\frac{\nu_4}{4}\int_0^\infty \!d\ell~
 \sum_{\vec \alpha}  \Bigg\{
\frac{2T_2^{(1)}T_2^{(2)}}{T_2}
\,\tr\big(\gamma_{ \vec\alpha}^{-1}(\Omega')\,
\gamma_{\vec\alpha}^T(\Omega')\big)\,\frac{Q_o+Q_v}{\eta^8}
\big(\ii\ell+\ft12\big)\,W_4(2\ell)\\
&~~~~~~~~~~+\frac{8}{T_2}\,\tr\big(\gamma_{ \vec\alpha}^{-1}(\Omega' \hat g)\,
\gamma_{\vec\alpha}^T(\Omega' \hat g)\big)\, \frac{Q_o-Q_v}{\eta^2
\vartheta_2^2}  \big(\ii\ell+\ft12\big) \Bigg\}\, P(2\ell)~,
\end{aligned}
\label{mob77ex}
\end{equation}
and
\begin{eqnarray}
 \mathcal M_{3/3}&=&-\frac{\nu_4}{4}\int_0^\infty\! d\ell~
\sum_{\vec \xi}  \Bigg\{\frac{8}{T_2}\,\tr\big(\gamma_{ \vec\xi}^{-1}(\Omega')\,
\gamma_{\vec\xi}^T(\Omega')\big)\, \frac{Q_o-Q_v}{\eta^2 \vartheta_2^2}
\big(\ii\ell+\ft12\big)\label{Mob33ex}\\
&&~~~~~~+\frac{2}{T_2^{(1)}T_2^{(2)}T_2}\,
\tr\big(\gamma_{ \vec\xi}^{-1}(\Omega'\hat g)\,
\gamma_{\vec\xi}^T(\Omega'\hat g)\big)\,\frac{Q_o+Q_v}{\eta^8}
\big(\ii\ell+\ft12\big)\,P_4(2\ell)\Bigg\}\, P(2\ell)~.
\nonumber
\end{eqnarray}
Finally, the exchange channel for the Klein bottle is reached by the
transformation $\tau_2=\frac{1}{2\ell}$, so that using Eqs.~(\ref{PtoYMob}) and
(\ref{etamod}) and the modular
properties of the characters, we get
\begin{equation}
\begin{aligned}
\cK  &= \frac{\nu_4}{4}
\int_0^\infty \!d\ell~
\Bigg\{\frac{Q_o+Q_v}{\eta^8}(\ii \ell)\,
\Big[\frac{32T_2^{(1)}T_2^{(2)}}{\alpha'T_2}W_4\big(2\ell\big)+
\frac{32}{T_2^{(1)}T_2^{(2)}T_2}P_4\big(2\ell\big)\Big] \\
&~~~~~~~~~~~~~+\frac{256}{T_2}\,\frac{Q_o-Q_v}{\eta^2 \vartheta_2^2}(\ii\ell)
\Bigg\}\, P\big(2\ell\big)~.
\end{aligned}  
\label{kleinfin1}
\end{equation}

The annulus, M\"obius and Klein bottle amplitudes exhibit divergences 
for $\ell\to \infty$ which are due to the exchange of massless closed string
states. 
The exchanged states can be identified from the corresponding character $Q_o$,
$Q_v$,
$Q_s$ and $Q_c$, while the $T_2^{(i)}$-dependence specifies the volume of the
D-brane/O-plane source. 
Since only massless states contribute to the divergences we can discard the
massive character $Q_c$. 
Contributions proportional to $Q_s$, which correspond to the exchange of a
twisted state, should
cancel identically since they appear only in the annulus amplitudes.
{From} Eqs.~(\ref{A77ex})~-~(\ref{A73ex}), we see that this requires
\begin{equation}
\tr\gamma_{\vec\alpha}(\hat g) = \tr\gamma_{\vec \xi}(\hat g)=0
\label{traceless}
\end{equation}
Defining $\cV_4=T^{(1)}_2T^{(2)}_2$, we see that the contributions 
proportional to $Q_o$ and $Q_v$ can be put in the form
\begin{equation}
\begin{aligned}
 \mathcal A_{o,v} &= \frac{\cV_4}{32 T_2}\Bigg[   
\sum_{\vec\alpha} N_{\vec\alpha }\pm \frac{1}{\cV_4}\sum_{ \vec\xi} \!M_{
\vec\xi} \,
\Bigg]^2 ~,\\
 \mathcal M_{o,v} &=-\frac{2 \cV_4}{T_2} \Bigg[\sum_{\vec\alpha} 
\tr\big(\gamma_{\vec\alpha}^{-1}(\Omega')\gamma_{\vec\alpha}^{T}(\Omega')\big)
 \pm   \frac{1}{\cV_4}  \sum_{\vec\alpha}
\tr\big(\gamma_{\vec\alpha}^{-1}(\Omega' \hat g) \,
 \gamma_{\vec\alpha}^T(\Omega' \hat g)\big) \\
&~~~~~~\pm \frac{1}{\cV_4} \sum_{ \vec\xi}
\tr\big(\gamma_{\vec\xi}^{-1}(\Omega')\,
 \gamma_{\vec\xi}^T(\Omega')\big) 
+\frac{1}{\cV_4^2} \sum_{ \vec\xi}\tr\big(\gamma_{\vec\xi}^{-1}(\Omega' \hat
g)\,
 \gamma_{\vec\xi}^T(\Omega' \hat g)\big) \Bigg]  ~,\\
 \mathcal K_{o,v} &= \frac{32 \cV_4}{T_2} \Bigg[ 1 \pm  \frac{1}{\cV_4}\Bigg]^2
~,
\end{aligned}
\label{tadamk}
\end{equation}
with the upper sign referring to $Q_o$ and the lower one to $Q_v$. The
open/close string consistency requires that the sum of these three amplitudes
should form a complete squares. This condition implies the following constraints
on the CP traces
\begin{equation}
\begin{aligned}
 N_{\vec \alpha} &= \phantom{\vdots} 
\tr\big(\gamma_{\vec\alpha}^{-1}(\Omega')\gamma_{\vec\alpha}^{T}(\Omega')\big)
=\tr\big(\gamma_{\vec\alpha}^{-1}(\Omega' \hat g) \gamma_{\vec\alpha}^T(\Omega'
\hat g)\big)~,\\
 M_{\vec\xi} &= \phantom{\vdots} \tr\big(\gamma_{\vec\xi}^{-1}(\Omega')\,
 \gamma_{\vec\xi}^T(\Omega')\big)= \tr\big(\gamma_{\vec\xi}^{-1}(\Omega' \hat g)
 \gamma_{\vec\xi}^T(\Omega' \hat g)\big)~,
\end{aligned}
\label{ntrmr}
\end{equation}
which are satisfied with the matrices given in Eqs.~(\ref{gamma7}) and
(\ref{gamma3}).
Plugging them into \eq{tadamk}, we then find
\begin{equation}
\mathcal A_{o,v} + \mathcal M_{o,v}+  \mathcal K_{o,v} =\frac{\cV_4}{32 T_2}\,
\Bigg[ \Big( \sum_{\vec\alpha} N_{\vec\alpha }-32 \Big) \pm 
\frac{1}{\cV_4} \Big(\sum_{ \vec\xi} M_{ \vec\xi}\, -  32\Big) \Bigg]^2 ~,
\label{squares}
\end{equation}
so that the cancellation of the tadpoles is globally achieved if 
\begin{equation}
\sum_{\vec \alpha} N_{\vec\alpha} =32~~~~\mbox{and}~~~~
\sum_{\vec\xi  } M_{\vec\xi}=32~.
\label{n32m32}
\end{equation}
However, for any $\vec \alpha$ we can also impose the stronger conditions 
\begin{equation}
N_{\vec\alpha} =8~~~~\mbox{and}~~~~\sum_{\vec\xi_4}M_{\vec\xi}=8
\label{n8m8}
\end{equation}
where the sum runs over all 6-vectors $\vec\xi$ of the form $(\vec \xi_4,\vec
\xi_2)$
for any fixed $\vec\xi_2$, 
which ensure local cancelation of the tadpoles along the last torus
$\cT_2^{(3)}$.

\subsection{1-loop magnetized diagrams}
\label{subapp:olm}
Here we discuss in turn the various 1-loop diagrams for open
strings (partly) attached to magnetized D7-branes that were considered in
Section
\ref{subsec:oneloopI}. 

As a preliminary step, using the Cayley matrix
$S=\frac{1}{\sqrt{2}}\Big(\begin{array}{rr}1&\ii\\
1&-\ii\end{array}\Big)$ we transform in the complex basis the 
matrices $\gamma$ acting on the D7-brane CP indices in order to be consistent
with what is done on the magnetization (see \eq{fdiag}).
Denoting by $\tilde\gamma = S \gamma S^{-1}$ these transformed
matrices, from \eq{gamma7} we easily find
\begin{equation}
 \label{CPactionsnew}
{\tilde\gamma}(\Omega') = \begin{pmatrix} ~\one~& ~0~ \\ 
~0~& ~\one~\end{pmatrix}  ~,~~~
  {\tilde\gamma}(\hat g) = {\tilde\gamma}(\Omega' \hat g) =  
  \begin{pmatrix}
   ~\one~ & ~0~ \\ 
   ~0~ & - \one~
  \end{pmatrix}
\end{equation}
where we have omitted the label $\vec\alpha$ since we are focusing on a given
fixed-point.

In presence of magnetic fluxes $h_i$ open strings satisfy twisted boundary
conditions. Taking the magnetic fluxes oriented along, say, the first complex
direction and denoting by $\nu$ the 
open string twist, the contribution of the worldsheet fermions to the partition
function becomes 
\begin{equation}
\begin{aligned}
\big(Q_o+Q_v\big)(\nu) &\,=\,
\frac{
\vartheta_3(\nu)\vartheta_3^3-\vartheta_4(\nu)\vartheta_4^3-\vartheta_2(\nu)
\vartheta_2^3} {2 \eta^4}\,=\,- \frac{\vartheta_1(\ft{\nu}{2})^4}{\eta^4}~,\\
\big(Q_o-Q_v\big)(\nu)  &\,=\,
\frac{
\vartheta_3(\nu)\vartheta_3\vartheta_4^2-\vartheta_4(\nu)\vartheta_4\vartheta_3^
2}{2 \eta^4}\,=\, -\frac{\vartheta_1(\ft{\nu}{2})^2
\,\vartheta_2(\ft{\nu}{2})^2}{\eta^4 }~,\\
\big(Q_s+Q_c\big)(\nu)  &\,=\,
\frac{
\vartheta_3(\nu)\vartheta_3\vartheta_2^2-\vartheta_2(\nu)\vartheta_2\vartheta_3^
2}{2 \eta^4}\,=\, 
- \frac{\vartheta_1(\ft{\nu}{2})^2 \,\vartheta_4(\ft{\nu}{2})^2}{\eta^4 }~,\\
\big(Q_s-Q_c\big)(\nu)  &\,=\,
\frac{
\vartheta_4(\nu)\vartheta_4\vartheta_2^2-\vartheta_2(\nu)\vartheta_2\vartheta_4^
2}{2 \eta^4}\,=\, 
-\frac{\vartheta_1(\ft{\nu}{2})^2 \,\vartheta_3(\ft{\nu}{2})^2}{\eta^4}~,
\end{aligned}
\label{qoqvh}
\end{equation}
where the right hand sides follow from the Riemann identity (\ref{riemann}).
In particular for an open string stretching between the $i$-th and $j$-th
D7-brane, the twist is given by $\nu=\frac{\nu_{ij} \tau_2}{2}$ with $\nu_{ij}$
related to the magnetic fluxes $h_i$ and $h_j$ 
at the two endpoints via \eq{twistnu}.
On the other hand, the contribution of a twisted complex bosonic coordinate is
\begin{equation}
-\frac{\ii(h_i-h_j)}{4\pi^2\alpha'}~\frac{\eta}{\vartheta_1\big(\ft{\ii\nu_{ij}
\tau_2}{2}\big)} ~.
\label{bosh}
\end{equation}

The 1-loop amplitudes for magnetized D7-branes can be read from the 
formulas written in the last subsection after replacing the characters 
$Q_o$, $Q_v$, $Q_s$ and $Q_c$ by their twisted versions (\ref{qoqvh}) and the
contribution 
of one complex bosonic direction by (\ref{bosh}). The quadratic structures in
the background field are then extracted from the $h^2$-terms in the expansion of
these string amplitudes. Notice that at this order the contributions
proportional to $(Q_o+Q_v)$ can be neglected since they are of order $h^4$. 
Thus, the magnetized version of the annulus amplitude (\ref{an77tot}) is
\begin{eqnarray}
\mathcal{A}_{7/7}(h) &=& \int_0^\infty \frac{d\tau_2}{2\tau_2}\,\sum_{i,j}
\Tr_{(h_i,h_j)}\Bigg(\frac{1}{2}\,\frac{1+\hat g}{2}\,
\frac{1+(-1)^F}{2}\,\ee^{-\pi \tau_2(L_0-\frac{c}{24})}\,\Bigg)
\label{annu1pg}\\
&=&\ii\,\nu_4\sum_{i,j}\big({\tilde \gamma}(\hat g)\big)^i_{\,i}\,\big({\tilde
\gamma}(\hat g)\big)^j_{\,j}\,(h_i-h_j) \int_0^\infty \frac{d\tau_2}{2\tau_2^2}
~
\frac{\vartheta_1\big(\ft{\ii\nu_{ij} \tau_2}{4}\big)^2\,
\vartheta_2\big(\ft{\ii\nu_{ij} \tau_2}{4}\big)^2}{\vartheta_1\big(
\ft{\ii\nu_{ij} \tau_2}{2} \big)\, \eta^3 \,  \vartheta_2^2 }
\, W(\tau_2)
\nonumber
\end{eqnarray}
where in the second line we have understood that the second argument of the
$\vartheta$-functions
is $\frac{\ii\tau_2}{2}$ and have written only those terms that can contribute
to the quadratic action.
Expanding up to $O(h^2)$, we find
\begin{equation}
\begin{aligned}
\mathcal{A}_{7/7}(h)&= \frac{V_4}{8\pi^2} \Bigg(\sum_{I,J}
\left(\frac{h_I-h_J}{2\pi\alpha'}\right)^2
- \sum_{I,J} \left(\frac{h_I+h_J}{2\pi\alpha'}\right)^2
\Bigg)\int_0^\infty \frac{d\tau_2}{2\tau_2}\,W(\tau_2)+ O(h^3)\\
& = \frac{V_4}{8\pi^2}~4 \left(\tr \mathcal H\right)^2 \int_0^\infty
\frac{d\tau_2}{2\tau_2}\,
W(\tau_2)+ O(h^3)
\end{aligned}
\label{aijres}
\end{equation}
Notice that the contributions of oscillator modes completely cancel at this
order.
This is a consequence of the fact that the quadratic terms in ${\mathcal N}=2$
gauge theories 
receive contributions only from BPS states \cite{Bianchi:2000vb}. 

Now let us consider a M\"obius strip with its boundary on the magnetized
D7-branes. Taking into account that, due to the presence of $\Omega'$ inside the
trace, only
the configurations with $h_j = -h_i$ give a non-vanishing contribution, from 
(\ref{moeb77fin}) we have
\begin{eqnarray}
\mathcal{M}_{7/7}(h) &=& \int_0^\infty \frac{d\tau_2}{2\tau_2}\,\sum_{i}
\Tr_{(h_i,-h_i)}\Bigg(\frac{\Omega'}{2}\,\frac{1+\hat g}{2}\,
\frac{1+(-1)^F}{2}\,\ee^{-\pi \tau_2(L_0-\frac{c}{24})}\,\Bigg) \label{M7cong}\\
&=&\ii\,\nu_4\sum_{i}
\big({\tilde \gamma}^{-1}(\Omega' \hat g)
{\tilde \gamma}^\dagger(\Omega' \hat g)\big)^i_{\,i}~
\,(2h_i) \int_0^\infty \frac{d\tau_2}{2\tau_2^2} ~
\frac{\vartheta_1\big(\ft{\ii\nu_{i} \tau_2}{2}\big)^2\,
\vartheta_2\big(\ft{\ii\nu_{i} \tau_2}{2}\big)^2}{\vartheta_1\big(\ii\nu_{i}
\tau_2 \big)\, \eta^3 \,  \vartheta_2^2 }
\, W(\tau_2)
\nonumber
\end{eqnarray}
where now the second argument of all modular functions is  $\frac{\ii
\tau_2}{2}+\frac 12$, and again
only those structures contributing to the quadratic terms have been written. 
Expanding to order $h^2$, as before we find a complete cancelation between the
modular forms in the numerator and denominator with the result
\begin{equation}
\begin{aligned}
\mathcal{M}_{7/7}(h) &= \frac{V_4}{8\pi^2} ~4 \sum_I
\left(\frac{h_I}{2\pi\alpha'}\right)^2 
\int_0^\infty \frac{d\tau_2}{2\tau_2} \,W(\tau_2)+ O(h^3)\\
& =  - \frac{V_4}{8\pi^2} ~4\, \tr \mathcal H^2\, \int_0^\infty
\frac{d\tau_2}{2\tau_2}\, 
W(\tau_2)+ O(h^3)~.
\end{aligned}
\label{mijres}
\end{equation}
Finally, we consider the annulus amplitudes with mixed 7/3 boundary
conditions. First we observe that since the magnetic fluxes are turned on only
on D7-branes,
these amplitudes are proportional either to ${\tr}_{\mathrm{D3}} (1)$ or to
${\tr}_{\mathrm{D3}} 
\tilde\gamma(\hat g)=0$, and therefore only the unprojected part contributes to
the result.
Indeed, we find
\begin{equation}
\begin{aligned}
 \mathcal{A}_{7/3}(h) + \mathcal{A}_{3/7}(h)
& =\int_0^\infty \frac{d\tau_2}{2\tau_2}\,\sum_{i,a}
\Tr_{(h_i,a)}\Bigg(\frac{1}{2}\,\frac{1+\hat g}{2}\,
\frac{1+(-1)^F}{2}\, q^{L_0-\frac{c}{24}}\Bigg) \\
&=-\frac{\ii\,\nu_4}{2}\sum_{i,a}
h_i \int_0^\infty \frac{d\tau_2}{2\tau_2^2} ~
\frac{\vartheta_1\big(\ft{\ii\nu_{i} \tau_2}{4}\big)^2\,
\vartheta_4\big(\ft{\ii\nu_{i} \tau_2}{4}\big)^2}{\vartheta_1\big(\ft{\ii\nu_{i}
\tau_2}{2} \big)\, \eta^3 \,  \vartheta_4^2 }
\, W(\tau_2)~.
\end{aligned}
\end{equation}
Expanding to second order in $h$, we obtain
\begin{equation}
\begin{aligned}
\mathcal{A}_{7/3}(h) + \mathcal{A}_{3/7}(h)&=-\frac{V_4}{16\pi^2} \sum_{I,a}
\left(\frac{h_I}{2\pi\alpha'}\right)^2 
\int_0^\infty  \frac{d\tau_2}{2\tau_2}\, W(\tau_2)+ O(h^3) \\
& = \frac{V_4}{8\pi^2} \,m\,\tr \mathcal H^2  \int_0^\infty  
\frac{d\tau_2}{2\tau_2}\, W(\tau_2)+ O(h^3) ~.
\end{aligned}
\label{73sineg}
\end{equation}
Collecting Eqs. (\ref{aijres}), (\ref{mijres}) and (\ref{73sineg}), 
the total 1-loop effective action is 
\begin{equation}
\label{s1loopb}
\begin{aligned}
S_{\mathrm{1-loop}} &= \phantom{\Bigg(}\!\!\mathcal A_{7/7}(h) +\mathcal
M_{7/7}(h) + 
\mathcal{A}_{7/3}(h)
+\mathcal{A}_{3/7}(h)\\
&= -\frac{V_4}{8\pi^2} \Big[(4 - m)\,\tr \mathcal H^2 -4 \,(\tr \mathcal H)^2
\Big]
\int_0^\infty  \frac{d\tau_2}{2\tau_2} \,W(\tau_2)+ O(h^3)
\end{aligned}
\end{equation}
as reported in \eq{s1loop} of the main text.

\section{D-instanton sums: explicit results up to 3 instantons}
\label{app:3inst}
In this appendix we present the results of calculations up to $k=3$ instantons,
including finite $\epsilon_{1,2}$ (gravitational) and $\epsilon_{3,4}$ 
(anti-symmetric hyper-multiplet mass) corrections. 

According to \eq{Z}, the 1-instanton partition function is given by 
\begin{equation}
Z_1  = \frac{\epsilon_1+\epsilon_2}{\epsilon_1 \epsilon_2 }
\int \frac{d\chi_1}{2\pi\ii} \, 
\frac{1}{(4\chi_1^2 -\epsilon^2_3) (4\chi_1^2 -\epsilon^2_4)} \, \prod_{r=1}^{m}
\frac{ (\chi_1 +b_r)^2-\frac{(\epsilon_3-\epsilon_4)^2}{4}}{(\chi_1 -b_r)^2-
\frac{(\epsilon_1+\epsilon_2)^2}{4}}\, 
\prod_{u=1}^{4} (\chi_1-a_u)~.
 \label{Z1}
\end{equation}
The pole prescription is specified by
$\mathrm{Im} b_r=0$, and $\mathrm{Im}\epsilon_1\gg \mathrm{Im} \epsilon_2\gg
\mathrm{Im} \epsilon_3
\gg \mathrm{Im}\epsilon_4 >0$, and the integral is
computed by closing the contour in the upper half-plane, $\mathrm{Im} \chi_1>0$.
The poles contributing to the integral (\ref{Z1}) are located then at  
$\chi_1=b_r+\frac{\epsilon_1+\epsilon_2}{2}$ ($r=1,\ldots,m$),
$\chi_1=\frac{\epsilon_3}{2}$ and $\chi_1= \frac{\epsilon_4}{2}$.
$Z_1$ is then the sum of residues at these points. A simple inspection of this
formula shows that no dependence on $b_r$ 
arises at the leading $\frac{1}{\epsilon_1\epsilon_2 \epsilon_3 \epsilon_4}$
order. 
This can be seen by noticing that 
the $b$-dependent factors cancel between numerator and denominator at $\chi_1=
\frac{\epsilon_3}{2},\frac{\epsilon_4}{2} \approx 0$, and that the $b$-dependent
poles $\chi_1 \approx b_r$ contributes only to the $\frac{1}{\epsilon_1
\epsilon_2}$-order.

For higher instanton numbers one should perform the integrations over
$\chi_1,\ldots,\chi_k$ one after  the other, subsequently evaluating the
residues at the poles satisfying the above mentioned rules. 
Unfortunately, the problem becomes algebraically 
more and more complicated as $k$ increases, and we have been able 
to explicitly perform the integrations up to $k=3$ only.
Since our main interest is the pure U(4) gauge theory living on the D7
world-volume, when
there are also D3-branes, we only present the result of the calculations for
$b=0$. 

The result for $m=0$ can be written as
\begin{eqnarray}
 \epsilon_1 \epsilon_2\!\!\!\!&&\! \!\!\!\log Z^{(m=0)}(a,\epsilon)
= \Bigg(\frac{4a_1a_2a_3a_4}{\epsilon_3\epsilon_4}-\sum_{i<j}a_ia_j-
\frac{\epsilon_1+\epsilon_2}{2} \sum_{i}a_i
-\frac{\epsilon_3^2+\epsilon_3\epsilon_4+\epsilon_4^2}{4}\Bigg)\,q\nonumber\\
&+&\Bigg(-\frac{1}{\epsilon_3\epsilon_4}\sum _{i<j} a_i{}^2a_j{}^2+
 \frac{\epsilon_1^2+\epsilon_1\epsilon_2+\epsilon_2^2
}{4\epsilon_3\epsilon_4}\sum_ia_i^2-
 \frac{\big(\epsilon_1^2+\epsilon_1\epsilon_2+\epsilon_2^2\big)^2}{
16\epsilon_3\epsilon_4}
 \nonumber\\
&&~~~+\sum_{i<j}a_ia_j-\frac{1}{4}\sum_ia_i^2 
+\frac{\epsilon_1+\epsilon_2}{4} \sum_ia_i+
\frac{3\big(\epsilon_3^2+\epsilon_3\epsilon_4+\epsilon_4^2\big)}{8}\Bigg)\,q^2\\
&+&\Bigg(\frac{16 a_1a_2a_3a_4}{3\epsilon_3\epsilon_4}
 -\frac{4}{3}\sum_{i<j}a_ia_j-\frac{\epsilon_1+\epsilon_2}{3}
 \sum_{i}a_i-\frac{\epsilon_3^2+\epsilon_3\epsilon_4+\epsilon_4^2}{3}\Bigg)\,
q^3+\cdots
\nonumber
\label{z30}
\end{eqnarray}
while for $m=1$ and $b=0$ we find
\begin{eqnarray}
 \epsilon_1 \epsilon_2\!\!\!\!&&\!\!\!\!\log Z^{(m=1)}(a,\epsilon)
\!=\! \Bigg(\frac{4a_1a_2a_3a_4}{\epsilon_3\epsilon_4}+3\sum_{i<j}a_ia_j-
\frac{\epsilon_1+\epsilon_2}{2} \sum_{i}a_i
+\frac{3\big(\epsilon_3^2+\epsilon_3\epsilon_4+\epsilon_4^2\big)}{4}\Bigg)\,
q\nonumber\\
&+&\Bigg(-\frac{1}{\epsilon_3\epsilon_4}\sum _{i<j} a_i{}^2a_j{}^2+
 \frac{\epsilon_1^2+\epsilon_1\epsilon_2+\epsilon_2^2
}{4\epsilon_3\epsilon_4}\sum_ia_i^2-
 \frac{\big(\epsilon_1^2+\epsilon_1\epsilon_2+\epsilon_2^2\big)^2}{
16\epsilon_3\epsilon_4}
 \nonumber\\
&&+~\sum_{i<j}a_ia_j+\frac{7}{4}\sum_ia_i^2 
+\frac{\epsilon_1+\epsilon_2}{4} \sum_ia_i-
\frac{\epsilon_3^2+\epsilon_3\epsilon_4+\epsilon_4^2}{8}
-\frac{\epsilon_1^2+\epsilon_1\epsilon_2+\epsilon_2^2}{2}\Bigg)\,q^2\\
&+&\Bigg(\frac{16 a_1a_2a_3a_4}{3\epsilon_3\epsilon_4}
 +4\sum_{i<j}a_ia_j-\frac{\epsilon_1+\epsilon_2}{3}
 \sum_{i}a_i+\big(\epsilon_3^2+\epsilon_3\epsilon_4+\epsilon_4^2\big)\Bigg)\,
q^3+\cdots
\nonumber
\label{z31}
\end{eqnarray}
It is interesting to notice that the two formulas (\ref{z30}) and (\ref{z31}) 
share their divergent $\frac{1}{\epsilon_3 \epsilon_4}$ term. 
A more extensive inspection for finite non-zero $b$ shows that indeed the full
$b$-dependence
cancels in this term, so that in any case ${\mathcal F}_{\mathrm{IV}}$ is given
by
\begin{equation}
\label{f4g}
\begin{aligned}
{\mathcal F}_{\mathrm{IV}}=&\,
4\, a_1a_2a_3a_4\,q + \Bigg(-\sum _{i<j} a_i{}^2a_j{}^2+
\frac{1}{4} ( \epsilon_1^2+\epsilon_1\epsilon_2+\epsilon_2^2) -
\frac{1}{16} (\epsilon_1^2+\epsilon_1\epsilon_2+\epsilon_2^2)^2\Bigg)\,q^2\\
&~~~~~+\frac{16}{3}\, a_1a_2a_3a_4 \,q^3+\cdots
\end{aligned}
\end{equation}
This expression matches precisely 
with $\frac12{\mathcal F}_{\mathrm{SO(8)}}$, including its gravitational
corrections.
After the subtraction of the quartic terms, the resulting quadratic
prepotentials 
are then given by the formulas (\ref{F2b}) in the main text.

\section{Details on the heterotic computation}
\label{app:dethet}
In this appendix we provide some details on the calculations presented
in Section~\ref{sec:prephet} for the heterotic model.

In the partition function (\ref{z00}) and its weighted version
(\ref{weightz2}), 
the trace over the right-moving oscillators of the heterotic string
can be written as a spin-structure sum as follows
\begin{equation}
\rho\sp{0}{0}\!(w,\tau) =\frac{1}{2}\sum_{a,b=0}^1 \frac{\ee^{\ii\pi(a+b+a b)}}
{\eta^{12}} \,\vartheta\sp{a}{b}\!(2\tau_2w)
\,\vartheta\sp{a}{b}^3
=\frac{\vartheta_1(\tau_2w)^4}{\eta^{12}}~,
\label{char00}
\end{equation}
and 
\begin{equation}
\begin{aligned}
\rho\sp{h_0}{g_0}\!(w,\tau)
&=\frac{4}{2}\sum_{a,b=0}^1 \frac{\ee^{\ii\pi(a+b+a b+bh_0)}}
{\eta^{6}\,\vartheta\sp{1+h_0}{1+g_0}^2} 
\,\vartheta\sp{a}{b}\!(2\tau_2w)\,\vartheta\sp{a}{b}\,\vartheta\sp{a+h_0}{b+g_0}
^2\\
&=-4\,\ee^{\ii\pi h_0}\,
\frac{\vartheta_1(\tau_2w)^2\,\vartheta\sp{1+h_0}{1+g_0}\!(\tau_2w)^2}
{\eta^{6}\,\vartheta\sp{1+h_0}{1+g_0}^2}
\end{aligned}
\label{rhow}
\end{equation}
for $(g_0,h_0)\neq (0,0)$. Here we have used the standard conventions for the
$\vartheta$-functions summarised in App.~\ref{app:theta}, and in particular the
Riemann identity to perform the
summation over $a$ and $b$. The trace over the left-moving modes, instead, is
given by
\begin{equation}
\chi\sp{0\,h_1\,h_2}{0\,g_1\,g_2}\!(\vec v,\bar \tau)=\frac{1}{2}\sum_{a,b=0}^1
\frac{1}{\bar\eta^{24}}\, 
\bar\vartheta\sp{a}{b}\!(v_1)\,\bar\vartheta\sp{a}{b}\!(v_2)\,\bar\vartheta\sp{a
}{b}^2\,
\bar\vartheta\sp{a+h_1}{b+g_1}^4\,
\bar\vartheta\sp{a+h_2}{b+g_2}^4\,\bar\vartheta\sp{a+h_1+h_2}{b+g_1+g_2}^4
~,\label{chiw0}
\end{equation}
and
\begin{equation}
\begin{aligned}
\chi\sp{h_0\,h_1\,h_2}{g_0\,g_1\,g_2}\!(\vec v,\bar
\tau)&=\frac{4}{2}\sum_{a,b=0}^1
\frac{\ee^{\ii\pi(ag_0+bh_0-\frac{1}{2}g_0h_0})}{\bar\eta^{18}\,\bar\vartheta\sp
{1+h_0}{1+g_0}^2}\, 
\bar\vartheta\sp{a+\frac{h_0}{2}}{b+\frac{g_0}{2}}\!(v_1)\,
\bar\vartheta\sp{a+\frac{h_0}{2}}{b+\frac{g_0}{2}}\!(v_2)\,
\bar\vartheta\sp{a+\frac{h_0}{2}}{b+\frac{g_0}{2}}^2
\\&\hspace{60pt}\times\,
\bar\vartheta\sp{a+\frac{h_0}{2}+h_1}{b+\frac{g_0}{2}+g_1}^4\,
\bar\vartheta\sp{a+\frac{h_0}{2}+h_2}{b+\frac{g_0}{2}+g_2}^4\,
\bar\vartheta\sp{a+\frac{h_0}{2}+h_1+h_2}{b+\frac{g_0}{2}+g_1+g_2}^4
\end{aligned}
\label{chiw}
\end{equation}
for $(g_0,h_0)\neq (0,0)$. The factors of 4 in Eqs.~(\ref{rhow}) and
(\ref{chiw}) account for
the 16 fixed points of the $\Z_2$ orbifold action in our heterotic model. 
Notice also that for $(g_0,h_0)\neq (0,0)$ the left-moving amplitudes 
$\chi\sp{h_0\,h_1\,h_2}{g_0\,g_1\,g_2}$ involve $\Z_4$ 
$\vartheta$-functions with half-integer characteristics but, as we will see,
they can always be rewritten in terms of standard $\vartheta$-functions, as
expected for a $\Z_2$-orbifold.
Furthermore, the relative phases between the different structures in all these
formulas are
fixed by modular invariance up to discrete torsions 
that we have chosen to be zero for simplicity. In particular, one can show
the following modular transformation properties
\begin{equation}
\begin{aligned}
&\!\rho\sp{h_0}{g_0}\!(w,\tau+1)
\,\chi\sp{h_0\,h_1\,h_2}{g_0\,g_1\,g_2}\!(\vec v,\bar \tau+1)= 
\rho\sp{~~h_0\,}{g_0+h_0}\!(w,\tau)
\,\chi\sp{\,~h_0~~~~\,h_1~~~~~\,h_2}{g_0+h_0\,g_1+h_1\,g_2+h_2}\!(\vec v,\bar
\tau)~,\\
&\\
&\!\rho\sp{h_0}{g_0}\!(w,-{1}/{\tau})
\,\chi\sp{h_0\,h_1\,h_2}{g_0\,g_1\,g_2}\!(\vec v,-{1}/{\bar \tau})= 
|\tau|^{-8+4(g_0+h_0-h_0g_0)} \rho\sp{g_0}{h_0}\!(w,{\tau})
\,\chi\sp{g_0\,g_1\,g_2}{h_0\,h_1\,h_2}\!(\vec v,{\bar \tau})~.
\end{aligned}
\label{rhochimod}
\end{equation}

In the calculation of the massless spectrum of this heterotic compactification,
one needs
the explicit expressions for the left-moving functions of the type
$\chi\sp{h_0\,0\,\,\,0}{g_0\,g_1\,g_2}(\vec 0,\bar\tau)$. They are given by
\begin{equation}
\begin{aligned}
&\chi\sp{0\,\,0\,\,0}{0\,\,0\,\,0}(\vec 0,\bar\tau) = \frac{\bar\vartheta_3^{16}
+\bar \vartheta_4^{16}+ \bar\vartheta_2^8}{2 \bar \eta^{24}}~,\\
&\chi\sp{0\,\,0\,\,\,0}{0\,g_1\,g_2}(\vec 0,\bar\tau)= \frac{\bar\vartheta_3^{8}
\,\bar \vartheta_4^{8}}{\bar \eta^{24}}~~~~
\mbox{for}~(g_1,g_2)\neq(0,0)~,
\\
&\chi\sp{0\,\,0\,\,\,0}{1\,g_1\,g_2}(\vec 0,\bar\tau)= \frac{\bar\vartheta_3^{2}
\,\bar \vartheta_4^{2}}{\bar \eta^{24}}\,
\Big(\bar\vartheta\sp{\,\,0}{1/2}^{16}-\bar\vartheta\sp{\,\,1}{1/2}^{16}\Big)=
\frac{\bar\vartheta_3^{8}
\,\bar \vartheta_4^{8}}{2\bar \eta^{24}}\,
\big(\bar\vartheta_3^4+\bar\vartheta_4^4\big)~, 
\\
&\chi\sp{1\,\,0\,\,0}{0\,\,0\,\,0}(\vec 0,\bar\tau)= \frac{\bar\vartheta_2^{2}
\,\bar \vartheta_3^{2}}{\bar \eta^{24}}\,
\Big(\bar\vartheta\sp{1/2}{\,\,0\,}^{16}-\bar\vartheta\sp{1/2}{\,\,1\,}^{16}
\Big)=
\frac{\bar\vartheta_2^{8}
\,\bar \vartheta_3^{8}}{2\bar \eta^{24}}\,
\big(\bar\vartheta_2^4+\bar\vartheta_3^4\big)~, 
\\
&\chi\sp{0\,\,0\,\,0}{1\,\,0\,\,0}(\vec 0,\bar\tau)=
\ee^{-\frac{\ii\pi}{2}}\,\frac{\bar\vartheta_2^{2}
\,\bar \vartheta_4^{2}}{\bar \eta^{24}}\,
\Big(\bar\vartheta\sp{1/2}{1/2}^{16}-\bar\vartheta\sp{\,1/2}{-1/2}^{16}\Big)=
\frac{\bar\vartheta_2^{8}
\,\bar \vartheta_4^{8}}{2\bar \eta^{24}}\,
\big(\bar\vartheta_2^4+\bar\vartheta_4^4\big)~,\\
&\chi\sp{1\,\,0\,\,\,0}{0\,g_1\,g_2}(\vec
0,\bar\tau)=\chi\sp{1\,\,0\,\,\,0}{1\,g_1\,g_2}(\vec
0,\bar\tau)=0~~\phantom{\Big(}~
\mbox{for}~(g_1,g_2)\neq(0,0)~,
\end{aligned}
\label{chiapp}
\end{equation}
where we have repeatedly used the identities (\ref{der}) to rewrite the $\Z_4$
$\vartheta$-functions
in terms of the standard Jacobi $\vartheta$-functions.

Finally, in the partition function (\ref{z00}) and its weighted version
(\ref{weightz2})
the function $\Gamma\sp{h_0\,h_1\,h_2}{g_0\,g_1\,g_2}\!(\tau,\bar\tau)$
represents the contribution
of the bosonic zero-modes in the internal compact directions, which is given by
\begin{equation}
\begin{aligned}
&\Gamma\sp{0\,h_1\,h_2}{0\,g_1\,g_2}=\frac{1}{\tau_2^3}\,\Gamma_{4,4}~\Gamma_{2,
2}\sp{h_1\,h_2}{g_1\,g_2}~,\\
&\Gamma\sp{h_0\,h_1\,h_2}{g_0\,g_1\,g_2}=\frac{1}{\tau_2}\Gamma_{2,2}\sp{h_1\,
h_2}{g_1\,g_2} \quad\mbox{for}~
(g_0,h_0)\neq (0,0)~.
\end{aligned}
\label{zeromodes}
\end{equation}
Here $\Gamma_{4,4}$ is the standard lattice sum over $\cT_4$, 
while $\Gamma_{2,2}\sp{h_1\,h_2}{g_1\,g_2}$ is the lattice sum for the various
sectors of the $\Z_2\times\Z_2$ freely acting orbifold over $\cT_2$, namely
\begin{equation}
\Gamma_{2,2}\sp{h_1\,h_2}{g_1\,g_2}(\tau,\bar\tau;T,U) =T_2
\sum_{ M } \ee^{2\pi \ii T \,\mathrm{det} M}\, \ee^{-
\frac{\pi T_2}{\alpha'\tau_2 U_2}\,|( 1  \, U)\,      M \,(_{-1}^\tau) |^2}
\label{momwin0}
\end{equation}
with the sum running over four integers specifying the wrapping numbers over
$\cT_2$ according to
\begin{equation}
M = \begin{pmatrix}
       w^1 + \frac{h_1}{2} & m^1 + \frac{g_1}{2} \\
       w^2 + \frac{h_2}{2} & m^2 +\frac{g_2}{2}
      \end{pmatrix} ~,~~~(m^i,w^i)\in\Z^2~.
\label{momwin}
\end{equation}

In the calculation of the 1-loop thresholds, the second derivatives of the 
$\chi\sp{h_0\,h_1\,h_2}{g_0\,g_1\,g_2}$ functions are needed. In particular, in
\eq{DeltaIJ0}
we have used the following expressions
\begin{equation}
\begin{aligned}
\!\!\!\chi_{11}(\bar
\tau)&=\frac{1}{8\pi^2}\,\partial^2_{v_1}\chi\sp{0\,0\,0}{1\,0\,0}\!(\vec 0,\bar
\tau)=
\frac{1}{8\pi^2}\frac{\bar\vartheta_3^2 \bar\vartheta_4^2}{\bar\eta^{24}} 
 \Big(\bar\vartheta\sp{\,\,0}{1/2}''\, 
\bar\vartheta\sp{\,\,0}{1/2}^{15}
-  \bar\vartheta\sp{\,\,1}{1/2}''\, 
\bar\vartheta\sp{\,\,1}{1/2}^{15}\Big)
~,\\
\!\!\!\chi_{12}(\bar \tau)&=\frac{1}{8\pi^2}\,\partial_{v_1}\partial_{v_2}
\chi\sp{0\,0\,0}{1\,0\,0}\!(\vec 0,\bar \tau)=
\frac{1}{8\pi^2}\frac{\bar\vartheta_3^2 \bar\vartheta_4^2}{\bar\eta^{24}} 
 \Big(\bar\vartheta\sp{\,\,0}{1/2}'^{\,2}\, 
\bar\vartheta\sp{\,\,0}{1/2}^{14}
-  \bar\vartheta\sp{\,\,1}{1/2}'^{\,2}\, 
\bar\vartheta\sp{\,\,1}{1/2}^{14}\Big) ~,
\end{aligned}
\label{chi1112d}
\end{equation}
while in \eq{DeltaIJh} we have introduced
\begin{equation}
\begin{aligned}
\chi_{11}^{~(h_1\,h_2)}(\bar \tau)
&=\frac{1}{8\pi^2}\,\partial^2_{v_1}\chi\sp{0\,h_1\,h_2}{1\,\,0\,\,\,0}\!(0,\bar
\tau)\\
&=
\frac{1}{8\pi^2}\frac{\bar\vartheta_3^2 \bar\vartheta_4^2}{\bar\eta^{24}} \,
\bar\vartheta\sp{\,\,0}{1/2}^7\, 
\bar\vartheta\sp{\,\,1}{1/2}^{7}\,
 \Big(\bar\vartheta\sp{\,\,0}{1/2}''\, 
\bar\vartheta\sp{\,\,1}{1/2}
-  \bar\vartheta\sp{\,\,1}{1/2}''\, 
\bar\vartheta\sp{\,\,0}{1/2}\Big)~,\\
\chi_{12}^{~(h_1\,h_2)}(\bar \tau)
&=\frac{1}{8\pi^2}\,\partial_{v_1}\partial_{v_2}\chi\sp{0\,h_1\,h_2}{1\,\,0\,\,\
,0}\!(0,\bar \tau)
\\
&=
\frac{1}{8\pi^2}\frac{\bar\vartheta_3^2 \bar\vartheta_4^2}{\bar\eta^{24}} \,
\bar\vartheta\sp{\,\,0}{1/2}^6\, 
\bar\vartheta\sp{\,\,1}{1/2}^{6}\,
 \Big(\bar\vartheta\sp{\,\,0}{1/2}'^{\,2}\, 
\bar\vartheta\sp{\,\,1}{1/2}^{2}
-  \bar\vartheta\sp{\,\,1}{1/2}'^{\,2}\, 
\bar\vartheta\sp{\,\,0}{1/2}^{2}\Big)
\end{aligned}
\label{chihd}
\end{equation}
for any $(h_1,h_2)\neq(0,0)$. 

Using the duplication formulas together with the modular properties of the
$\vartheta$-functions, 
and the relations collected in App.~\ref{app:theta}, one can show that
\begin{equation}
\begin{aligned}
&\chi_{11}\big(\ft{\bar\tau}{2}\big) =\frac{1}{24\,\bar\eta^{12}}\,\Big[
\bar\vartheta_3^8\big(\bar\vartheta_3^4+\bar\vartheta_4^4-2\hat E_2\big)
-\bar\vartheta_2^8\big(\bar\vartheta_2^4-\bar\vartheta_4^4-2\hat
E_2\big)\Big]~,\\
&\chi_{11}\big(\ft{\bar\tau+1}{2}\big)=-\frac{1}{24\,\bar\eta^{12}}\,\Big[
\bar\vartheta_4^8\big(\bar\vartheta_4^4+\bar\vartheta_3^4-2\hat E_2\big)
+\bar\vartheta_2^8\big(\bar\vartheta_2^4+\bar\vartheta_3^4+2\hat
E_2\big)\Big]~,\\
&\chi_{11}\big(\ft{-1}{2\bar\tau}\big)=\frac{1}{24\,\bar\eta^{12}}\,\Big[
\bar\vartheta_3^8\big(\bar\vartheta_3^4+\bar\vartheta_2^4+2\hat E_2\big)
-\bar\vartheta_4^8\big(\bar\vartheta_4^4-\bar\vartheta_2^4+2\hat
E_2\big)\Big]~.\\
\end{aligned}
\label{x11}
\end{equation}
Here, in the right-hand sides the $\vartheta$-functions are evaluated at
$\bar\tau$, and as customary
\cite{Kiritsis:1997hf,Kiritsis:2000zi} 
we have replaced the second Eisenstein series $E_2$ with the modular form $\hat
E_2= E_2-\frac{3}{\pi\tau_2}$. Likewise, we have
\begin{equation}
\chi_{12}\big(\ft{\bar\tau}{2}\big) = \chi_{12}\big(\ft{\bar\tau+1}{2}\big)=
\chi_{12}\big(\ft{-1}{2\bar\tau}\big) =\frac{1}{8\,\bar\eta^{12}}\,
\bar\vartheta_2^4\,\bar\vartheta_3^4\,\bar\vartheta_4^4  = 2~.
\label{x12}
\end{equation}
{From} these expressions, the Hecke transforms of $\chi_{IJ}$ read
\begin{equation}
\begin{aligned}
c_{11}(\bar\tau)=&
 \left[\chi_{11}\Big(\ft{\bar\tau}{2} \Big)+\chi_{11}\Big(\ft{\bar\tau+1}{2}
 \Big)+\chi_{11}\Big(\ft{-1}{2\bar\tau} \Big)\right]=6~,\\
c_{12}(\bar\tau)=&
 \left[\chi_{12}\Big(\ft{\bar\tau}{2} \Big)+\chi_{12}\Big(\ft{\bar\tau+1}{2}
 \Big)+\chi_{12}\Big(\ft{-1}{2\bar\tau} \Big)\right]=6~.
\end{aligned}
\label{c1112}
\end{equation}
Finally, from \eq{chihd} and the $\vartheta$-functions properties we have
\begin{equation}
\begin{aligned}
 \chi_{11}^{~(h_1\,h_2)}(\bar \tau)&=
\frac{1}{128\,\bar\eta^{24}}\,\bar\vartheta_2^8\,\bar\vartheta_3^8\,
\bar\vartheta_4^8= 2~,\\
\chi_{12}^{~(h_1\,h_2)}(\bar \tau)&=
-\frac{1}{128\,\bar\eta^{24}}\,\bar\vartheta_2^8\,\bar\vartheta_3^8\,
\bar\vartheta_4^8= -2~,
\end{aligned}
\label{chiresd}
\end{equation}
as reported in \eq{chihres} of the main text.

\section{Holomorphic couplings}
\label{app:hol}
Here we briefly review the relation between string thresholds and holomorphic
gauge couplings required by $\cN\ge 1$ supersymmetry in four dimensions
following 
Refs.~\cite{Dixon:1990pc,Kaplunovsky:1995jw}; our model exhibits
$\cN=2$ supersymmetry, and in the main text we focused directly on this case,
but we find 
convenient to recap here how the $\cN=2$ relations arise from the general
$\cN=1$ case.

In a supersymmetric string vacuum, the 1-loop
gauge coupling constant $g(\mu)$ for a gauge group $G$
is given by
\begin{equation}
\left.\frac{4\pi}{g^2(\mu)}\right|_{\mathrm{1-loop}}= \frac{4\pi}{g_0^2} +
\frac{1}{4\pi}\,\Bigg[ b\, \log\Big(\frac{4\pi\mu^2}{g_0^2 M_{\mathrm{Pl}}^2}
\Big)+\Delta\Bigg] 
\label{gstringd}
\end{equation}
where $g_0$ is the bare tree-level coupling, $\Delta$ is the threshold
correction and $b$ is the 1-loop
$\beta$-function coefficient given by
\begin{equation}
b= 3\, T(G) - \sum_r  \widetilde{n}_{r}\, T(r)~.
\label{betac}
\end{equation}
Here, $\widetilde{n}_{r}$ is the number of $\mathcal N=1$ chiral multiplets 
transforming in the representation $r$ of $G$ whose index is denoted by $T(r)$,
with $T(G)$ standing for $T(\mathrm{adj})$ (see \eq{indexT}). Formula
(\ref{gstringd})
applies to any supersymmetric string vacuum, either heterotic or type II or type
I.

On the other hand, as explained for example in Ref.~\cite{Kaplunovsky:1995jw},
the general form of the 1-loop coupling constant required by supersymmetry in
the four-dimensional effective quantum field theory is
\begin{equation}
\begin{aligned}
\left. \frac{4\pi}{g^2(\mu)}\right|_{\mathrm{1-loop}}=& \re f_{(0)}
+\frac{1}{4\pi}\,\Bigg[
\re f_{(1)} +b \log\Big(\frac{\mu^2}{M_{\mathrm{Pl}}^2}\Big) +
2 T(G) \log \big( \re f_{(0)} \big)
\\& 
\hspace{75pt}-c \,K- 2\sum_r \widetilde{n}_{r}\,T(r)\,\log Z_r\Bigg]
\end{aligned}
\label{geqftall}
\end{equation}
where $f_{(0)}$ and $f_{(1)}$ are, respectively, the tree-level and 1-loop
contributions 
to the Wilsonian holomorphic gauge coupling function $f$, $K$ is the tree-level 
K\"ahler potential, and $Z_r$ is the tree-level K\"ahler metric for the
chiral superfield in the representation $r$. Finally, the coefficient $c$ is
defined as
\begin{equation}
c= T(G) - \sum_r \widetilde{n}_{r} \,T(r) = b - 2 \,T(G)~.
\label{ca}
\end{equation}

In comparing Eqs.~(\ref{gstringd}) and (\ref{geqftall}) we must take into
account the fact
that the tree-level coupling obtained from string theory does not necessarily
coincide
with $ \re f_{(0)} $ since there may be 1-loop effects that spoil this
identification
\cite{Kaplunovsky:1995jw}. Indeed, in general we have 
\begin{equation}
\frac{4\pi}{g_0^2}=  \re f_{(0)} + \frac{1}{4\pi}\Delta_{\mathrm{univ}}(M,\bar
M)
\label{gfd}
\end{equation}
where $\Delta_{\mathrm{univ}}(M, \bar M)$ is a universal term related to the
1-loop correction of the K\"ahler potential, which is a (real) function of the
compactification
moduli $M$ (different from those parametrizing the gauge coupling) that can mix
with the
dilaton. Using this information and equating the string theory expression
(\ref{gstringd})
with the field theory one (\ref{geqftall}), we easily obtain
\begin{equation}
\re f_{(1)} = \Delta + \Delta_{\mathrm{univ}}
+c\, \log\big( \re f_{(0)} \big) + c\, K  +2\sum_r \widetilde{n}_r\,T(r)\,\log
Z_r~.
\label{fa1}
\end{equation}
The terms in the right hand side contain non-holomorphic pieces 
that, for consistency, must compensate each other in order to yield a
holomorphic result for 
$f_{(1)}$.

When the $\mathcal N=1$ chiral multiplets organize into
$\mathcal N=2$ hyper-multiplets, like in our case, 
it is possible to show that \cite{deWit:1995zg}
\begin{equation}
c\, \log\big( \re f_{(0)} \big) + c\, K  +2\sum_r \widetilde{n}_r\,T(r)\,\log
Z_r
= b\,\widehat K~,
\label{relnew1}
\end{equation}
where $\widehat K$ is related to the K\"ahler metric of the adjoint chiral
multiplet
which becomes part of the $\cN=2$ vector multiplet according to  
$\widehat K  = \log Z_{\mathrm{adj}}$. Thus, for $\cN=2$ vacua
the relation (\ref{fa1}) reduces to
\begin{equation}
\re f_{(1)} = \Delta+\Delta_{\mathrm{univ}} + b\, \widehat K ~.
\label{fa12}
\end{equation}

\eq{relnew1} can be explicitly verified in our specific type I$^\prime$ and
heterotic
setups. To this purpose let us recall that the various quantities to be used
are:
\begin{itemize}
\item in type I$^\prime$
\begin{equation}
\begin{aligned}
 &\re f_{(0)}=t_2~~~,~~~\widehat K =-\log(\lambda_2 U_2)~~~,~~~c=b-8=-8~,\\
&Z_{\Yfund} =(t_2^{(1)} U_2^{(1)}t_2^{(2)}
U_2^{(2)})^{-1/2}~~~,\phantom{\vdots}~~
Z_{\mathrm{adj}} =(\lambda_2U_2)^{-1}~,\\
&Z_{\Yasymm_1}=Z_{\overline{\Yasymm}_1} =(t_2^{(1)} U_2^{(1)})^{-1}~~~,~~~
Z_{\Yasymm_2}=Z_{\overline{\Yasymm}_2} =(t_2^{(2)} U_2^{(2)})^{-1}~,\\
&\widetilde{n}_{\mathrm{adj}} = \widetilde{n}_{{\Yasymm}_{1,2}} =
\widetilde{n}_{\overline{\Yasymm}_{1,2}}=1~~~,~~~\widetilde{n}_{\Yfund}=8~;
\end{aligned}
\label{ingrI}
\end{equation}
\item in heterotic
\begin{equation}
\begin{aligned}
~~~~~~~~~ &\re f_{(0)}=S_2~~~,~~~\widehat K =-\log(T_2 U_2)~~~,~~~c=b-8=-8~,\\
~~~~~~~~~&Z_{\Yfund} =(T_2^{(1)} U_2^{(1)}T_2^{(2)}
U_2^{(2)})^{-1/2}~~~,\phantom{\vdots}~~
Z_{\mathrm{adj}} =(T_2U_2)^{-1}~,\\
~~~~~~~~~&Z_{\Yasymm_1}=Z_{\overline{\Yasymm}_1} =(T_2^{(1)}
U_2^{(1)})^{-1}~~~,~~~
Z_{\Yasymm_2}=Z_{\overline{\Yasymm}_2} =(T_2^{(2)} U_2^{(2)})^{-1}~,\\
~~~~~~~~~&\widetilde{n}_{\mathrm{adj}} = \widetilde{n}_{{\Yasymm}_{1,2}} =
\widetilde{n}_{\overline{\Yasymm}_{1,2}}=1~~~,~~~\widetilde{n}_{\Yfund}=8~.
\end{aligned}
\label{ingrhet}
\end{equation}
\end{itemize}
The K\"ahler metrics for the various multiplets in the type I$^\prime$ model
written in
\eq{ingrI} can be deduced from those reported for example in 
Refs.~\cite{Lust:2004fi,Bertolini:2005qh,Billo:2007sw}; those for the heterotic
model
written in \eq{ingrhet} can be obtained upon replacing the type I$^\prime$
variables with the
corresponding heterotic ones, or can be deduced from results existing in the
literature,
for example in Refs.~\cite{Dixon:1990pc,Kaplunovsky:1995jw,Antoniadis:1996vw}.

\section{Theta functions}
\label{app:theta}
In this Appendix we collect some useful formulas on the Jacobi
$\vartheta$-functions and the
Dedekind $\eta$-function.
We adopt the standard definitions
\begin{equation}
 \vartheta\sp{a}{b}(v|\t)=\sum_{n\in \Z} q^{\frac{1}{2}(n-\frac{a}{2})^2}
\, \ee^{2\p \ii (n-\frac{a}{2})(v-\frac{b}{2})}~,~~~
\eta(q)= q^\frac{1}{24} \prod_{n=1}^\infty (1-q^n)
\label{th}
\end{equation}
with $q=\ee^{2\pi\ii\tau}$. We also take $\vartheta_1\equiv\vartheta\sp{1}{1}$,
$\vartheta_2\equiv\vartheta\sp{1}{0}$,
$\vartheta_3\equiv\vartheta\sp{0}{0}$, $\vartheta_4\equiv\vartheta\sp{0}{1}$.
Most of the times,
when we do not write explicitly their arguments, we understand that the
$\vartheta$-functions 
are evaluated at $v=0$ with modular parameter $\tau$. 
Sometimes, we indicate only its first argument $v$. When ambiguities may arise, 
we write explicitly both arguments. In the calculations we also need the
properties of the
second Eisenstein series $E_2$ defined by
\begin{equation}
E_2(\tau) = \frac{12}{\ii\pi}\,\partial_\tau\log \eta(\tau) =
1-24\sum_{n=1}^\infty\frac{nq^n}{1-q^n}~.
\label{eisen2}
\end{equation}
$E_2$ does not enjoy nice modular properties, but $\hat E_2= E_2
-\frac{3}{\pi\tau_2}$ is a modular
form of weight 2, even if it is not holomorphic.

\begin{itemize}
\item Jacobi/Riemann identities and other relations:
\begin{equation}
\begin{aligned}
& \vartheta_3^4-\vartheta_2^4-\vartheta_4^2=0~,~~~
\vartheta_3^{12}-\vartheta_2^{12}-\vartheta_4^{12}=48 \,\eta^{12}~,~~~
\vartheta_2 \,\vartheta_3\, \vartheta_4=2\,\eta^3\phantom{\vdots},\\
&\frac12 \sum_{a,b} (-1)^{a+b+ab}   \vartheta\sp{a}{b}(w)\, 
\vartheta\sp{a}{b}\,\vartheta\sp{a+h}{b+g}(w)^2=
-\vartheta_1\big(\ft{w}{2}\big)^2 \,
\vartheta\sp{1+h}{1+g}\big(\ft{w}{2}\big)^2 ~.
\end{aligned}
\label{riemann}
\end{equation}
\item Duplication formulas:
\begin{equation}
\begin{aligned}
&  \vartheta_2(0|2\tau)=\sqrt{\frac{\vartheta_3^2-\vartheta_4^2}{2}}~,~~~
\vartheta_4(0|2\tau)=\sqrt{\vartheta_3\,\vartheta_4}~,\\
&  \vartheta_3(0|2\tau)=\sqrt{\frac{\vartheta_3^2+\vartheta_4^2}{2}}~,~~~
\eta(2\tau) = \sqrt{\frac{\vartheta_3\,\eta}{2}}~,\\
&E_2(2\tau)= \frac{1}{4}\big(2E_2+\vartheta_3^4+\vartheta_4^4\big)~,\\
& \vartheta_3\big(0|\ft{\tau}{2}\big)=\sqrt{\vartheta_3^2+\vartheta_2^2}~,~~~
\vartheta_2\big(0|\ft{\tau}{2}\big)=\sqrt{2\,\vartheta_2\,\vartheta_3}~,\\
& \vartheta_4\big(0|\ft{\tau}{2}\big)=\sqrt{\vartheta_3^2-\vartheta_2^2}~,~~~
\eta\big(\ft{\tau}{2}\big)=\sqrt{\vartheta_4\,\eta}~,\\
&E_2\big(\ft{\tau}{2}\big)=2E_2-\vartheta_2^4-\vartheta_3^4\phantom{\vdots}.
\end{aligned}
\label{dup}
\end{equation}
\item Modular transformations:
\begin{equation}
\begin{aligned}
&T~:~~~  \vartheta\sp{a}{b}(v|\t+1)= \ee^{\frac{\ii\pi}{4}a(2-a)}
\vartheta\sp{~~a~}{a+b-1}(v|\t)~,~~~\eta(\tau+1) =
\ee^{\frac{\ii\pi}{12}}\,\eta(\tau)~,\\
&S~:~~~  \vartheta\sp{a}{b}\big(\ft{v}{\t}\big|\ft{-1}{\t}\big)= \sqrt{-\ii\t}\,
\ee^{\frac{\ii\pi}{2}(ab+\frac{2v^2}{\t})}
\,\vartheta\sp{~b~}{-a}(v|\t)~,~~~\eta\big(\ft{-1}{\t}\big) =
\sqrt{-\ii\t}\,\eta(\tau)~,
\end{aligned}
\label{modtran}
\end{equation}
which imply
\begin{equation}
\begin{aligned}
&T~:~~~~ \vartheta_3 \leftrightarrow \theta_4 ~,~~~ \vartheta_2 \to 
\ee^{\frac{\ii\pi}{4}}\, \vartheta_2~,~~~\hat E_2 \to  \hat E_2~,\\
&S~:~~~~ \frac{\vartheta_2}{\eta}  \leftrightarrow 
\frac{\vartheta_4}{\eta}~,~~~  
\frac{\vartheta_3}{\eta}  \to \frac{\vartheta_3}{\eta}~,~~~\hat E_2 \to \tau^2\,
\hat E_2~.
\end{aligned}
\label{modtran1}
\end{equation}
\item $\Z_4$ $\vartheta$-functions:
\begin{equation}
\begin{aligned}
 \vartheta\sp{\,~0}{\pm 1/2}(v|\tau) &= \vartheta_1(2v|4\tau)+
\vartheta_4(2v|4\tau)~,\\
\vartheta\sp{\,~1}{\pm 1/2}(v|\tau)& = \frac{\ee^{-\frac{\ii
\pi}{8}}}{\sqrt{2}} 
\Big[ \vartheta_1\big(v|\tau+\ft12\big)+
\vartheta_2\big(v|\tau+\ft12\big) \Big]~.
\end{aligned}
\label{theta4}
\end{equation}
\item Derivatives of $\vartheta$-functions (the $'$ denotes a derivative with
respect to the first argument of the $\vartheta$-functions):
\begin{equation}
\begin{aligned}
&\frac{\vartheta''_2}{\vartheta_2}=-\frac{\pi^2}{3} \,\big(E_2+
\vartheta_3^4+\vartheta_4^4\big)
~,~~~\frac{\vartheta''_3}{\vartheta_3}=-\frac{\pi^2}{3} \,\big(E_2+
\vartheta_2^4-\vartheta_4^4\big)
~,\\
&\frac{\vartheta''_4}{\vartheta_4}=-\frac{\pi^2}{3} \,\big(E_2-
\vartheta_2^4-\vartheta_3^4\big)
~,~~~
\vartheta_1^\prime=2\pi \eta^3~,~~~\frac{\vartheta_1'''}{\vartheta_1'}=-\pi^2
E_2~.
\end{aligned}
\label{dtheta4}
\end{equation}
\item Useful identities (all $\vartheta$-functions have vanishing first
argument):
\begin{equation}
\begin{aligned}
&\vartheta\sp{\,~0}{\pm 1/2}^4=\frac12\,
\vartheta_3\,\vartheta_4\,\big(\vartheta_3^2+\vartheta_4^2\big)~,~~~
\vartheta\sp{\,~1}{\pm 1/2}^4=\frac12\,
\vartheta_3\,\vartheta_4\,\big(\vartheta_3^2-\vartheta_4^2\big)~,\\
& \frac{\vartheta\sp{\,~0}{\pm 1/2}''}{\vartheta\sp{\,~0}{\pm 1/2}}
=-\frac{\pi^2}{12}\,\big(\vartheta_3^4-6\vartheta_3^2\,
\vartheta_4^2+\vartheta_4^4 +4\hat E_2\big)~,\\
& \frac{\vartheta\sp{\,~1}{\pm 1/2}''}{\vartheta\sp{\,~1}{\pm 1/2}}
=-\frac{\pi^2}{12}\,\big(\vartheta_3^4+6\vartheta_3^2\,
\vartheta_4^2+\vartheta_4^4 +4\hat E_2\big)~,\\
& \frac{\vartheta\sp{\,~0}{\pm 1/2}'}{\vartheta\sp{\,~0}{\pm 1/2}}
=\frac{\pi}{2}\,\big(\vartheta_3^2+\vartheta_4^2\big)~,~~~
\frac{\vartheta\sp{\,~1}{\pm 1/2}'}{\vartheta\sp{\,~0}{\pm 1/2}}
=\frac{\pi}{2}\,\big(\vartheta_3^2-\vartheta_4^2\big)~,\\
&\vartheta\sp{\,~0}{\pm
1/2}^4\big(\ft{\tau}{2}\big)=\vartheta_4^2\,\vartheta_3^2~,~~~
\vartheta\sp{\,~1}{\pm
1/2}^4\big(\ft{\tau}{2}\big)=\vartheta_4^2\,\vartheta_2^2~,\\
&\frac{\vartheta\sp{\,~0}{\pm
1/2}''\big(\ft{\tau}{2}\big)}{\vartheta\sp{\,~0}{\pm 1/2}\big(\ft{\tau}{2}\big)}
=4\,\frac{\vartheta_4''(2\tau)}{\vartheta_4(2\tau)}=
\frac{\pi^2}{3}\,\big(\vartheta_3^4 +\vartheta_4^4 -2 E_2\big)~,\\
&\frac{\vartheta\sp{\,~1}{\pm
1/2}''\big(\ft{\tau}{2}\big)}{\vartheta\sp{\,~1}{\pm
1/2}\big(\ft{\tau}{2}\big)} 
=\frac{\vartheta_2''\big(\ft{\tau+1}{2}\big)}{\vartheta_2\big(\ft{\tau+1}{2}
\big)}=
\frac{\pi^2}{3}\,\big(\vartheta_2^4 -\vartheta_4^4 -2 E_2\big)~,\\
&\frac{\vartheta\sp{\,~0}{\pm
1/2}'\big(\ft{\tau}{2}\big)}{\vartheta\sp{\,~0}{\pm 1/2}\big(\ft{\tau}{2}\big)}
= 2\,\frac{\vartheta_1'(2\tau)}{\vartheta_4(2\tau)}
=\pi \vartheta_2^2~,~~~
\frac{\vartheta\sp{\,~1}{\pm 1/2}'\big(\ft{\tau}{2}\big)}{\vartheta\sp{\,~1}{\pm
1/2}\big(\ft{\tau}{2}\big)} =
\frac{\vartheta_1'\big(\ft{\tau+1}{2}\big)}{\vartheta_2\big(\ft{\tau+1}{2}\big)}
=\pi \vartheta_3^2~.
\end{aligned}
\label{der}
\end{equation}
\end{itemize}

\providecommand{\href}[2]{#2}\begingroup\raggedright

%
\end{document}